\documentclass[plaindraft]{jpp-AAS}

\usepackage{graphicx}
\usepackage{amsmath}
\usepackage{upgreek}
\usepackage{cancel}
\usepackage{natbib}
\usepackage[pdftex,colorlinks,citecolor=blue]{hyperref}
\usepackage{cleveref}

\ifCUPmtlplainloaded \else
  \checkfont{eurm10}
  \iffontfound
    \IfFileExists{upmath.sty}
      {\typeout{^^JFound AMS Euler Roman fonts on the system,
                   using the 'upmath' package.^^J}%
       \usepackage{upmath}}
      {\typeout{^^JFound AMS Euler Roman fonts on the system, but you
                   dont seem to have the}%
       \typeout{'upmath' package installed. JPP.cls can take advantage
                 of these fonts, if you use 'upmath' package.^^J}%
       \providecommand\upi{\pi}%
      }
  \else
    \providecommand\upi{\pi}%
  \fi
\fi

\ifCUPmtlplainloaded \else
  \checkfont{msam10}
  \iffontfound
    \IfFileExists{amssymb.sty}
      {\typeout{^^JFound AMS Symbol fonts on the system, using the
                'amssymb' package.^^J}%
       \usepackage{amssymb}%
       \let\le=\leqslant  
       \let\ge=\geqslant  
      }{}
  \fi
\fi

\ifCUPmtlplainloaded \else
  \IfFileExists{amsbsy.sty}
    {\typeout{^^JFound the 'amsbsy' package on the system, using it.^^J}%
     \usepackage{amsbsy}}
    {\providecommand\boldsymbol[1]{\mbox{\boldmath $##1$}}}
\fi

\newcommand{\pD}[2]{\frac{\partial #2}{\partial #1}}

\newcommand{\D}[2]{\frac{{\rm d} #2}{{\rm d} #1}}

\newcommand{\od}{\operatorname{d}\!}
\newcommand\bs[1]{{\boldsymbol{#1}}}
\newcommand\bb[1]{{\boldsymbol{#1}}}
\newcommand\grad{\bb{\nabla}}
\renewcommand\bcdot{\,\bb{\cdot}\,}
\newcommand{\bscdot}{\,\bs{\cdot}\,}
\newcommand{\btimes}{\,\bb{\times}\,}
\newcommand{\bstimes}{\bs{\times}}
\newcommand{\mc}[1]{\mathcal{#1}}
\newcommand{\mr}[1]{\mathrm{#1}}

\newcommand{\msb}[1]{\bb{\mathsf{#1}}}

\newcommand{\dblbrck}[1]{[#1]}
\newcommand{\ansatz}{ansatz}

\newcommand{\const}{{\rm const}}
\newcommand{\imag}{{\rm i}}
\newcommand{\rmd}{{\rm d}}
\newcommand{\rme}{{\rm e}}
\newcommand{\eb}{\bb{\hat{b}}}
\newcommand{\ez}{\bb{\hat{z}}}
\newcommand{\ex}{\bb{\hat{x}}}

\newcommand{\eig}{\bb{\hat{e}}}

\newcommand{\ethree}{\boldsymbol{\hat{e}}_3}
\newcommand{\etwo}{\boldsymbol{\hat{e}}_2}
\newcommand{\eone}{\boldsymbol{\hat{e}}_1}

\newcommand{\vth}[1]{v_{{\rm th {#1}}}}
\newcommand{\ROS}{\eb\eb\, \bb{:}\, \grad \bb{u}}
\newcommand{\drive}{\widetilde{\bb{f}}}
\newcommand{\Reeff}{\mathrm{Re}_{\parallel \mathrm{eff}}}
\newcommand{\Reprl}{\mathrm{Re}_\parallel}

\newcommand{\Deltap}{{\Updelta p}}
\newcommand{\tcorrf}{{t_\mathrm{corr,f}}}
\newcommand{\visc}{\mu}

\newcommand{\Omegai}{\ensuremath{\mathit{\Omega}_\mr{i}}}
\newcommand{\vthi}{\ensuremath{v_\mr{thi}}}

\newcommand{\rhoio}{\ensuremath{\rho_\mr{i0}}}
\newcommand{\Omegaio}{\ensuremath{\mathit{\Omega}_\mr{i0}}}

\newcommand{\rndb}{\tilde{{b}}}
\newcommand{\nrndb}{{b}}
\newcommand{\rndB}{\tilde{B}}
\newcommand{\rndk}{\tilde{k}}
\newcommand{\rndu}{\tilde{u}}
\newcommand{\rnds}{\tilde{\sigma}}
\newcommand{\ea}[1]{\overline{#1}}

\newcommand{\Punorm}{\widehat{P}}

\DeclareMathAlphabet{\mathsfbi}{OT1}{\sfdefault}{bx}{sl}
\DeclareMathVersion{sfletters}
\SetSymbolFont{letters}{sfletters}{OML}{ntxsfmi}{b}{it}

\defcitealias{StOnge_2017}{Paper~I}

\title[Fluctuation dynamo in a weakly collisional plasma]{Fluctuation dynamo in a weakly collisional plasma}

\author[D.~A.~St-Onge and others]%
{D.~A.~St-Onge\ls$^{1,2,4}$
  \thanks{Email address for correspondence: denis.st-onge@physics.ox.ac.uk} 
, M.~W.~Kunz$^{1,2}$, J.~Squire$^{3}$ and A.~A.~Schekochihin$^{4,5}$}

\affiliation{$^1$Department of Astrophysical Sciences, Princeton University, Peyton Hall, Princeton, NJ 08544, USA\\[\affilskip]
$^2$Princeton Plasma Physics Laboratory, PO Box 451, Princeton, NJ 08543, USA\\[\affilskip]
$^3$Department of Physics, University of Otago, 730 Cumberland St, North Dunedin, Dunedin 9016, New Zealand\\[\affilskip]
$^4$The Rudolf Peierls Centre for Theoretical Physics, University of Oxford, Clarendon
Laboratory, Parks Road, Oxford, OX1 3PU, UK\\[\affilskip]
$^5$Merton College, Oxford OX1 4JD, UK}

\pubyear{2020}
\volume{}
\pagerange{}
\date{\today}

\begin{document}

\maketitle

\begin{abstract}
The turbulent amplification of cosmic magnetic fields depends upon the material properties of the host plasma. In many hot, dilute astrophysical systems, such as the intracluster medium (ICM) of galaxy clusters, the rarity of particle--particle collisions allows departures from local thermodynamic equilibrium. These departures -- pressure anisotropies -- exert anisotropic viscous stresses on the plasma motions that inhibit their ability to stretch magnetic-field lines. We present an extensive numerical study of the fluctuation dynamo in a weakly collisional plasma using magnetohydrodynamic (MHD) equations endowed with a field-parallel viscous (Braginskii) stress. When the stress is limited to values consistent with a pressure anisotropy regulated by firehose and mirror instabilities, the Braginskii-MHD dynamo largely resembles its MHD counterpart, particularly when the magnetic field is dynamically weak. If instead the parallel viscous stress is left unabated -- a situation relevant to recent kinetic simulations of the fluctuation dynamo and, we argue, to the early stages of the dynamo in a magnetized ICM -- the dynamo changes its character, amplifying the magnetic field while exhibiting many characteristics reminiscent of the saturated state of the large-Prandtl-number (${\rm Pm}\gtrsim{1}$) MHD dynamo. We construct an analytic model for the Braginskii-MHD dynamo in this regime, which successfully matches simulated dynamo growth rates and magnetic-energy spectra. A prediction of this model, confirmed by our numerical simulations, is that a Braginskii-MHD plasma without pressure-anisotropy limiters will not support a dynamo if the ratio of perpendicular and parallel viscosities is too small. This ratio reflects the relative allowed rates of field-line stretching and mixing, the latter of which promotes resistive dissipation of the magnetic field. In all cases that do exhibit a viable dynamo, the generated magnetic field is organized into folds that persist into the saturated state and bias the chaotic flow to acquire a scale-dependent spectral anisotropy.
\end{abstract}

%
%
\section{Introduction}

An outstanding problem in the study of astrophysical plasmas is the origin and sustenance of dynamically important magnetic fields found throughout the Universe. It is believed that, for systems without any large-scale features such as rotational shear or net helicity, weak seed fields are amplified and sustained by the fluctuation (or small-scale) dynamo \citep{KulsrudZweibel}, in which a random turbulent flow amplifies a weak seed magnetic field to near-equipartition energies \citep{Batchelor,Zeldovich,Childressgilbert,Tobias2019,Rincon_2019}. This process is thought to be particularly important in the intracluster medium (ICM), a hot (temperatures $T \sim 1$--$10$ keV) and diffuse (densities $n \sim 10^{-4}$--$10^{-2}$ cm$^{-3}$) ionized plasma that sits in the dark-matter-dominated gravitational potential of a galaxy cluster \citep{fabian94}. Observations of Faraday rotation and synchrotron emission \citep{Carilli02,Bonafede,Govoni2004} show that intracluster magnetic fields have strengths ${\sim}\upmu$G, with an energy density that is comparable to that of the bulk turbulent motions. The latter have recently been directly measured in the Perseus cluster by the \citet{Hitomi1}. As the ICM is not known to contain a significant mean field or net helicity, it is generally assumed that the present intracluster magnetic field is a result of the fluctuation dynamo. 

Previous theoretical work on the fluctuation dynamo has typically adopted a collisional MHD description in ${\rm Pm}\gtrsim{1}$ (`high-Pm') limit, where Pm is the ratio of the plasma viscosity and magnetic diffusivity. In these situations, the shearing motions of the turbulent background organize the magnetic field into folded sheets, whose lengths are comparable to the viscous scale and whose reversal scale is controlled by the magnetic diffusivity \citep{Scheko_sim,maron04}. This results in a magnetic field whose energy is concentrated near the resistive scale $k_\eta^{-1}$ \citep{Kazantsev,Kulsrud1992}, which is ${\sim}10^4~{\rm km}$ in the ICM assuming Spitzer values \citep[see table 1 of][]{SchekoCowley06b}. By contrast, observations suggest that the energy spectra of intracluster magnetic fields peaks on scales ${\sim}1$--$10~{\rm kpc}$ 
\citep{Murgia04,Vogt05,vacca12,govoni17}. While it has been proposed that, during the nonlinear phase of the dynamo, the magnetic energy would migrate to larger scales \citep{Kulsrud1997,Subramanian1998} -- thus potentially becoming compatible with ICM observations -- this scenario has yet to be borne out in simulations.

The solution to this discrepancy may be related the fact that the ICM is not rigourously a magnetohydrodynamic (MHD) fluid \citep{Scheko_2005,KulsrudZweibel}. First, the ion--ion Coulomb collision frequency
\begin{equation}
    \nu_\mr{i} \approx 0.2 \,\biggl(\frac{n}{10^{-3}~{\rm cm}^{-3}}\biggr) \biggl(\frac{T}{5~{\rm keV}}\biggr)^{-3/2}~{\rm Myr}^{-1} 
\end{equation} 
is only a factor of ${\sim}100$ larger than the inverse dynamical time of the turbulent fluid motions at the largest scales,\footnote{\,Here we have normalized $n$ and $T$ to typical cluster values, for which the Coulomb logarithm ${\approx}40$.  The representative values given for the outer scale $\ell_0$ and its characteristic turbulent velocity $u_0$ are motivated by a variety of observational constraints on gas motions in nearby clusters \citep[e.g.][]{Hitomi1,Zhuravleva2018,Simionescu2019}.}
\begin{equation}
    t^{-1}_{\rm dyn} \approx 0.002 \,\biggl(\frac{u_0}{200~{\rm km~s}^{-1}}\biggr) \biggl(\frac{\ell_0}{100~{\rm kpc}}\biggr)^{-1}~{\rm Myr}^{-1} .
\end{equation}
Thus, ${\sim}1\%$ deviations in the particle distribution function from local thermodynamic equilibrium are to be expected -- i.e., the ICM plasma is {\em weakly collisional}. That the energy density of these deviations is comparable to that stored in the observed turbulent motions and magnetic-field fluctuations indicates that ${\sim}1\%$, while small, is nevertheless enough to be dynamically important. Secondly, already magnetic-field strengths as small as
\begin{equation}\label{eqn:Bmin}
    B \gtrsim 10^{-18} \,\biggl(\frac{n}{10^{-3}\textrm{ cm}^{-3}}\biggr)\biggl(\frac{T}{5\textrm{ keV}}\biggr)^{-3/2}~{\rm G}
\end{equation}
are sufficient to ensure that the ICM plasma is {\em magnetized} -- i.e., that the ion gyrofrequency $\Omegai \doteq eB/m_\mr{i}c$ is larger than $\nu_\mr{i}$. As seed magnetic fields are thought to be produced by various processes in the era preceding galaxy formation with magnitudes ${\sim}10^{-22}$--$10^{-19}~{\rm G}$ \citep[e.g.][]{Biermann1950,KulsrudZweibel},
the amplification of the intracluster magnetic field via the fluctuation dynamo occurs almost exclusively in the weakly collisional, magnetized regime, and is thus not \emph{a priori} appropriately described by MHD with isotropic transport. 

At magnetic-field strengths larger than  \eqref{eqn:Bmin}, departures of the plasma from local thermodynamic equilibrium are biased with respect to the magnetic-field direction \citep{CGL}, and the transport of momentum and energy across magnetic-field lines becomes stifled by the smallness of the particles' Larmor radii. In weakly collisional plasmas like the ICM, these departures manifest themselves {\it inter alia} as a pressure anisotropy $\rmDelta p \doteq p_\perp - p_\parallel$, where $p_\perp$ ($p_\parallel$) is the thermal pressure across (along) the magnetic-field direction. This anisotropy exerts an anisotropic viscous stress on the fluid motion \citep{Braginskii}, which in an incompressible plasma primarily damps the parallel component of the rate of strain, $\ROS$, where $\eb \doteq \bb{B}/B$ is the unit vector in the direction of the magnetic field $\bb{B}$ and $\bb{u}$ is the flow velocity. In the limit where  $\nu_\mr{i}\gg\ROS$, this pressure anisotropy can be expressed as a balance between adiabatic production (via changes in the magnetic-field strength, since the Lagrangian time rate of change $\rmd\ln B/\rmd t = \ROS$ in a perfectly conducting, incompressible plasma) and collisional isotropization of the pressure tensor, {\it viz.}
\begin{equation}\label{eqn:bragvisc}
    \rmDelta p = 3 \mu_\mathrm{B} \ROS,
\end{equation}
where $\mu_\mathrm{B} = p/\nu_\mathrm{i}$ is the field-aligned Braginskii viscosity \citep{Braginskii}.\footnote{\,Factors of order unity derived by \citet{Braginskii}, which are different for different collision operators, are subsumed into the definition of the collision frequency.}

In practice, the Braginskii parallel viscous stress
\begin{equation}
    \msb{\Uppi}_\parallel \doteq -\left(\eb\eb - \frac{1}{3} \msb{I} \right) \rmDelta p 
\end{equation}
with the pressure anisotropy $\rmDelta p$ given by \eqref{eqn:bragvisc} is only suitable for plasmas with small to order-unity values of $\beta \doteq 8 \upi p/ B^2$, the ratio of the thermal and magnetic pressures. The reason is that high-$\beta$, weakly collisional plasmas are susceptible to several kinetic instabilities when $|\rmDelta p/p| \gtrsim 1/\beta$, such as the firehose \citep{Rosenbluth56,Chandrasekhar58,HellingerMatsumoto00} and mirror \citep{Barnes66,Hasegawa69,SouthwoodKivelson93,Hellinger07} instabilities. 
These instabilities regulate the pressure anisotropy to values near the instability thresholds, an effect that has been diagnosed in various kinetic particle-in-cell simulations \citep{Kunz_kin,Riquelme_2015,Hellinger_2015,Melville} and directly observed using \emph{in situ} measurements of particle distribution functions and magnetic fluctuations in the solar wind \citep{kasper02,hellinger06,Bale,chen16,Chen16b}. 
Thus, because the magnitudes of pressure anisotropy (and thus parallel viscosity) specified by (\ref{eqn:bragvisc}) are often unphysically large in weakly collisional, high-$\beta$ plasmas, any fluid model of the fluctuation dynamo must adopt some form of microphysical closure to account for the enhanced (but not necessarily complete -- see \S\ref{sec:comparison}) regulation of the pressure anisotropy due to microscale kinetic instabilities.

\subsection{Theoretical considerations on pressure anisotropy and dynamo}\label{sec:paniso}

\subsubsection{Hard-wall limiters}

One popular choice of closures in fluid simulations of weakly collisional, high-$\beta$ plasmas is to limit, by hand, the pressure anisotropy to remain within the firehose and mirror instability thresholds:
\begin{equation}
\label{stability}    
-\frac{B^2}{4\upi}\le \rmDelta p \le \frac{B^2}{8\upi}.
\end{equation}
The resulting `hard-wall' limiters, which have their origin in pioneering work on the kinetic magnetorotational instability by \citet{Sharma06} and have since been used in Braginskii-MHD simulations of magnetothermal~\citep{Kunz_2012} and magnetorotational~\citep{Kempski_2019} turbulence, take the form
\begin{align}\label{eqn:mirror_hard}
    \rmDelta p = \mr{min}\left(\frac{B^2}{8\upi},\, 3\mu_\mr{B}\ROS\right)
\end{align}
on the mirror ($\rmDelta p>0$) side and 
\begin{align}\label{eqn:firehose_hard}
    \rmDelta p = \mr{max}\left(-\frac{B^2}{4\upi},\, 3\mu_\mr{B}\ROS\right)
\end{align}
on the firehose ($\rmDelta p<0)$ side (again, assuming incompressibility). Similarly effective limiters, in the form of a large anomalous collision frequency enacted in regions of firehose/mirror instability, were employed by \citet{SantosLima} in simulations of turbulent dynamo using the double-adiabatic \citet{CGL} equations.

\subsubsection{Effective Reynolds number}\label{sec:Reeff}

Such pressure-anisotropy limiters effectively imply an enhanced collisionality in the unstable regions given by
\begin{equation} \label{eqn:nu_eff}
\nu_\mathrm{eff} \sim \beta\ROS .
\end{equation}
Following the reasoning presented in \S\,4.2.2 of \citet{Melville}, we estimate the effective parallel viscosity $\mu_{\parallel{\rm eff}} \doteq \vth{i}^2/\nu_\mathrm{eff}$, and thus the effective parallel-viscous Reynolds number $\Reprl$ associated with the enhanced collisionality \eqref{eqn:nu_eff}, as follows. (Here, $v_\mathrm{thi} \doteq \sqrt{2 T_{\rm i} / m_\mr{i}}$ is the ion thermal velocity and $m_\mr{i}$ is the ion mass.) The Kolmogorov scaling $u_{\ell_\parallel} \propto \ell_\parallel^{1/3}$ for the field-stretching turbulent velocity at parallel scale $\ell_\parallel$ implies that the magnitude of the field-parallel rate of strain $|\ROS| \sim u_{\ell_\parallel}/{\ell_\parallel} \propto {\ell_\parallel}^{-2/3}$ is largest at the effective parallel-viscous scale $\ell_{\mu_{\parallel{\rm eff}}}$, where such motions are dissipated. The value of $\Reeff$ corresponding to \eqref{eqn:nu_eff} is then
\begin{align}\label{eqn:Re_eff}
\Reeff \doteq \frac{u_0\ell_0}{\mu_{\parallel{\rm eff}}} \sim u_0 \ell_0 \frac{\nu_\mathrm{eff}}{v_\mathrm{thi}^2}\sim \beta M^2 \Reeff^{1/2} \quad \Longrightarrow\quad \Reeff \sim \beta^2 M^4 ,
\end{align}
where $u_0$ is the characteristic speed of the outer-scale bulk fluid motions, $\ell_0$ is the energy injection (outer) scale, and $M \doteq u_0/v_\mathrm{thi}$ is the Mach number. The resulting effective particle mean free path is
\begin{equation}\label{eqn:mfp_eff}
    \lambda_{\rm mfp,eff} \doteq \frac{\vth{i}}{\nu_{\rm eff}} \sim \ell_0 \, \beta^{-2} M^{-3} .
\end{equation}
The result is a parallel-viscous cutoff
\begin{equation}\label{eqn:ellvisc}
    \ell_{\mu_{\parallel{\rm eff}}} \sim \ell_0 \, \Reeff^{-3/4} \sim \ell_0 \, \beta^{-3/2} M^{-3}
\end{equation}
that is smaller than the cutoff caused by Coulomb collisions, $\ell_{\rm Coulomb} \sim \ell_0 \, {\rm Re}^{-3/4}_\parallel$ with ${\rm Re}_{\parallel} \doteq u_0 \ell_0 / (\vthi^2/\nu_{\rm i})$, if the magnetic-field strength satisfies
\begin{equation}\label{eqn:collisional}
B \lesssim 2 \,\biggl(\frac{n}{10^{-3}\textrm{ cm}^{-3}}\biggr)^{1/4}\biggl(\frac{T}{5\textrm{ keV}}\biggr)\biggl(\frac{M}{0.2}\biggr)^{3/4}\biggl(\frac{\ell_0}{100\textrm{ kpc}}\biggr)^{-1/4}~\upmu\mathrm{G} .
\end{equation}
(In \eqref{eqn:collisional}, we have chosen a normalization based on typical values found in the bulk ICM.) In this case, a larger maximal shear rate can be achieved:
\begin{equation}\label{eqn:Svisc}
    S_{\mu_{\parallel{\rm eff}}} \sim \frac{u_0}{\ell_0} \, \Reeff^{1/2} \sim \frac{u_0}{\ell_0}\, \beta M^2 .
\end{equation}
Note that $\ell_{\mu_{\parallel{\rm eff}}}\propto B^3$ increases and $S_{\mu_{\parallel{\rm eff}}} \propto B^{-2}$ decreases with increasing magnetic-field strength. This suggests that the dynamo subject to `hard-wall' limiters is \emph{likely faster than its unlimited counterpart}, but it slows down on average as the magnetic energy grows.

\subsubsection{Suitability of the hard-wall limiters}\label{sec:comparison}

The use of hard-wall limiters implicitly assumes that the enhanced collisionality given by \eqref{eqn:nu_eff} is achievable in the plasma, {\em viz.}, that the collision frequency needed to pin the pressure anisotropy to either marginal-stability threshold, ${\sim}\beta\ROS$ \dblbrck{see \eqref{eqn:nu_eff}}, is smaller than the ion-Larmor frequency $\Omega_{\rm i}$. Using \eqref{eqn:Svisc} as an estimate for the maximal value of $\ROS$, this criterion becomes
\begin{subequations}\label{eqn:ultrahighbeta}
\begin{equation}\label{eqn:ultrahighbeta_A}
    \beta \gtrsim \biggl(\frac{\ell_0}{d_{\rm i}}\biggr)^{2/5} M^{-6/5} ,
\end{equation}
where $d_{\rm i}$ is the (field-strength-independent) ion inertial scale. Equation \eqref{eqn:ultrahighbeta_A} is referred to as the `ultra-high-$\beta$' limit in \citet{Melville}. For characteristic parameters of the ICM, equation \eqref{eqn:ultrahighbeta_A} implies
%
%
\begin{equation}\label{eqn:ultrahighbeta_B}
B \lesssim 6 \, \biggl(\frac{n}{10^{-3}\textrm{ cm}^{-3}}\biggr)^{2/5}\biggl(\frac{T}{5\textrm{ keV}}\biggr)^{1/2}\biggl(\frac{M}{0.2}\biggr)^{3/5}\biggl(\frac{\ell_0}{100\textrm{ kpc}}\biggr)^{-1/5}~\mathrm{nG} .
\end{equation}
\end{subequations}
For magnetic-field strengths satisfying \eqref{eqn:ultrahighbeta_B}, the hard-wall limiters are unsuitable and the parallel Reynolds number is smaller than that given by \eqref{eqn:Re_eff}.

\subsubsection{Three dynamo regimes}\label{sec:regimes}

The conditions \eqref{eqn:Bmin} and \eqref{eqn:ultrahighbeta} suggest that there exist three distinct regimes of magnetization and collisionality: 
\vspace{1ex}
\begin{enumerate}[itemindent=-15pt]
    \item\label{unmag} the {\em unmagnetized regime}, when $B \lesssim 10^{-18}$ G; 
    \item\label{kinmag} the {\em magnetized `kinetic' regime} (ultra-high-$\beta$), when $10^{-18}\textrm{ G} \lesssim B \lesssim 6 \textrm{ nG}$ and for which the hard-wall limiters \eqref{eqn:mirror_hard} and \eqref{eqn:firehose_hard} are unsuitable; and 
    \item\label{flumag} the {\em magnetized `fluid' regime}, when $B \gtrsim 6\textrm{ nG}$ and for which the pressure anisotropy can be well regulated by the instabilities (i.e.~$\nu_{\rm eff} \lesssim \Omegai$). 
\end{enumerate}
\vspace{1ex}
The nomenclature of `kinetic' and `fluid' used here is a reflection of our result (see \S\,\ref{sec:resultsoutline}) that, when the pressure anisotropy $\Delta p/p$ is limited by the firehose and mirror instabilities to values ${\sim}\beta^{-1}\ll{1}$ \dblbrck{regime~\eqref{flumag}}, the plasma dynamo largely resembles its fluid (MHD) counterpart, particularly when the magnetic field is dynamically weak. If instead the pressure anisotropy is poorly regulated by the instabilities \dblbrck{regime~\eqref{kinmag}}, the associated parallel viscous stress is dynamically important throughout the entire evolution of the dynamo, a feature that qualitatively changes the character of the fluctuation dynamo.

The saturated state of the dynamo, in which the magnetic and kinetic energies are comparable, would be obtained when
\begin{equation}
B_{\rm sat} \sim 3 \,\biggl(\frac{n}{10^{-3}\textrm{ cm}^{-3}}\biggr)^{1/2}\biggl(\frac{T}{5\textrm{ keV}}\biggr)^{1/2}\biggl(\frac{M}{0.2}\biggr) ~\upmu\mr{G} 
\end{equation}
and thus occurs in the magnetized fluid regime. Perhaps coincidentally, this value is close to the field strength at which the effective scattering frequency due to efficient firehose/mirror is comparable to that due to Coulomb collisions, equation \eqref{eqn:collisional}. Namely, for $B\gtrsim 2~\upmu{\rm G}$, the effective viscous length scale given by \eqref{eqn:ellvisc} satisfies $\ell_{\mu_{\parallel{\rm eff}}} \lesssim \ell_\mathrm{Coulomb}$, and so Coulomb collisions are frequent enough to maintain a kinetically stable pressure anisotropy. At smaller magnetic-field strengths, the mean free path is effectively set by the kinetic instabilities \dblbrck{see \eqref{eqn:mfp_eff}}.

An intriguing feature of these regimes is that, while the Reynolds number in the unmagnetized regime is set by Coulomb collisions, resulting in $\Reprl \sim 1 \textrm{--} 100$ \citep{Scheko_2005}, at the transition from the second (magnetized kinetic) regime to the third (magnetized fluid) regime we find $\Reeff \sim \beta^2 M^4 \gg 1$. This suggests that $\Reeff$ must experience a large increase at some time between these two epochs. Since the viscous-scale rate of strain increases as $\Reeff^{1/2}$, the dynamo in this intermediate second regime should, at some point, be {\em self-accelerating}, with the field-stretching eddies becoming smaller and faster as the magnetic field is amplified. Similar scenarios have previously formed the basis for theories of explosive dynamo in collisionless plasmas \citep{SchekoCowley06a,SchekoCowley06b,Melville,Mogavero}, a topic that will be the subject of a separate publication.\footnote{\, Pre-existing magnetic fields that were injected into the intergalactic medium or early ICM by its galactic residents could serve as seeds for the fluctuation dynamo \citep[e.g.,][]{rs68,rees87,rephaeli88,fl01}, a conjecture supported indirectly by the observed early enrichment of galaxy clusters by metals. For example, a very recent paper by \citet{mantz20} uses XMM-Newton  observations of a $M \sim 2\times 10^{14}~{\rm M}_\odot$ cluster at redshift $z\simeq 1.7$ that reveal metal enrichment of ${\sim}1/3$ Solar to argue that cluster metallicities were already mostly set when the Universe was less than ${\sim}4~{\rm Gyr}$ old. If such pollutants were to have been accompanied by ${\sim}\upmu{\rm G}$ galactic magnetic fields, it is possible that what we refer to as the magnetized kinetic regime would have been short lived, or perhaps even non-existent, depending of course on the efficiency with which such fields were dispersed and diluted throughout the ${\sim}$Mpc cluster volume.}

 \subsubsection{Previous simulations and numerical results}
 
The first two dynamo regimes have been previously studied through the use of hybrid-kinetic numerical simulations by \citet{Rincon_2016} and \citet[][hereafter Paper I]{StOnge_2017}. These studies observed the generation of firehose and mirror instabilities as the dynamo entered the magnetized regime, aiding in breaking adiabatic invariance and regulating the pressure anisotropy.  \citet{SantosLima} studied the effects of pressure-anisotropy regulation during the fluctuation dynamo using a collisionless double-adiabatic closure to evolve $p_\perp$ and $p_\parallel$ \citep{CGL} supplemented by a non-zero collision frequency $\nu_\mr{eff}$ that was activated in spatial regions of kinetic instability. It was found that simulations with instantaneous pressure-anisotropy relaxation exhibited magnetic-field growth rates similar to those in isotropic MHD, with some minor details differing in the saturated state. As the collision frequency was lowered, the dynamo growth rates and the final value of the saturated magnetic energy decreased. In the entirely collisionless case, the pressure anisotropy was allowed to grow arbitrarily large, and no growth of the magnetic energy was observed. This is consistent with theoretical considerations of magnetic-field amplification occurring under adiabatic invariance in collisionless plasmas by \citet{helander_constraints}, who concluded that dynamo action always requires collisions or some small-scale kinetic mechanism for breaking the adiabatic invariance of the magnetic moment~\citep[cf.][]{Kulsrud1997b}.

At the moment, the third (magnetized fluid) regime of the dynamo is prohibitively expensive to investigate using kinetic simulation, except near the saturated state (which was the focus of \S\,3.3 in \citetalias{StOnge_2017}). To appreciate this difficulty, let us imagine that one wishes to resolve two orders of magnitude of magnetic-energy growth (equivalently, one order of magnitude in growth of the magnetic-field strength) in this regime in a single simulation. The constraints on the initial plasma beta $\beta_0$ required to simulate this regime are, in terms of the controllable simulation parameters,
\begin{equation}\label{eq:comp_argument}
    \frac{A}{M^2} \lesssim \beta_0 \lesssim \frac{1}{M^{3/2}}  \left(\frac{\ell_0}{\rhoio}\right)^{1/2},
\end{equation}
where $A$ is the desired magnetic-energy amplification factor and $\rhoio$ is the ion Larmor radius at the transition from the second to this third regime. The first inequality in \eqref{eq:comp_argument} follows from the dynamo  not yet being saturated (i.e.~$\beta M^{-2} \lesssim 1$), while the second inequality follows from the requirement $\nu_{\rm eff} \lesssim\Omegaio$. For such a range of $\beta_0$ to exist, 
\begin{align}
    A \lesssim M^{1/2} \left(\frac{\ell_0}{\rhoio}\right)^{1/2} .
\end{align}
To allow for an appreciable range of field amplification in this regime, we must then maximize $M$ and $\ell_0/\rhoio$. If we demand $M \lesssim 0.2$ (so as to maintain relevance to the sub-sonic turbulence observed in the ICM), then resolving two orders of magnitude of energy growth ($A \sim 10^2$) requires $\ell_0/\rhoio \gtrsim 50,000$. If we then wish to resolve the ion Larmor radius at the end of this growth period by a minimum of two cells, then the number of cells needed in each spatial direction is ${\sim}10^6$! For a problem that is intrinsically three-dimensional \citep{Cowling1933,zeldovich57}, this requirement is well beyond current computational capabilities. 

\subsubsection{This work}

The goal of the work presented here is twofold. First, we study the plasma dynamo in the third (magnetized fluid) regime using a reduced set of fluid equations that takes the effects of kinetic instabilities into account through the use of microphysical closures that limit the pressure anisotropy. For this, we use the incompressible MHD equations including the parallel component of the Braginskii viscosity tensor, the latter being hard-wall limited to kinetically stable values [see (\ref{stability}) and \S\,\ref{sec:equations}]. Secondly, in order to understand better the effects of this regulation and the nature of the dynamo in the second (magnetized kinetic) regime, we also perform a number of Braginskii-MHD simulations without any  hard-wall regulation of the pressure anisotropy, relying instead on incomplete regulation by a constant (e.g., Coulomb) collisionality. This approach is justified by measurements of imperfect pressure-anisotropy regulation during this regime in the hybrid-kinetic simulations presented in \citetalias{StOnge_2017}.

\subsection{Main results and organization of this paper}\label{sec:resultsoutline}

Given the rather technical nature of the presentation that follows, we first provide an overview of our main results as a preamble, with pointers to the relevant sections and figures in the main text. 
We arrive at three main conclusions (which are re-summarized in \S\,\ref{sec:discussion} alongside a recapitulation of ancillary results and a preview of future work):
\begin{enumerate}
    \item \label{conc_1} With hard-wall limiters on the pressure anisotropy that prevent $\rmDelta p$ from growing beyond its kinetically stable values, the Braginskii-MHD dynamo is in most respects identical to the standard high-${\rm Pm}$ MHD dynamo. This is to be expected, because the limited pressure anisotropy becomes dynamically important only once the Lorentz force does, {\it viz.}, as the dynamo starts saturating. Certain minor differences with isotropic MHD dynamo do appear in the saturated state.
    \item \label{conc_2} When no pressure-anisotropy limiters are used (as relevant to regimes in which an effective collision frequency $\nu_{\rm eff} \gtrsim \Omegai$ would be required to keep $\rmDelta p$ near marginal stability), the Braginskii-MHD dynamo behaves quite differently to the MHD dynamo. Nearly all of these differences can be understood by noting that, in its growing phase, the structure and statistics of the magnetic field are remarkably similar to those found in the saturated state of the (high-${\rm Pm}$) MHD dynamo.  This occurs because the form of the Braginskii-viscous stress, $\grad\bcdot(\eb\eb\Delta p)$ \dblbrck{the third term on the right-hand side of~\eqref{brag_MHD:mom}}, is mathematically similar to that of the Lorentz force,  $\bb{B}\bcdot\grad\bb{B}=\grad\bcdot(\eb\eb B^2/4\upi)$ \dblbrck{the first term on the right-hand side of~\eqref{brag_MHD:mom}}, if one makes the substitution $B^2/4\upi \rightarrow \rmDelta p \sim \mu_\mr{B} \, \rmd\ln B/\rmd t = \mu_{\rm B} \, \ROS$ (neglecting resistivity).
    \item \label{conc_3} Without pressure anisotropy limiters, Braginskii MHD no longer supports a dynamo if the ratio of the isotropic viscosity ($\mu$) and the Braginskii viscosity ($\mu_\mr{B}$) is too small. This behaviour may be understood heuristically by noting that the Braginskii viscosity, by targeting only those fluid motions that stretch the magnetic field ($\ROS\ne{0}$), curbs the growth of that field while simultaneously promoting its resistive decay by allowing motions that mix the field lines. Finite isotropic viscosity moderates the mixing motions, thereby allowing the field to grow if this viscosity is sufficiently large. In the limit where the Braginskii viscosity is so strong that the outer-scale fluid motions become two-dimensionalized with respect to the magnetic-field direction, the dynamo shuts down unless the isotropic viscosity is large enough to diffuse velocity gradients into the field-perpendicular direction, thus once again rendering the dynamics three-dimensional.
\end{enumerate}
\vspace{1ex}
The outline of this paper is as follows. We present the Braginskii-MHD equations and some definitions of relevant dimensionless numbers in \S\,\ref{sec:equations}. We then describe the numerical method of solution and several useful diagnostics  in \S\S\,\ref{sec:simulation} and \ref{sec:diagnostics}. The full list of runs is given in table \ref{tab:runs}. We open the results section (\S\,\ref{sec:results}) with a brief overview of the fluctuation dynamo, broken down into its four evolutionary stages (\S\,\ref{sec:overview}). We then present evidence from our simulations in favour of our first conclusion, that the limited Braginskii-MHD dynamo is similar to the isotropic-MHD dynamo in a ${\rm Re}\gg{1}$, ${\rm Pm}\gtrsim{1}$ fluid (\S\,\ref{sec:limited}). This discussion covers the basic structure of the fields and flows in the kinematic and saturated stages (see figures \ref{fig:printout_lim}, \ref{fig:lim_spec}--\ref{fig:wavenumbers_lim}) and the dynamo growth rates (figure \ref{fig:energy_lim}). Section \ref{sec:unlimited} provides evidence for our second conclusion, that the structure and statistics of the unlimited Braginskii-MHD dynamo imitate those in the saturated state of the more standard ${\rm Pm}\gtrsim{1}$ MHD dynamo (see figures \ref{fig:printout_unlim}--\ref{fig:curvature}). Much of the rest of the paper is devoted to exploring the consequences of this similarity. This leads us to a treatment of the anisotropization of the fluid flow by the Braginskii viscosity (\S\,\ref{sec:anisotropization}), in which we study the structure of the rate-of-strain tensor (figure \ref{fig:ROS_all}) and the alignment of the eigenvectors of the rate-of-strain tensor with the magnetic-field direction $\eb$ (figures \ref{fig:angle_some}--\ref{fig:eigen_some}).\footnote{\,In appendix \ref{sec:magnetoimmutability}, we examine further the anisotropization of the rate-of-strain tensor by the Braginskii viscosity in the context of `magneto-immutability,' a concept introduced and previously examined in the context of guide-field Alfv\'enic turbulence by \citet{Jono_magnetoimmutability}. We do this by studying the distribution of $\ROS$ (figures \ref{fig:pdfROS} and \ref{fig:brazil}) and the energy transfer spectra (figure \ref{fig:unlim_trans}). 
} 
These studies also reveal minor differences in the properties of the saturated state of the limited Braginskii-MHD dynamo compared to isotropic MHD, alluded to in conclusion (\ref{conc_1}). Motivated by these results, in \S\,\ref{sec:kazantsev} we formulate an analytic model for the kinematic stage of the unlimited Braginskii-MHD dynamo, which is based on an extension of the Kazantsev--Kraichnan model to anisotropic magnetic-field statistics first proposed to describe the saturated MHD dynamo by \citet{Scheko_saturated} (a detailed derivation of the model, which was omitted in their paper, is given in appendix \ref{ap:kazantsev}). The model, although necessarily simplistic, appears to give a reasonable match to the spectra (figure \ref{fig:kaz_spectra}) seen in simulations, and predicts that the dynamo shuts off for sufficiently small isotropic viscosities (figure \ref{fig:kazantsev}). It also provides a reasonable explanation for the behaviour of a separate set of simulations in the Braginskii `Stokes-flow' regime ($\Reeff\lesssim 1$), described in \S\,\ref{sec:stokes}, in which we see the dynamo shut off for sufficiently small $\Reeff\lesssim 1$ (unless the isotropic viscosity is also very large, in which case the dynamo can operate). These results support our conclusion (\ref{conc_3}). We finish  with a more thorough discussion of our conclusions, as well as suggestions for future work, in \S\,\ref{sec:discussion}.

\section{Method of solution}\label{sec:method}

\subsection{Equations and free parameters}\label{sec:equations}

The incompressible MHD equations including the field-aligned component of the Braginskii viscosity are
\begin{subequations}\label{brag_MHD}
\begin{align}\label{brag_MHD:mom}
\D{t}{\bb{u}} &\doteq \left(\pD{t}{} + \bb{u}\bcdot\grad\right)\bb{u} =  \bb{B}\bcdot\grad\bb{B} - \grad p + \grad \bcdot(\eb\eb\rmDelta p) - (-1)^h \mu_h \nabla^{2h}\bb{u}+ \drive, \\*
\D{t}{\bb{B}} &\doteq \left(\pD{t}{} + \bb{u}\bcdot\grad\right)\bb{B} = \bb{B}\bcdot \grad \bb{u} - (-1)^h \eta_h\nabla^{2h} \bb{B} ,\label{brag_MHD:ind}
\end{align}
\end{subequations}
where the magnetic field $\bb{B}$ is expressed in Alfv\'{e}nic units and the mass density has been scaled out. The last term on the right-hand side of \eqref{brag_MHD:mom}, $\drive$, is a random driving body force, further described in \S\,\ref{sec:simulation}. The additional diffusive terms in \eqref{brag_MHD}, featuring $\mu_h$ and $\eta_h$, are Laplacian ($h=1$) or hyper ($h=2$) viscosity and diffusivity, respectively; these are introduced to truncate the cascades of kinetic and magnetic energy near the smallest wavelengths captured in our simulations. The pressure anisotropy is given by \eqref{eqn:bragvisc} except when limited by heuristic micro-instability limiters, as per \eqref{eqn:mirror_hard} and \eqref{eqn:firehose_hard}.\footnote{\,In effect, we are assuming that the standard diffusive form of the parallel viscous stress remains valid, notwithstanding any changes to its form that might result from the microphysical regulation of the pressure anisotropy.} Because the ratio of initial magnetic energy to kinetic energy is extremely small, much of the dynamo evolution occurs with the majority of the plasma residing in the firehose- or mirror-unstable regions. Depending upon the material properties of the plasma, this ratio can remain small even in the saturated state. As a result, the pressure-anisotropy limiters are likely to play an important dynamical role throughout the entire evolution of the plasma dynamo. 

Equations \eqref{brag_MHD} have four free parameters: the two isotropic diffusivities $\mu_h$ and $\eta_h$, the anisotropic Braginskii viscosity $\mu_\mathrm{B}$, and the specifications of the random forcing $\tilde{\bb{f}}$, described in \S\,\ref{sec:simulation}. For our analysis, it is useful to define the following Reynolds and Prandtl numbers:
\begin{equation}\label{eqn:Reynolds}
\mathrm{Re} \doteq \frac{u_0 \ell_0}{\mu}, \quad
\mathrm{Re}_\parallel \doteq \frac{u_0 \ell_0}{\mu_\mathrm{B}}, \quad
\mathrm{Rm} \doteq  \frac{u_0 \ell_0}{\eta}, \quad
\mathrm{Pm} \doteq \frac{\mathrm{Rm}}{\mathrm{Re}},
\end{equation}
where $u_0$ is the root-mean-square fluid velocity, $\ell_0 \doteq L/ 2 \upi  \sim k_\mr{f}^{-1}$, $L$ is the size of the simulation domain, and $k_\mr{f}$ is the forcing wavenumber. In the unlimited case, the effective parallel-viscous Reynolds number introduced in \eqref{eqn:Re_eff} satisfies $\Reeff = \mathrm{Re}_\parallel$; in the limited case, $\mathrm{Re}_\parallel \le \Reeff\le \mathrm{Re}$. The definitions \eqref{eqn:Reynolds} are suitable for Laplacian dissipation, but generalized Reynolds numbers $\mathrm{Re}_h$ ($\mathrm{Rm}_h$) can be formulated for higher-order dissipation. To do so, we  make the substitutions $\mu\rightarrow \ell_\mu^{-2(h-1)}\mu_h$ and $\eta \rightarrow \ell_\eta^{-2(h-1)}\eta_\mathrm{h}$ in the standard definitions~\eqref{eqn:Reynolds}. This replacement is done so that, for a given value of $\mathrm{Re}_h$ ($\mathrm{Rm}_h$), $\ell_\mu/\ell_0$ ($\ell_\eta/\ell_0$) is held fixed for all values of $h$. This allows one to compare two systems by directly comparing their generalized Reynolds numbers. Assuming Kolmogorov scalings ({\it viz.}~$|\grad \bb{u}| \propto \mathrm{Re}_h^{2/3}$) to compute the dissipation scales $k_\mu$ and $k_\eta$, we have
\begin{subequations}
\begin{align}
    \mathrm{Re}_h &\doteq \frac{u_0 \ell_0}{\ell_\mu^{-2(h-1)} \mu_{h}} =\left(\frac{u_0 \ell_0^{2h-1}}{\mu_h}\right)^{2/(3h-1)},\\
    \mathrm{Rm}_h &\doteq \frac{u_0\ell_0}{\ell_\eta^{-2(h-1)} \eta_{h}} = \left(\frac{ u_0 \ell_0^{2h-1}}{\eta_h \Reeff^{(h-1)/2}}\right)^{1/h}.
\end{align}
\end{subequations}

While more general definitions can be devised without assuming Kolmogorov scalings, they will generically depend on some characteristic of the underlying fields that must be determined \emph{a posteriori}.

\subsection{Numerical simulations}\label{sec:simulation}

To solve \eqref{brag_MHD}, we use a version of the pseudospectral incompressible-MHD code \texttt{Snoopy} \citep{Lesur2007} equipped with the parallel Braginskii viscosity. All simulations are run on a triply periodic, $224^3$ grid with $2/3$ de-aliasing (with the exception of the `Stokes-flow' runs discussed in \S\,\ref{sec:stokes}, which use $112^3$ collocation points).\footnote{\,The parallel Braginskii stress involves a quintic nonlinearity and, in a spectral code like \texttt{Snoopy}, formally introduces aliasing due to division by $B^2$. However, numerical tests show that, if such aliasing effects are present, they produce no quantifiable difference in magnetic-energy growth rates, field statistics, or turbulent spectra between simulations with $1/3$ de-aliasing and a grid of size $448^3$ and those with $2/3$ de-aliasing and a grid of size $224^3$.} Turbulence is driven using an incompressible, zero-mean-helicity,\footnote{\, The time-correlated nature of our random forcing implies that, at any given time in the simulation, it has volume-averaged net helicity. However, the time average of its helicity is zero. Comparisons between simulations with white-in-time forcing (with zero pointwise helical forcing) and time-correlated forcing reveal no substantive differences in terms of magnetic-field statistics or geometry.} random forcing at wavenumbers $k_\mathrm{f} \in [2 \upi/L, 4\upi/L]$; its power is distributed evenly among these Fourier modes. The forcing is time-correlated by an Ornstein--Uhlenbeck process. Code units are based on the box size $L=1$ and energy injection rate $\varepsilon =1$. The latter appears in the Ornstein--Uhlenbeck process via
\begin{align}
\drive(t+\rmDelta t) = \drive(t) \,\rme^{-\rmDelta t/\tcorrf} + \left[\frac{\varepsilon}{\tcorrf}\left(1-\rme^{-2 \rmDelta t /\tcorrf}\right)\right]^{1/2}\bb{\widetilde{g}}, 
\end{align}
where $\rmDelta t$ is the simulation timestep, $\tcorrf$ is the correlation time of the forcing, and $\bb{\widetilde{g}}$ is a Gaussian noise at wavenumber $k_{\rm f}$ \citep{gillespie}. These normalizations of $L$ and $\varepsilon$ lead to saturated turbulence amplitudes of order unity ($u_\mathrm{rms} \sim 1$). The correlation time is chosen to be  $\tcorrf=(2\pi)^{-1} \approx \ell_0/u_{\rm rms}$, which is comparable to the inverse decorrelation rate at the outer scale for $\mathrm{Re} \ge 1$ turbulence. Simulations are initiated with $\bb{u} =0$ and a random zero-net-flux magnetic field with power at $k \in [2 \upi/L, 4\upi/L]$. All runs have an initial rms field strength $B_\mr{0,rms} = 10^{-3}$. Because we have assumed incompressibility, the thermal velocity $v_\mr{thi}$, and thus $\beta$, have been eliminated from the equations. Accordingly, we formulate the stability thresholds \eqref{stability} in terms of $\rmDelta p$ and $B^2$ directly and subsume $p$ and $\nu_\mr{i}$ into the definition of $\mu_\mr{B}$.

For comparisons, we have also performed simulations of the dynamo in standard incompressible MHD with isotropic diffusion (i.e.~$\mu_\mr{B}=0$) for Reynolds numbers ranging from order unity to large values.

A list of all the simulations used in this work, along with some parameters of note, is given in table \ref{tab:runs}. We limit our study to isotropic $\mathrm{Pm} \gtrsim 1$ and modestly large $\mathrm{Rm}$; future work should examine whether our conclusions hold in other limits, e.g., $\mathrm{Re}\gg \mathrm{Rm} \gg \Reprl$ or $\mathrm{Rm} \gg \mathrm{Re} \gg \Reprl$.

\begin{table}
    \centering
    \small
    \begin{tabular}{cccccccccccc}
    \hline 
    \hline
            Run & Res. & $\varepsilon$ &  $\visc_\mr{B}^{-1}$ & $\visc^{-1}$  & $\eta^{-1}$ & $\!\!\langle u_\mathrm{rms}^2 \rangle^{1/2}_t\!\!\!\!$ & $\langle B_0^2\rangle^{1/2}\!\!$ &Re$_\parallel$ & Re & Rm & limiter   \\
    \hline
    MHD1 & $224^3$& 1 & $\infty$ & 20 &  1500  & 0.56 & $10^{-3}$ &  --- & 1.8 & 130 & --- \\
    MHD2 & $224^3$& 1 & $\infty$ &100 &  1500  & 1.07  & $10^{-3}$ &  --- & 17 & 260  & --- \\
    MHD3 & $224^3$ & 1 & $\infty$ & 500 & 1500 & 1.35 & $10^{-3}$&  --- & 110  & 320 & ---\\
    MHD4 & $224^3$ & 1 &  $\infty$ & 1500 &1500 & 1.43 & $10^{-3}$&  --- &340  &  340 & ---\\
    MHDH & $224^3$ & 1 &  $\infty$ &(H) & (H) & 1.50 & $10^{-3}$&  --- & 100  &  100 & ---\\
    \\
    U1 & $224^3$&1 &   20 &1500 & 1500 & 1.21 & $10^{-3}$ & 3.9 & 290 & 290 & unlimited \\
    U2 & $224^3$ &1 &   20 & 600 &1500 & 1.18 & $10^{-3}$& 3.8 & 110 & 280 & unlimited\\
    U3 & $224^3$ &1 &   20 & 240 &1500 & 1.07 & $10^{-3}$ & 3.4 &40 &  255 & unlimited\\
    U4 & $224^3$ & 1 & 20 &96 &  1500 &  0.90& $10^{-3}$ & 2.9   & 14 & 210  & unlimited\\
    U1a & $224^3$& 1 & 100 &1500 &  1500 & 1.38 & $10^{-3}$ & 20 &330 & 330 & unlimited \\
    U1b & $224^3$& 1 & 500 &1500 &  1500 & 1.45 & $10^{-3}$ & 115 & 350 & 350 & unlimited \\
    U1H & $224^3$& 1 &  20 &(H) & (H) & 1.21   & $10^{-3}$ & 3.9 & 100 & 300 & unlimited \\

    \\
    L1 & $224^3$&1 &   20 &1500 & 1500 & 1.47 & $10^{-3}$ & 4.7 & 350 &350 & hard-wall\\
    L2 & $224^3$&1 &  20 &600 &  1500 & 1.43 & $10^{-3}$ &4.6 & 140 & 340 & hard-wall\\
    L3 & $224^3$&1 &   20 &240 & 1500 & 1.33  & $10^{-3}$& 4.2 & 50 & 320 & hard-wall\\
    L4 & $224^3$& 1 &  20 &96 & 1500 & 1.12 & $10^{-3}$ & 3.6 & 17  & 270 & hard-wall\\
    L1m & $224^3$ &1 &  20 &  1500 &1500 & 1.42 & $10^{-3}$& 4.5 &340  &340  & mirror \\
    L1a & $224^3$ & 1 &   100 &1500 & 1500 & 1.43 & $10^{-3}$ &23  & 340 &  340 & hard-wall\\
    L1b & $224^3$ & 1 & 500 & 1500 & 1500 & 1.44 & $10^{-3}$ &  115 & 340 & 340 & hard-wall\\
     L1H & $224^3$& 1 &   20 &(H) & (H) & 1.47 & $10^{-3}$ & 4.7 & 100 & 100 & hard-wall\\
     \\
    MHDSa & $112^3$ & 1 & $\infty$ &  20 & (H)  & 0.60 & $10^{-3}$ &  --- & 2 & 180 & --- \\
    MHDSb & $112^3$ & 20 &  $\infty$ &4 & (H)  & 0.81  & $10^{-3}$ &  --- & 0.5 & 290  & --- \\
    MHDSc & $112^3$ & 500 &  $\infty$ & 0.5 &(H) & 0.67 & $10^{-3}$&  --- &0.05  & 460  & ---\\
    USa & $112^3$ & 1 &  20  &1500 & (H)  & 1.06 & $10^{-3}$ &  3.4 & 250 &  200 & unlimited \\
    USb & $112^3$ & 1 & 10  &1500 &  (H)  & 1.01& $10^{-3}$ &  1.6 & 240 & 240 & unlimited \\
    USc & $112^3$ & 1 &  6  &1500 & (H)  & 0.96 & $10^{-3}$ &  0.9 & 230 & 270 & unlimited \\
    USd & $112^3$ & 2 &  4  &1500 &(H)  & 1.15 & $10^{-3}$ &  0.7 & 275 & 310 & unlimited \\
    USe & $112^3$ & 3 & 2  & 1500 & (H)  & 1.16 & $10^{-3}$ &  0.4 & 275 & 370 & unlimited \\
    USf & $112^3$ & 4 &  1  &1500 & (H)  & 1.11 & $10^{-3}$ &  0.18 & 265 &  440 & unlimited \\            
    USg & $112^3$ & 5 &  0.5 & 1500 &(H)  &  1.07 & $10^{-3}$ &  0.09 & 260 & 515 & unlimited \\
    USg$^*$  & $112^3$ & 50 &  0.5 &4 & (H)  &  0.69 & $10^{-3}$ &  0.05 & 0.4 & 470 & unlimited \\
    \hline
    \end{tabular}
    \caption{Index of runs, sorted into those using isotropic MHD (prefix `MHD'), unlimited Braginskii MHD (prefix `U'), and limited Braginskii MHD (prefix `L'). Run names adorned with an `S' employ viscosities approaching and entering the `Stokes-flow' regime (\S\,\ref{sec:stokes}). \emph{Note:}  `Res.' denotes the number of collocation points in each simulation, with the effective resolution reduced by a factor of $(2/3)^3$ due to de-aliasing.  Viscosity and diffusivity values with an `(H)' indicate simulations with hyper-diffusion with value $\mu_\mathrm{H}^{-1}$, $\eta_\mathrm{H}^{-1} =1.8\times 10^7$. `mirror'  denotes a simulation with a hard-wall limiter at the mirror threshold and no limiter at the firehose threshold. Time averages for $\langle u_\mathrm{rms}^2 \rangle^{1/2}_t$ are taken over the kinematic stage.}
    \label{tab:runs}
\end{table}

\subsection{Diagnostics}\label{sec:diagnostics}

Before presenting our results, we define various diagnostics that will be used to study the structure and statistics of the turbulent velocity and magnetic fields. In what follows, $\langle \, \cdot\,\rangle$ ($\langle \, \cdot\, \rangle_t$) denotes a volume (time) average. The root-mean-square (rms) value of a quantity $A$ is given by $A_\mr{rms} \doteq \langle A^2 \rangle^{1/2}$. 

\subsubsection{Characteristic wavenumbers}

The following assortment of characteristic wavenumbers are useful for diagnosing the structure of the magnetic field (see \citealt{Scheko_sim}):
\begin{subequations}\label{char-wavenumbers}
\begin{gather}
k_\parallel \doteq \left( \frac{\left\langle|\bb{B}\bcdot\grad\bb{B}|^2\right\rangle}{\langle B^4\rangle}\right)^{1/2} ,  ~ 
k_{\bs{B}\bstimes\bs{J}} \doteq \left( \frac{\left\langle|\bb{B}\btimes\bb{J}|^2\right\rangle}{\langle B^4\rangle}\right)^{1/2}, \tag{\theequation {\it a,b}}\\*
k_{\bs{B}\bscdot\bs{J}} \doteq \left( \frac{\left\langle|\bb{B}\bcdot\bb{J}|^2\right\rangle}{\langle B^4\rangle} \right)^{1/2}, ~ 
k_{\mr{rms}} \doteq \left( \frac{\left\langle|\grad\bb{B}|^2\right\rangle}{\langle B^2\rangle} \right)^{1/2} . \tag{\theequation {\it c,d}}
\end{gather}
\end{subequations}
These quantities have simple interpretations: $k_\parallel$ measures the variation of the magnetic field along itself, and is typically set by the smallest-scale field-aligned stretching motions; 
$k_{\bs{B}\bstimes\bs{J}}$ measures the variation of the magnetic field across itself, and captures the field reversals in folds, which are ultimately limited by resistive dissipation; $k_{\bs{B}\bscdot\bs{J}}$ measures the variation of the field in the direction orthogonal both to $\bb{B}$ and to $\bb{B}\btimes \bb{J}$, which tends to be the direction of the greatest compression \citep{Zeldovich,Scheko_sim}; and $k_{\mr{rms}}$ provides a general measure of the overall variation of the magnetic field, but tends to follow $k_{\bs{B}\bstimes\bs{J}}$. For magnetic fields that are arranged in folded sheets -- a typical realization during the kinematic stage of the ${\rm Pm}\gg{1}$ MHD dynamo -- the relative ordering of these wavenumbers is $k_\parallel \lesssim k_{\bs{B}\bscdot \bs{J}} \ll k_{\bs{B}\bstimes\bs{J}} \sim k_\mr{rms} \sim k_\eta$, where $k_\eta$ is the spectral cut-off due to resistivity. For a magnetic field arranged in folded ribbons -- a typical realization during the saturated state of the ${\rm Pm}\gg{1}$ MHD dynamo -- $k_\parallel \ll  k_{\bs{B}\bscdot \bs{J}} \lesssim k_{\bs{B}\bstimes\bs{J}}\approx k_\mr{rms} \sim k_\eta$.

\subsubsection{Transfer functions}

We define the shell-filtered kinetic-energy transfer function $\mc{T}_k[\bb{A}]$ of an arbitrary vector field $\bb{A}$ via
\begin{equation}\label{eq:shell_trans}
    \mc{T}_k[\bb{A}] \doteq \frac{2}{u_\mathrm{rms}^{2}} \sum_{q \in (2^{-1/4}k, 2^{1/4}k]} \bb{u}_\bb{q}^* \bcdot \bb{A}_\bb{q},
\end{equation}
where the star denotes the complex conjugate and $\bb{A}_\bb{q}$ denotes the Fourier amplitude of $\bb{A}$ with wavenumber $\bb{q}$. If $\bb{A}$ has the units of length squared over time \dblbrck{such as any term from the momentum equation \eqref{brag_MHD:mom}}, then $\mc{T}_k[\bb{A}]$ has the units of inverse time, and $\mc{T}_k[\bb{A}]$ represents the rate of the kinetic energy flowing due to $\bb{A}$ into the Fourier shell at $k$ of width $2^{1/2}k$. For example, $\mc{T}_k[\bb{B}\bcdot \grad \bb{B}]$ denotes the energy flowing into the velocity field at shell $k$ due to the Lorentz force. This diagnostic can be used to probe the scale-by-scale energy balance in \eqref{brag_MHD}. We also define the root-mean-square shell-filtered kinetic-energy transfer function,
\begin{equation}\label{eq:shell_trans_rms}
    \mc{T}^{\mathrm{rms}}_k[\bb{A}] \doteq \frac{2}{\langle u^4\rangle^{1/2}} \left[ \sum_{q \in (2^{-1/4}k, 2^{1/4}k]}\left(\frac{1}{2}\Re( \bb{u}_\bb{q}^* \bcdot \bb{A}_\bb{q})\right)^2\right]^{1/2}.
\end{equation}
This diagnostic serves as an alternative to the fourth-order spectra previously used by \citet{Scheko_sim} and features the added benefit of enabling quantitative comparison between the nonlinear terms in the momentum equation.

Finally, we define a shell-filtering procedure on a vector field $\bb{A}$ as
\begin{equation}\label{eq:shell_u}
    \bb{A}^{[\mathrm{range}]} \doteq \int_{k \in \mathrm{range}} \frac{\od^3 \bb{k}}{(2\upi)^3} \,\bb{A}_\bb{k} \,\rme^{\imag \bs{k\bcdot x}},
\end{equation}
where the integration is taken over the specified range in $k$ space. We utilize three ranges: $[k]$ denotes modes in the shell of width $\sqrt{2}$ with range $(2^{-1/4}k,2^{1/4}k]$; $[{<}k]$ denotes all modes with wavenumber magnitude less than $2^{-1/4}k$; and $[{>}k]$ denotes all modes with wavenumber magnitude greater than $2^{1/4}k$. The shell-filtered quantity $\bb{A}^{[\mathrm{range}]}$ can be used to determine the amount of energy transfer from one region of $k$ space to another. For instance, the quantity $\mc{T}_k[\bb{u}\bcdot \grad \bb{u}^{[<K]}]$ denotes the net transfer of kinetic energy from all modes with wavenumber magnitudes ${<}2^{1/4}K$ to modes in the shell $k \in (2^{-1/4}K,2^{1/4}K]$. Such shell-filtered quantities have been used in analyses of spectral energy transfer in MHD guide-field turbulence \citep[e.g.][]{Alexakis2005,Grete2017} and have also been used alongside H\"older's inequality to establish constraints on non-local transport in the fluctuation dynamo \citep{Beresnyak12}. In this paper, we use the shell-filtered quantity $\bb{u}^{[\mathrm{range}]}$ to compare the relative strengths of the hydrodynamic nonlinearity and the viscous stresses (figures \ref{fig:trans_unlim} and \ref{fig:stokes_trans}), as well as to determine how the  motions at scale $k$ affect the growth of the magnetic energy (figure \ref{fig:unlim_trans}).

\subsubsection{Structure functions}

To probe the structure of the turbulent velocity field, it is useful to introduce structure functions, which relay information about the scale-by-scale structure and spatial anisotropy with respect to the \emph{local} magnetic-field direction. In particular, we employ three-point, second-order structure functions for increment $\bb{\ell}$, defined by
\begin{align}\label{eqn:strf}
    \mathrm{SF}_2[\bb{u}](\bb{\ell}) &\doteq \langle | \bb{u}(\bb{x}+\bb{\ell}) - 2\bb{u}(\bb{x}) + \bb{u}(\bb{x}-\bb{\ell}) |^2\rangle.
\end{align}
These can be used to extract information about variations of a given field along and across the local magnetic-field direction by conditioning the box average on the alignment of the point-separation vector $\bb{\ell}$ with the local magnetic field, defined by $\bb{B}_\bs{\ell} \doteq [\bb{B}(\bb{x}+\bb{\ell})+\bb{B}(\bb{x})+\bb{B}(\bb{x}-\bb{\ell})]/3$. This conditioning yields the parallel and perpendicular structure functions (cf.~\citealt{chen12})
\begin{subequations}\label{eqn:strf_prl_prp}
\begin{align}
    \mathrm{SF}_2[\bb{u}](\ell_\parallel) &\doteq \langle | \bb{u}(\bb{x}+\bb{\ell}) - 2\bb{u}(\bb{x}) + \bb{u}(\bb{x}-\bb{\ell}) |^2 ; \:0 \le \theta < \upi/18 \rangle, \\
    \mathrm{SF}_2[\bb{u}](\ell_\perp) &\doteq \langle | \bb{u}(\bb{x}+\bb{\ell}) - 2\bb{u}(\bb{x}) + \bb{u}(\bb{x}-\bb{\ell}) |^2 ;\: 8\upi/18 < \theta \le \upi/2\rangle,
\end{align}
\end{subequations}
respectively, where $\theta \doteq \arccos | \bb{B}_\bs{\ell}\bcdot\bb{\ell}/B_\ell \ell|$ is the angle between the point separation vector and the local magnetic field. 

The parallel and perpendicular structure functions may be combined to calculate the scale-dependent anisotropy of the fluctuations. For example, equating the two,
\begin{equation}\label{eqn:scale_aniso}
   \mathrm{SF}_2[\bb{u}](\ell_\parallel) =  \mathrm{SF}_2[\bb{u}](\ell_\perp),
\end{equation}
provides $\ell_{\parallel,\bs{u}}$ as a function $\ell_{\perp,\bs{u}}$, or vice versa.

Power laws with exponent $\alpha$ that appear in structure functions translate to Fourier spectra with spectral index $-\alpha-1$. The use of a three-point stencil allows one to resolve spectral indices that are less steep than $-5$, or structure functions with exponents less than $4$~\citep{lazarianPogosyan}.\footnote{\,A five-point stencil can do even better, resolving spectral indices as steep as $-9$~\citep{cho09}, although such steep spectra are not encountered here. A three-point structure function exhibiting an exponent near 4 indicates a spectral index of $-5$ or steeper. }


\section{Results}\label{sec:results}

\subsection{Stages of fluctuation dynamo}\label{sec:overview}

There are four stages in the typical evolution of the fluctuation dynamo:
\vspace{1ex}
\begin{enumerate}
    \item The {\em diffusion-free stage}, during which the magnetic diffusion due to resistivity has yet to become large enough to influence the growth of the magnetic field. This stage occurs only if the Prandtl number ${\rm Pm}$ is sufficiently large and the scale of the initial magnetic field is much larger than the resistive scale. Once the magnetic field has become folded enough by the flow that the bulk of the magnetic energy reaches the resistive scale, magnetic diffusion becomes important and the dynamo enters\dots
    \item \dots the {\em kinematic stage}, in which the magnetic energy continues to grow despite the resistivity (if the flow is indeed a dynamo and ${\rm Rm}$ is above threshold) but the Lorentz force ($\bb{B}\bcdot \grad \bb{B}$) is still too feeble to exert any dynamical feedback on the field-amplifying turbulence. In MHD, the kinematic dynamo is linear in $\bb{B}$ (though  nonlinear in the random fields), resulting in exponential growth of $B_\mathrm{rms}$ \citep{Kazantsev,moffatt78,Kulsrud1992}. The Braginskii viscosity introduces a dependence in the velocity equation on the unit vector $\eb$. If the parallel viscous stress (which is just the Maxwell stress with $B^2/4\upi$ replaced by $\rmDelta p$) is sufficiently large, the `kinematic' phase is then fundamentally nonlinear: even though the magnetic field is dynamically weak, its structure influences the properties of the flow, which in turn affects induction in a nonlinear way. On the other hand, if the Braginskii viscosity is subject to hard-wall limiters, then its efficacy is reduced to be comparable with that of the Maxwell stress, rendering this stage truly `kinematic'. Eventually, the magnetic field becomes strong enough for the Lorentz force to exert a back reaction on the smallest-scale eddies, suppressing their ability to amplify the magnetic field. The dynamo then enters\dots
    \item \dots the {\em nonlinear stage}, in which the magnetic-field amplification is driven by progressively larger (and slower) eddies and the dynamo begins to slow down \citep[e.g.][]{Scheko_theory,Scheko_sim,maron04,choDynamo,Beresnyak12}. An estimate of the growth rate in this stage may be obtained by positing that the Lorentz force disables all eddies with energy less than the total magnetic energy. If one assumes Kolmogorov scalings with eddies at scale $\ell$ having energy $E(\ell) \propto \ell^{2/3}$, then the scale of the smallest eddies that appreciably stretch and grow the field is given by $E(\ell) \sim \langle B^2\rangle$. This leads to secular evolution:
    \begin{align}
       \frac{1}{2}\D{t}{\langle B^2\rangle} = \langle B^2\ROS \rangle \sim \frac{\langle B^2 \rangle  [E(\ell)]^{1/2}}{\ell}  \sim \frac{[E(\ell)]^{3/2}}{\ell} \sim \mathrm{const}.
    \end{align}
    Therefore, as the field-stretching scales shift, exponential growth gives way to a linear-in-time growth of magnetic energy.
    \item The fourth and final stage of the dynamo is {\em saturation}, which is achieved when the magnetic and kinetic energies become comparable (though not necessarily scale by scale -- see, e.g., \citealt{Scheko_theory,Scheko_sim}).
\end{enumerate}
\vspace{1ex}
In what follows, the evolution and characteristics of the dynamo in each of these stages are examined using results from the hard-wall-limited Braginskii-MHD, unlimited Braginskii-MHD, and isotropic-MHD simulations. We begin with a comparison of the limited Braginskii-MHD and $\mathrm{Pm} = 1$, isotropic-MHD dynamo, which we find to be similar to one another in almost every respect.

%
%
\subsection{Limited Braginskii-MHD dynamo is similar to ${\rm Re}\gg{1}$, ${\rm Pm}\gtrsim{1}$ MHD dynamo}\label{sec:limited}

The first two stages of the dynamo take place while the magnetic field is dynamically weak. As a result, unless the Braginskii viscosity is negligibly small, a majority (by volume) of the plasma will have pressure anisotropies that  exceed the firehose and mirror instability thresholds, {\em viz.}~$ \mu_\mr{B}|\ROS| \gtrsim B^2$. Applying the hard-wall limiters then effectively disables the Braginskii viscosity in most of the plasma volume, effectively rendering  viscous transport mostly isotropic, at least until the saturated state is reached and  the magnetic field becomes dynamically strong. In what follows, we demonstrate this point both qualitatively and quantitatively through a series of diagnostics.

\subsubsection{Visual appearance of the velocity and magnetic fields}

%
%
\begin{figure}
    \centering
    \includegraphics[scale=0.71]{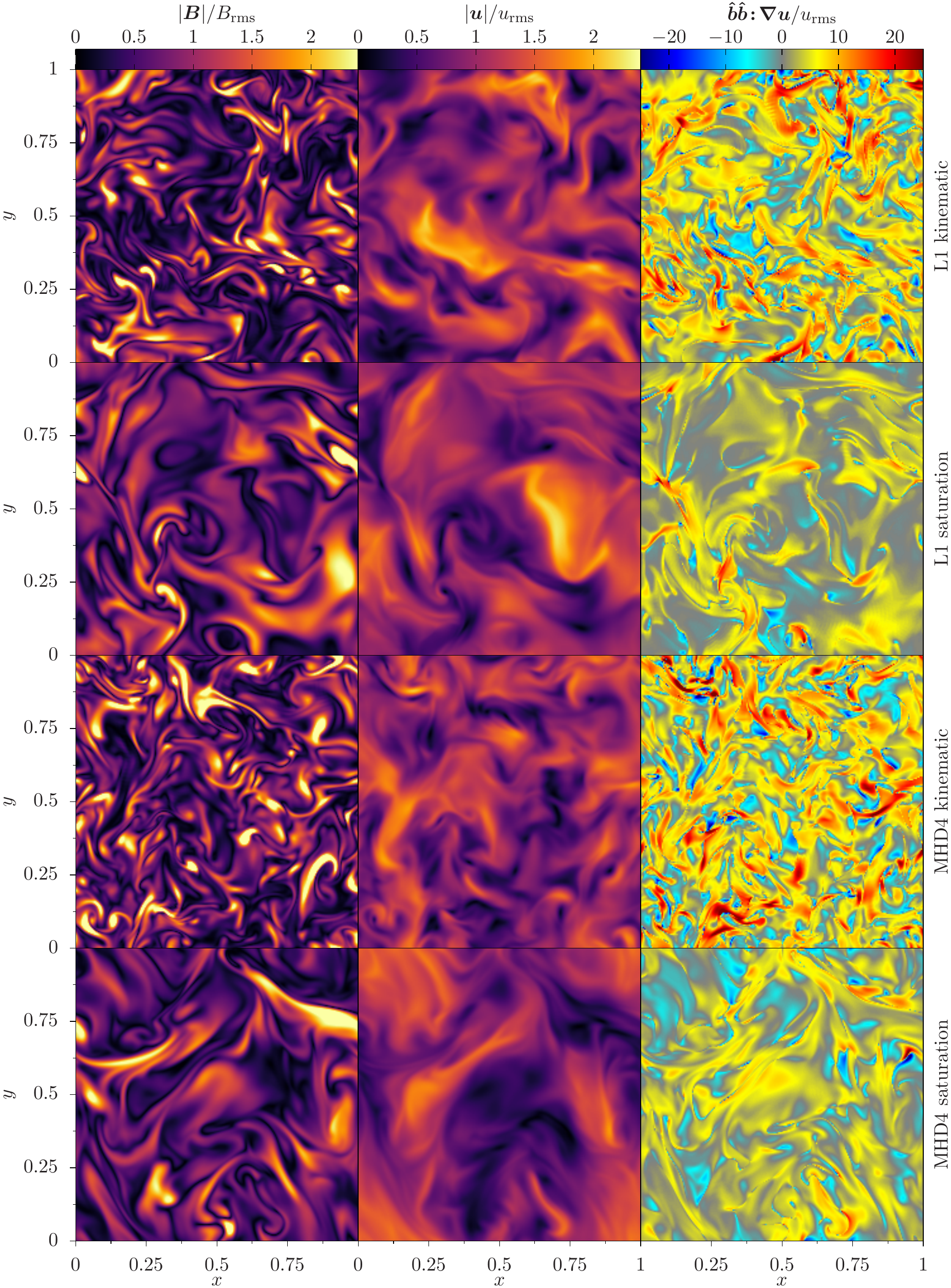}
    \caption{A two-dimensional cross-section of the magnetic field and flow in a hard-wall-limited Braginskii-MHD simulation (run L1) and in a comparable MHD simulation (run MHD4). Left (centre) [right] panels display the magnetic-field strength (velocity magnitude) [parallel rate of strain]. The top two rows display results from run L1 (parameters $\mu_\mr{B}^{-1}=20$, $\mu^{-1}=\eta^{-1}=1500$): the first row in the kinematic stage, the second row in the saturated state. The bottom two rows display results from run MHD4 (parameters $\mu^{-1}=\eta^{-1}=1500$): the first row in the kinematic stage, the second row in the saturated state. All plots are on a linear scale, with brighter regions denoting higher magnitudes.}
    \label{fig:printout_lim}
\end{figure} 

We first demonstrate that most of the qualitative features of the Braginskii-MHD dynamo with hard-wall pressure-anisotropy limiters are similar to those found in isotropic MHD. Figure \ref{fig:printout_lim} displays two-dimensional cross-sections in the $x$-$y$ plane of the magnetic-field strength, the velocity magnitude, and the parallel component of the rate-of-strain tensor from a limited Braginskii-MHD simulation (run L1) and from a comparable ${\rm Re}\gg{1}$, ${\rm Pm}=1$ simulation using isotropic MHD (run MHD4), both in the kinematic stage and in saturation. The only difference between these simulations is that $\mu^{-1}_{\rm B} = 20$ in the former. Despite this difference, the cross-sections of all displayed quantities are difficult to distinguish between the two systems. In both runs, the magnetic field is dominated by small-scale fluctuations that grow to somewhat larger scales in the saturated state. The usual folded structure of the magnetic field, with direction reversals at the resistive scale and field lines curved at the scale of the flow \citep[e.g.][]{Scheko_sim}, is manifest in both the kinematic and nonlinear stages. Both cases also feature a ${\rm Re}\gg{1}$ flow characterized by chaotic structures across multiple scales, with a tendency for the flow to shift to somewhat larger scales in the saturated state when the now  dynamically important Lorentz force is able to exert an influence on the dynamics. In principle, the Braginskii viscous stress could also influence the flow structure and dynamics, but its regulation to values comparable to the Maxwell stress by the hard-wall limiters merely changes the Maxwell stress by a factor of order unity.

\subsubsection{Evolution of magnetic energy}

%
%
\begin{figure}
    \centering
    \includegraphics[scale=0.86]{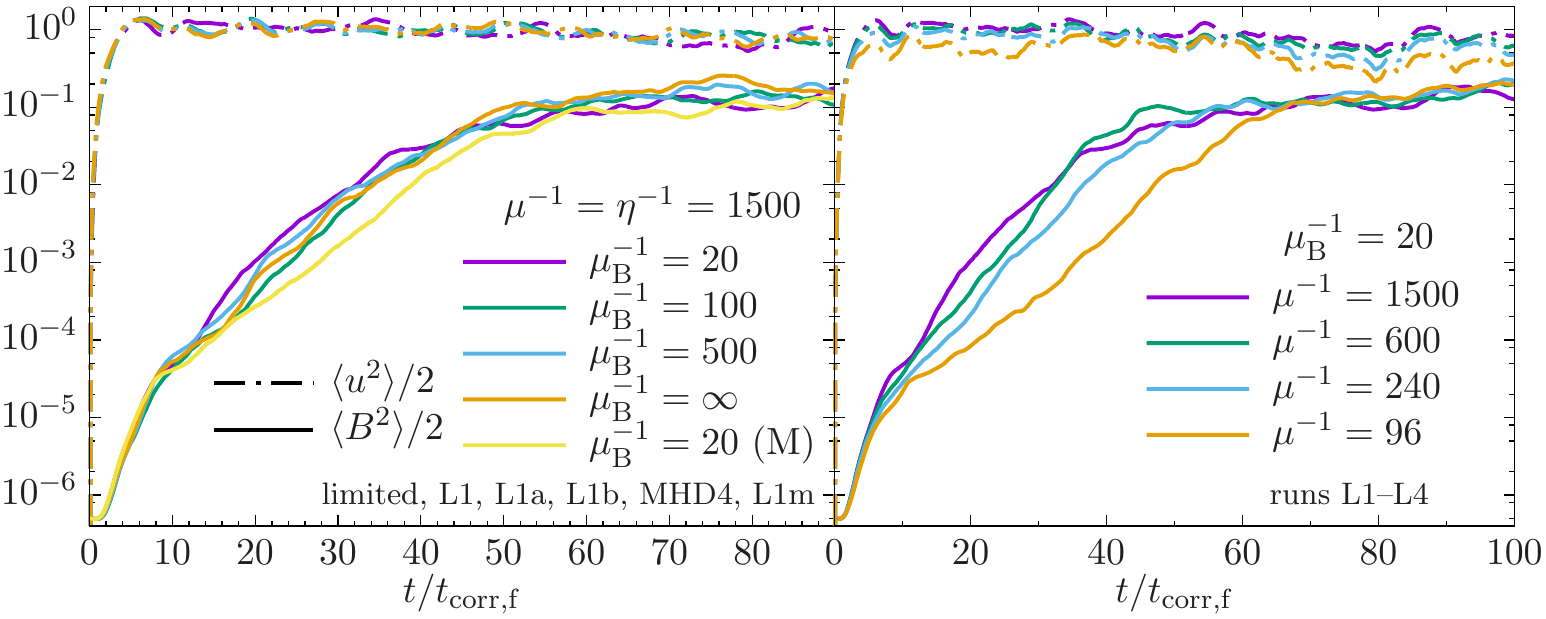}
    \caption{Evolution of the kinetic (dash-dotted lines) and magnetic (solid lines) energies for simulations employing hard-wall pressure-anisotropy limiters with varying Braginskii viscosity $\mu_{\rm B}$ and fixed isotropic viscosity $\mu$ (left) and  varying isotropic viscosity $\mu$ and fixed Braginskii viscosity $\mu_{\rm B}$ (right). A Braginskii-MHD simulation employing hard-wall limiters only on the mirror side, marked with an `(M)', is included in the left panel.}
    \label{fig:energy_lim} 
\end{figure}

Figure \ref{fig:energy_lim} displays the evolution of the boxed-averaged magnetic energy $\langle B^2 \rangle /2$ for a series of Braginskii-MHD simulations with hard-wall limiters and isotropic-MHD simulations. In the left panel, various Braginskii viscosities $\mu_\mr{B}$ and fixed Laplacian diffusivities ($\mu=\eta = 5\times10^{-4}$) are used. When the magnetic field is very weak, the hard-wall limiters affect the majority of the plasma, thus effectively disabling the parallel viscous stress. As a result, the growth rate of the magnetic energy in the kinematic stage is largely independent of $\mu_\mr{B}$ and exhibits magnetic-field growth closely resembling its isotropic MHD counterpart (the orange line). In the right panel, the Braginskii viscosity is fixed at $\mu^{-1}_{\rm B}=20$ while the isotropic viscosity $\mu$ is varied and the magnetic diffusivity is held fixed at $\eta^{-1}=1500$. As in the left panel, the parallel viscosity is effectively disabled at $\langle B^2\rangle\ll 1$ by the hard-wall limiters, and so the scale of the fastest eddies is determined instead by the isotropic viscosity. Accordingly, as $\mathrm{Pm} \sim 1$ is approached the growth rate of the magnetic energy converges to a maximal value (the growth rate would begin to \emph{decrease} once $\mathrm{Pm}$ decreased to be ${<}1$; see \citealt{Vincenzi2002} and \citealt{Iskakov2007}).

\subsubsection{\label{sec:powerspec} Power spectra of the velocity and magnetic fields}

%
%
\begin{figure}
    \centering
    \includegraphics[scale=0.79]{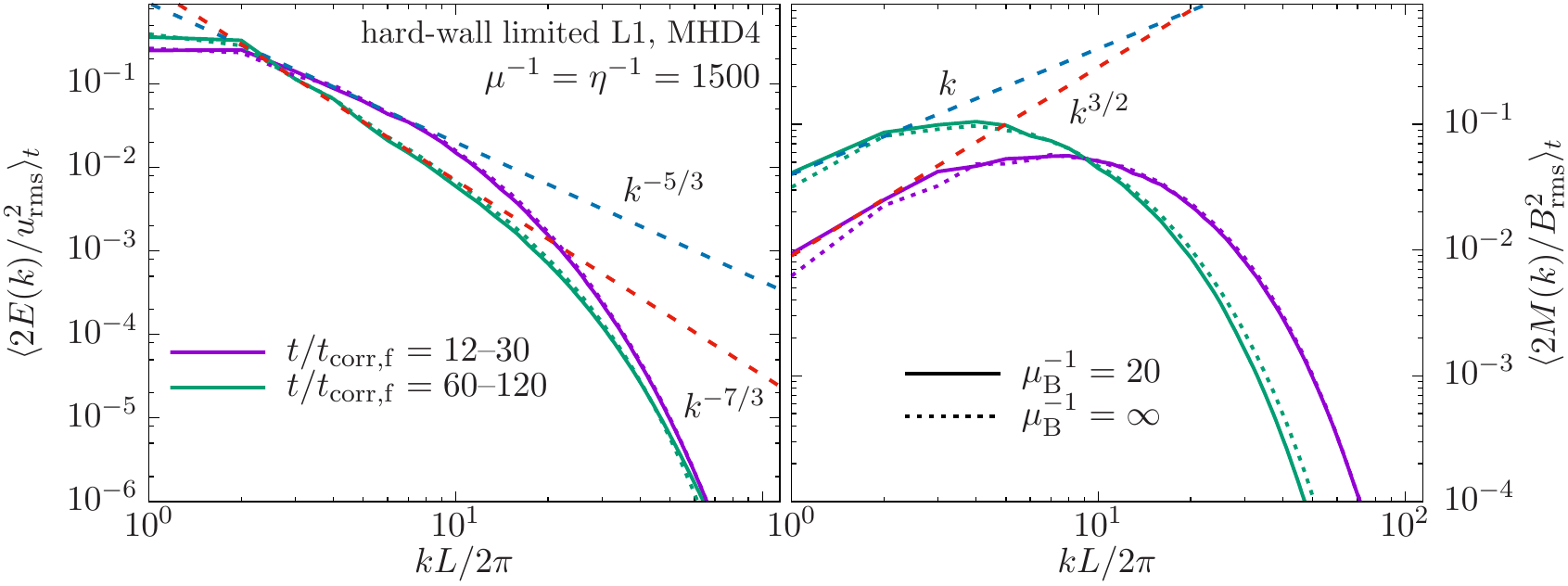}
    \caption{Time-averaged wavenumber spectra of  kinetic energy (left) and  magnetic energy (right) from run L1  with hard-wall-limited pressure anisotropy (solid lines) and run MHD4 with ${\rm Pm}=1$ (dotted lines) in the exponential phase (purple lines) and in saturation (green lines).}
    \label{fig:lim_spec}
\end{figure}

The similarity between MHD and limited Braginskii-MHD is further illustrated in figure \ref{fig:lim_spec}, which displays kinetic and magnetic energy spectra for the hard-wall-limited Braginskii simulation L1 and for the $\mathrm{Pm} = 1$ isotropic-MHD simulation MHD4, both in the kinematic stage (purple lines) and in saturation (green lines). For reference, both the Kasantsev $k^{3/2}$ scaling and a $k$ scaling are shown in the magnetic energy spectra.\footnote{\,The $M(k) \propto k$ scaling is a prediction of \citet{Malyshkin} for an unlimited Braginskii-MHD dynamo powered by a white-in-time velocity field, whose structure is biased by a Braginskii viscous stress acting along a straight but alternating magnetic field (similar to the geometry of a magnetic fold). An additional assumption made in their calculation is that there are no motions, either perpendicular or parallel, below the parallel-viscous scale. This assumption requires an {\it ad hoc} closure to regularize the perpendicular velocity variations that are otherwise undamped by the Braginskii viscosity; those authors argue that those variations couple nonlinearly to the Braginskii-damped motions (an effect those authors name `effective rotational damping') on parallel-viscous scales and thus take on the same $k$-space spectrum \dblbrck{proportional to $J_{0k}$, their equation (13)}. In contrast, the turbulent velocity fields found in both our limited and unlimited simulations contain dynamically important motions below the parallel-viscous scale (namely, motions that mix field lines and promote their resistive diffusion).} 
The kinetic-energy spectra from both runs have all the characteristics of a high-Re turbulent dynamo: a Kolmogorov $-5/3$ spectrum in the kinematic stage and a steeper $-7/3$ spectrum in the saturated state (as seen at these moderate Reynolds numbers but not explained by \citealt{Scheko_sim}). The magnetic spectrum in the kinematic stage is consistent with the \citet{Kazantsev} $k^{3/2}$ scaling at small wavenumbers, while in saturation it is shallower. The peaks of the magnetic spectra for both simulations move to larger scales as saturation is reached. In the isotropic-MHD dynamo, this migration to larger scales has been explained as a consequence of `selective decay' \citep{Scheko_theory,Scheko_sim} -- the increased importance of resistive dissipation on smaller-scale magnetic-field fluctuations as the Lorentz force begins to suppress field-stretching motions. Because limited Braginskii viscosity in regions of positive parallel rate of strain satisfying $3\mu_{\rm B}\ROS \gtrsim 1/\beta$ effectively enhances the magnetic tension by only a factor of 3/2 (since $B^2/4\upi \rightarrow B^2/4\upi + \Deltap = (3/2) B^2/4\upi$ at the mirror threshold), the similarities between the spectra in the saturated state of the limited-Braginskii and isotropic-MHD runs are not particularly surprising. The pressure anisotropy in regions of negative parallel rate of strain ($\ROS \lesssim 0$) is instead adiabatically pushed towards the firehose threshold, at which the limited Braginskii stress exactly nullifies the magnetic tension. The effect of this cancellation on the magnetic spectrum in saturation appears to be minimal, however, likely because the volume-filling factor of regions whose pressure anisotropy lies beyond the firehose threshold is small -- see the top panels of figure~\ref{fig:brazil}. (The paucity of regions with $3\mu_{\rm B}\ROS \lesssim -2/\beta$ does not necessarily imply that the magnetic energy is growing in the majority of the volume. Rather, resistive dissipation competes with suppressed field-stretching motions to give saturation, despite the majority of the volume having $\ROS > 0$.)

\subsubsection{Scale-by-scale energy transfer}

%
%
\begin{figure}
    \centering
    \includegraphics[scale=0.9]{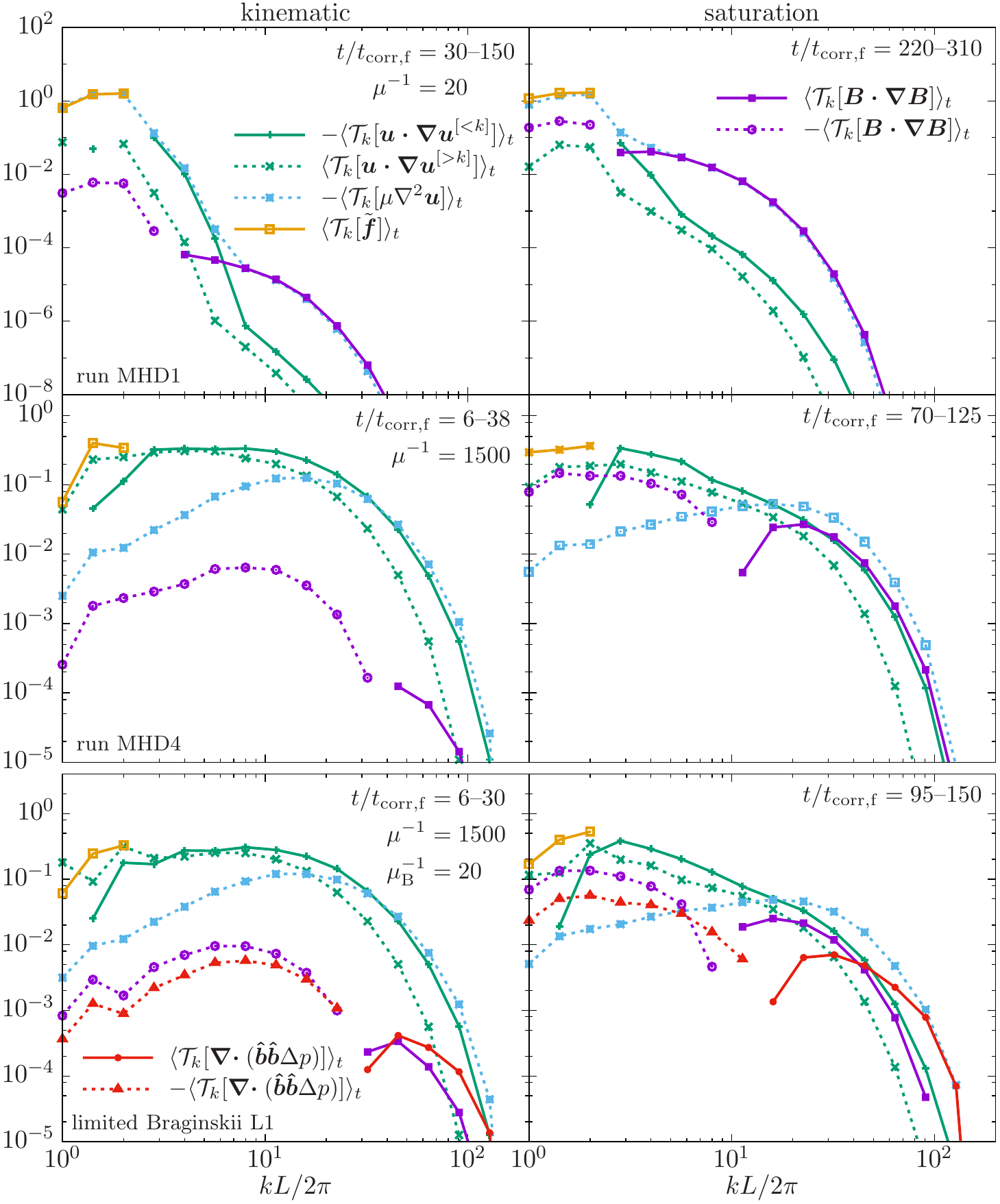}
    \caption{Shell-filtered kinetic-energy transfer functions $\mc{T}_k$ [see \eqref{eq:shell_trans}] for MHD (top two rows) and hardwall-limited Braginskii MHD (bottom row) in the kinematic (left) and saturated (right) stages.  Solid (dotted) lines denote energy flowing into (out of) the shell centred at wavenumber~$k$.}
    \label{fig:trans_mhd_lim}
\end{figure}

The relative importance of the various terms of the momentum equation \eqref{brag_MHD:mom} in shaping the kinetic-energy spectrum may be calculated using the energy-transfer functions introduced in \S\,\ref{sec:diagnostics}. Figure~\ref{fig:trans_mhd_lim} shows the time-averaged kinetic-energy transfer function $\mc{T}_k$ [see \eqref{eq:shell_trans}] for isotropic MHD with $\mathrm{Pm} \gg 1$ (run MHD1; top row) and $\mathrm{Pm} = 1$ (run MHD4; middle row), as well as for Braginskii MHD employing hard-wall limiters (run L1; bottom row). These plots quantify the flow of kinetic energy into and out of the Fourier shell centred geometrically around $k$, with solid (dotted) lines signifying energy flowing into (out of) each shell. 

During the kinematic regime of the $\mr{Pm} = 1$ isotropic-MHD dynamo, the magnetic tension (purple lines) remains subdominant to the other terms in the momentum equation, and thus the energy balance is predominantly between the hydrodynamic nonlinearity (green lines) and energy injection at large scales (orange lines), and the hydrodynamic nonlinearity and viscous dissipation (blue lines) at small scales. In the case of $\mr{Pm} \gg 1$ (run MHD1), the kinetic energy cascade is terminated by viscosity at large scales, and so the magnetic tension is balanced by viscous dissipation. This results in a steep sub-viscous velocity spectrum (previously explained by~\citealt{Scheko_sim}, see figure 18 in that reference). These motions play no dynamical role in the evolution of the turbulence or magnetic field.

As the isotropic-MHD dynamo approaches saturation, the magnetic field becomes dynamically important and its tension begins to influence the turbulent motions. Interestingly, the magnetic tension serves to \emph{redistribute} turbulent energy from large to small scales, a feature not previously noted in the literature. It is shown in \S\,\ref{sec:unlimited} that the Braginskii-MHD dynamo in the unlimited regime also exhibits this behaviour throughout its entire evolution. We demonstrate in appendix~\ref{sec:m-i_null} that the small-scale turbulent motions driven by the magnetic tension (or Braginskii stress) serve to nullify the parallel rate of strain $\ROS$, thereby counteracting the overall growth of the magnetic field and allowing the dynamo to saturate.

The hard-wall-limited Braginskii-MHD dynamo exhibits features that are largely similar to the $\mr{Pm} = 1$ isotropic-MHD dynamo. This is not surprising, as the limiters reduce the strength of the parallel viscous stress to be comparable to the magnetic tension. Importantly, the parallel viscous stress (red lines) mimics the magnetic tension in both the `kinematic' and saturated regimes, whereby it serves to redistribute kinetic energy from large to small scales (see appendix~\ref{sec:m-i_null}).

\subsubsection{Structure functions of the velocity field}

%
%
\begin{figure}
    \centering
    \includegraphics[scale=0.9]{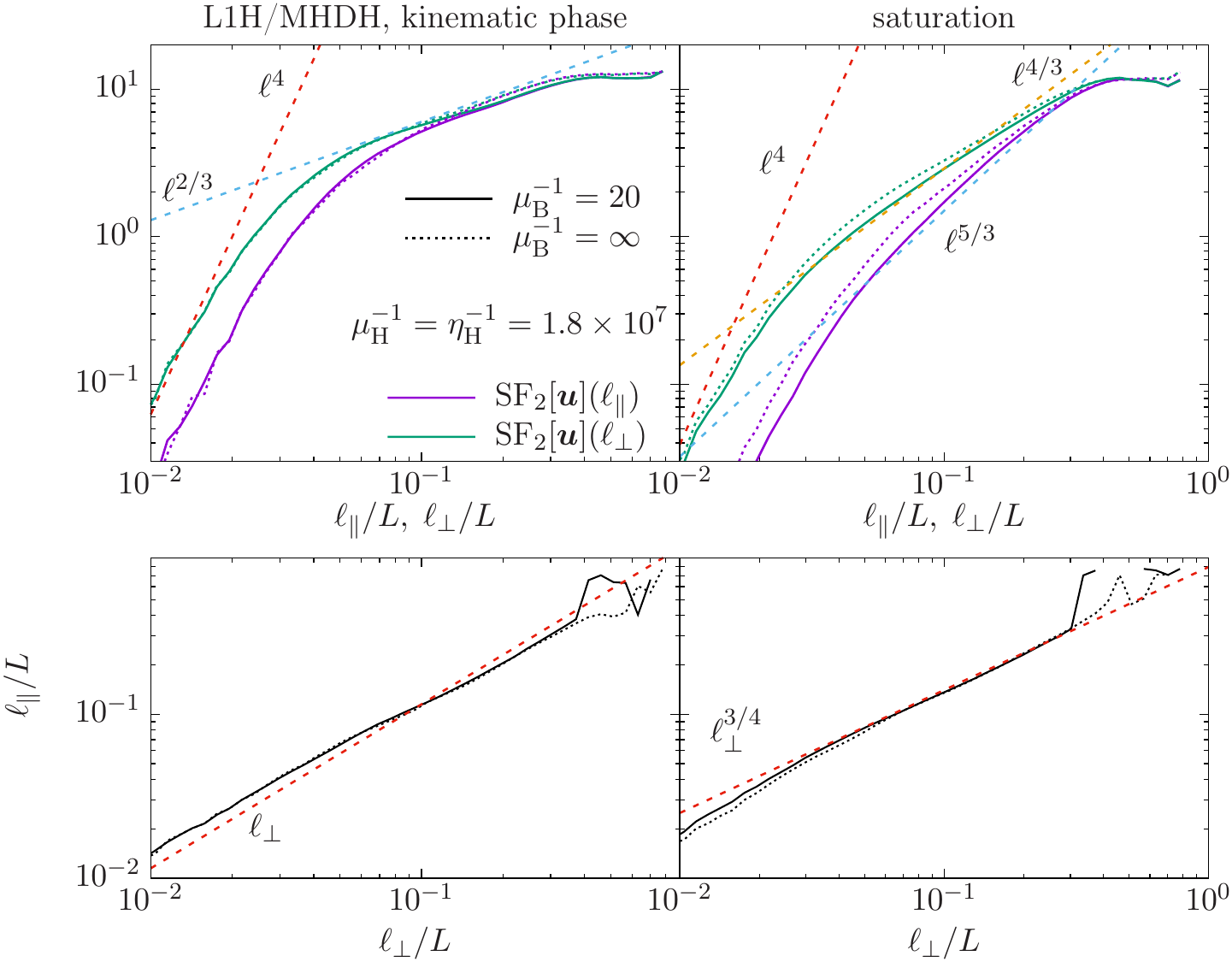}
    \caption{Various three-point, second-order structure functions for the Braginskii-MHD system with hard-wall-limited pressure anisotropy (L1H) as well as the MHD simulation employing hyper dissipation (MHDH) in the kinematic (left) and saturation (right) stages. Solid (dotted) line denotes the Braginskii-MHD (MHD) simulation.  `$\parallel$' (`$\perp$') refers to the direction parallel (perpendicular) to the local scale-dependent magnetic field [see \eqref{eqn:strf_prl_prp}]. Bottom panels show the scale-dependent anisotropy scaling of the parallel coherence length $\ell_\parallel$, as defined by \eqref{eqn:scale_aniso}.}
    \label{fig:struct_lim}
\end{figure}

The energy spectra shown in figure \ref{fig:lim_spec} do not provide information about the relative contributions of field-parallel/perpendicular gradients of field-parallel/perpendicular quantities to the energetics. Because it is only the field-aligned component of the rate-of-strain tensor that is responsible for increasing the magnetic energy ($\ROS$), it is worthwhile to separate out the field-parallel and -perpendicular components of the velocity field and study their individual contributions to the kinetic-energy spectra. To do so, we calculate the structure functions \eqref{eqn:strf} and \eqref{eqn:strf_prl_prp} of the velocity field in the hard-wall-limited, $\mu^{-1}_\mr{B}=20$ Braginskii-MHD simulation (L1H), as well as for the MHD simulation employing hyper-diffusion (MHDH). These simulations use hyper-diffusion in order to maximize the inertial range; this facilitates cleaner measurements of scale-dependent anisotropy than those obtained using only Laplacian dissipation.\footnote{\,One potential side effect of employing higher-order diffusion is the `bottle-neck' effect, which inhibits the transfer of energy to larger scales. 
While this is known to occur in some simulations of the large-scale dynamo~\citep{Brandenburg_hyper}, this effect is not seen in our simulations of the small-scale dynamo. Our results show no qualitative differences when we employ hyper-diffusion rather than Laplacian diffusion.} The top row of figure \ref{fig:struct_lim} displays the resulting curves in the kinematic stage (left) and in saturation (right). The structure functions for both systems during their kinematic stage are largely isotropic, exhibiting a $\ell^{2/3}$ power law (corresponding to a spectral index of $-5/3$) in the inertial range. Beyond the viscous cutoff, all structure functions steepen to a slope close to ${\rm SF}_2 \propto \ell^4$, which is the maximal slope measurable by structure functions using a three-point stencil.\footnote{\,Recall that a second-order structure function using a three-point stencil exhibits an $\ell^4$ power law for smooth flows, rather than an $\ell^2$ power law resulting from using a two-point stencil.} In the saturated state, both runs exhibit an anisotropization of the turbulence, with the parallel and perpendicular structure functions exhibiting different scalings. The perpendicular structure functions are steeper than Kolmogorov, being roughly proportional to $\ell^{4/3}_\perp$ (corresponding to a $-7/3$ spectral index). The $\ell_\perp^{4/3}$ scaling was previously observed by \citet{Yousef2007}, who studied the effects of disparate-scale interactions between turbulence and dynamo-generated magnetic fields on the exact scaling laws of structure functions typically found in isotropic MHD turbulence. We find the slopes of the parallel structure functions to be even steeper, with a scaling of approximately $\ell^{5/3}_\parallel$ (corresponding to a $-8/3$ spectral index). The structure functions of the limited-Braginskii and isotropic-MHD systems differ only in the saturated state, with those from run L1H being slightly steeper than those from run MHDH. This is likely due to addition of the parallel viscous stress to the magnetic tension force, resulting in a stronger influence of the magnetic field on the flow.

The difference in perpendicular and parallel scalings implies a scale-dependent anisotropy in the saturated state of the dynamo, which we quantify using \eqref{eqn:scale_aniso} and display in the bottom-right panel of figure \ref{fig:struct_lim}. Both runs exhibit a scaling close to $\ell_\parallel \sim \ell_\perp^{3/4}$.\footnote{\,This is to be contrasted with the $\ell_\parallel \sim \ell_\perp^{2/3}$ and $\ell_\parallel \sim \ell_\perp^{1/2}$ scalings predicted for guide-field MHD turbulence respectively without \citep{Goldreich1995} and with \citep{Boldyrev2006,Chandran2015,Mallet2017} scale-dependent dynamic alignment and intermittency corrections.} By contrast, in the kinematic stage (bottom-left panel of figure \ref{fig:struct_lim}) the anisotropy scaling is linear in both systems, indicating isotropic turbulence.

\subsubsection{Characteristic scales of the magnetic field}\label{sec:limscales}

%
%
\begin{figure}
    \centering
    \includegraphics[scale=0.9]{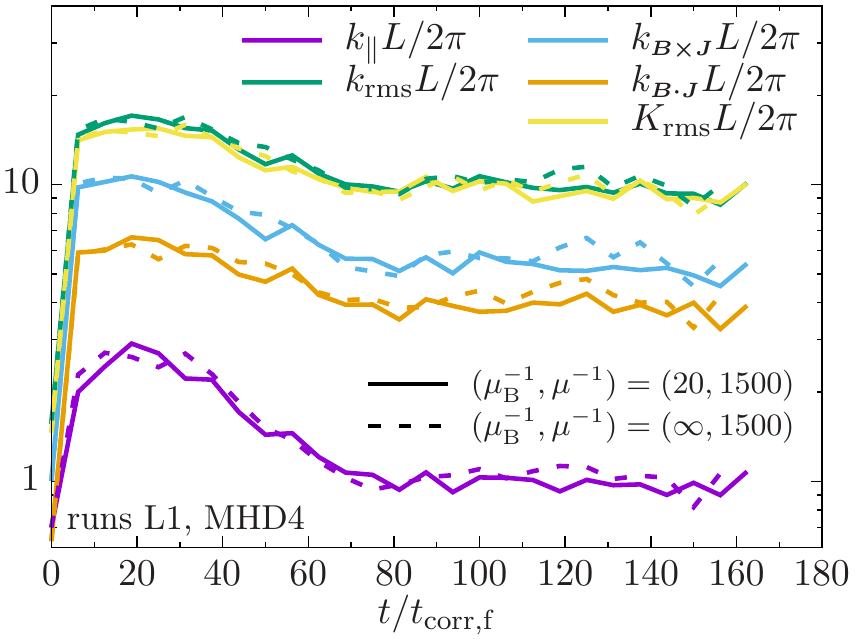}
    \caption{Evolution of the characteristic wavenumbers  \eqref{char-wavenumbers} for isotropic MHD (dashed) and Braginskii MHD with hard-wall pressure-anisotropy limiters (solid lines).}
    \label{fig:wavenumbers_lim}
\end{figure}

As a final point of contact with results from isotropic MHD, we present in figure~\ref{fig:wavenumbers_lim} the characteristic wavenumbers \eqref{char-wavenumbers} quantifying the geometry of the magnetic field in the runs L1 and MHD4. The similarity between the two systems is quite good, even in the saturated state, in which the limited Braginskii viscosity is dynamically important. As the magnetic field is stretched and folded by the flow, it is organized into long thin structures (folds). As a result, the wavenumbers in the kinematic phase satisfy the ordering $k_\parallel < k_{\bs{B\cdot J}} < k_{\bs{B\times J}}$, with each of these wavenumbers decreasing and the latter two  becoming more comparable in the saturated state, as the magnetic folds become more ribbon-like.\footnote{\,The limited scale separation afforded by these moderate-Reynolds-number simulations means that the `$\ll$' orderings given in \S\,\ref{sec:diagnostics} for magnetic folds and ribbons (namely, $k_\parallel \lesssim k_{\bs{B}\bscdot \bs{J}} \ll k_{\bs{B}\bstimes\bs{J}}$ and $k_\parallel \ll  k_{\bs{B}\bscdot \bs{J}} \lesssim k_{\bs{B}\bstimes\bs{J}}$, respectively) are not fully realized.}  The PDF of $K$ (not shown) is nearly identical to that found in run MHD4 (the blue line in figure \ref{fig:curvature} below), having a peak concentrated near the viscous scale and a power-law tail ${\propto}K^{-13/7}$ (as predicted for the high-Pm MHD dynamo by \citealt{Scheko_theory2}).  The shallowness of these tails implies that all non-zero moments of the curvature PDF diverge in the absence of resistive regularization. As a result, the root-mean-square value of the magnetic-field-line curvature $K\doteq|\eb\bcdot\grad\eb|$ is comparable to $k_{\rm rms} \sim k_\eta$~\citep{Malyshkin_curv}.
The curvature PDF is then representative of a three-dimensional field whose regions of large curvature occupy only a small fraction of the volume.

%
%
\subsection{Unlimited Braginskii-MHD dynamo is similar to saturated MHD dynamo}\label{sec:unlimited}

Having provided several lines of evidence that pressure-anisotropy limiters revert the Braginski-MHD dynamo to its more traditional ${\rm Re}\gg{1}$, ${\rm Pm}\gtrsim{1}$ counterpart, and motivated by the observation of imperfect regulation of pressure anisotropy in the hybrid-kinetic simulations presented in \citetalias{StOnge_2017}, we now turn off those limiters and let the full Braginskii viscous stress act unabated on the flow. In this case, the dynamo takes on a very different character. Left unchecked by limiters, and without a rapidly growing mirror instability in Braginskii MHD to reign it in (see appendix \ref{ap:linear}), the pressure anisotropy is proportional to the parallel rate of strain and to $\mu_{\rm B}$. For large values of $\mu_{\rm B}$, it spills over both the firehose and mirror thresholds. No longer bound to the relatively meager Lorentz force in the kinematic stage, the large parallel viscous stress exerts a strong dynamical feedback on those fluid motions responsible for amplifying the magnetic field. The result is a strong suppression of $\ROS$, leading to a viscous anisotropization of the fluid flow. Because \eqref{brag_MHD:ind} implies $\rmd\ln B/\rmd t = \ROS$ in the absence of resistivity, this anisotropization ultimately results in a less efficient dynamo. Since the Braginskii viscous stress is similar in form to the magnetic tension force $\bb{B}\bcdot \grad \bb{B} = \grad \bcdot (B^2 \eb\eb)$, with the pressure anisotropy playing the role of the magnetic-field strength ({\it viz.}~$B^2 \rightarrow \Deltap \propto \od_t B^2$), one may expect similarities between the unlimited Braginskii-MHD dynamo and the isotropic-MHD dynamo in its saturated state. As in \S\,\ref{sec:limited}, we confirm these expectations by using a variety of diagnostics to compare the unlimited Braginskii-MHD runs with the isotropic-MHD runs. Some notable differences are also highlighted.

\subsubsection{Visual appearance of the velocity and magnetic fields}

%
%
\begin{figure}
    \centering
    \includegraphics[scale=0.71]{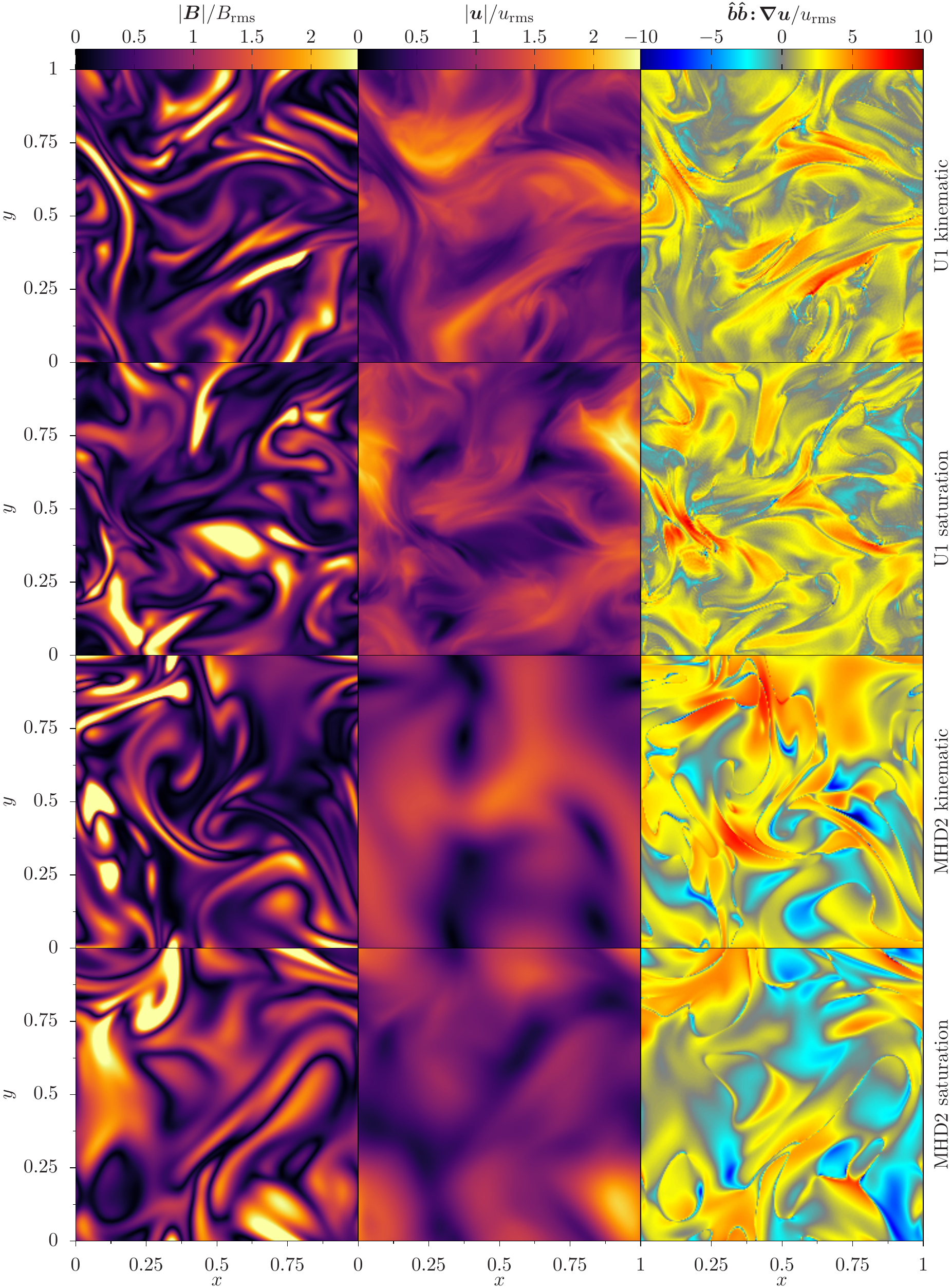}
    \caption{Same as figure \ref{fig:printout_lim} but for an unlimited Braginskii-MHD simulation (U1) and a comparable MHD run (MHD2). Left (centre) [right] panels display the magnetic-field strength (velocity magnitude) [parallel rate of strain]. The top two rows display results from the unlimited simulation U1 (parameters $\mu^{-1}=1500$, $\mu_\mr{B}^{-1}=20$, $\eta^{-1}=1500$): the first row in the kinematic stage, the second row in the saturated state. The bottom two rows display results from the MHD simulation MHD2 (parameters $\mu^{-1}=100$,  $\eta^{-1}=1500$): the first row in the kinematic stage, the second row in the saturated state. All plots are on a linear scale, with brighter regions denoting higher magnitudes.}
    \label{fig:printout_unlim}
\end{figure}
 
Figure \ref{fig:printout_unlim} displays the same quantities as in figure \ref{fig:printout_lim} but now for a Braginskii-MHD simulation without limiters (U1) and an isotropic-MHD simulation with $\mu=\mu_\mathrm{B}/5$ (MHD2). This choice of isotropic viscosity for comparison with Braginskii MHD was advocated by \citet{Malyshkin} as an effective closure for systems with a magnetic field that is isotropically tangled on sub-viscous scales. Indeed, the velocity fields in these two runs appear to be similar on scales larger than those on which the magnetic field is tangled. However, a major difference is that the velocity field in the unlimited Braginskii run features thin striations with sharp gradients across the local magnetic-field direction. The isotropic-MHD system instead exhibits only large-scale motions typical of order-unity Reynolds number flow. As a result, the Braginskii turbulent state throughout the evolution of the dynamo resembles more closely the saturated state of the high-Re MHD system (cf.~bottom centre panel of figure \ref{fig:printout_lim}), in which large-scale features of the velocity field are accompanied by small-scale turbulence. Another notable difference between the unlimited-Braginskii and isotropic-MHD runs is that, in the unlimited-Braginskii run, the magnetic field, velocity, and parallel rate of strain all exhibit remarkably little change in going from the kinematic stage to the saturated state -- a feature to which we return repeatedly throughout this section. 

A more subtle difference between runs U1 and MHD2 concerns the parallel rate of strain, displayed in the rightmost panels of figure \ref{fig:printout_unlim}. The MHD simulation features larger-scale patches of $\ROS$, with more extreme values, than found in the Braginskii-MHD simulation. This is because, in the unlimited Braginskii system, there is a dynamical feedback whereby the full pressure anisotropy driven by the field-stretching motions ({\it viz.}~$\ROS\ne 0$) dynamically suppresses those very same motions. There are also fewer regions that exhibit strong negative values of $\ROS$ in the unlimited run, most likely because the act of decreasing the magnetic-field strength with $\ROS<0$ is unstable to the production of firehose fluctuations that grow small-scale magnetic fields (and thus contribute a positive $\ROS$; see \citealt{Scheko_2008}, \citealt{Rosin_2011}, and \citealt{Melville} for further discussion of this point in the context of the parallel firehose instability). Further discussion of the contributions of small and large scales to the total parallel rate of strain and the suppression of $\ROS$ is deferred to appendix \ref{sec:magnetoimmutability}.

\subsubsection{Evolution of magnetic energy}\label{sec:unlim_evolution}

%
%
\begin{figure}
    \centering
    \includegraphics[scale=0.86]{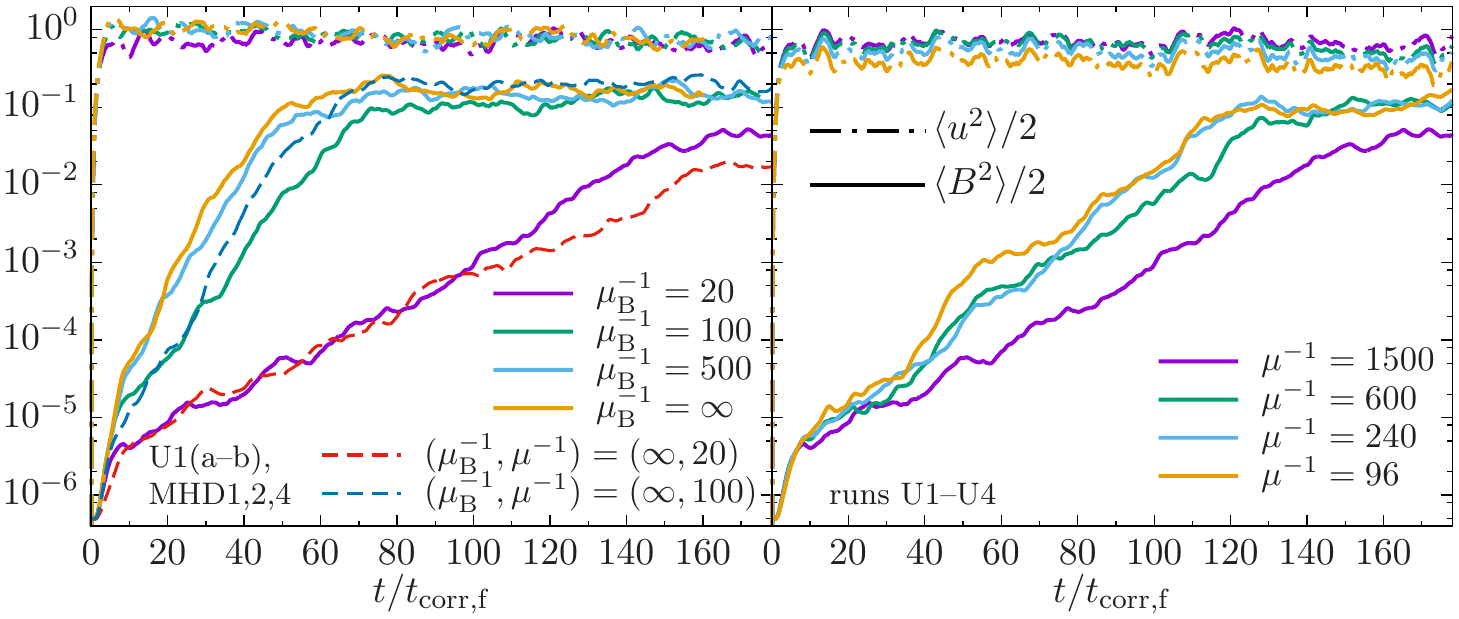}
    \caption{Evolution of the kinetic (dash-dotted lines) and magnetic (solid lines) energies for simulations that do not employ pressure-anisotropy limiters with  varying Braginskii viscosity and fixed isotropic viscosity (left)  and varying isotropic viscosity and fixed Braginskii viscosity (right). The evolution of magnetic energy in the MHD1 and MHD2 simulations ($\mu^{-1} = 20$ and $100$, respectively) is included for reference in the left panel (dashed lines). Solid lines in the left (right) panel have $\mu^{-1} = 1500$ ($\mu_\mathrm{B}^{-1} = 20$).}
    \label{fig:energy_unlim}
\end{figure}

The dynamical feedback of the full Braginskii viscosity on the flow affects the time evolution of the magnetic energy, shown in figure \ref{fig:energy_unlim}. As the Braginskii viscosity is increased, the viscous scale of the field-stretching motions becomes larger and the dynamo growth rate decreases accordingly (left panel), indicating that in the unlimited regime the dynamo growth rate is ultimately controlled by the Braginskii viscosity. Indeed, the growth of the average magnetic energy in the $\mu^{-1}_{\rm B} = 20$ and $100$ unlimited-Braginskii runs (purple and green solid lines, respectively) is almost identical to that in the $\mu^{-1} = 20$ and $100$ isotropic-MHD runs (red and blue dashed lines, respectively, in the same plot). Note that the value $\mu=\mu_\mathrm{B}/5$ advocated by \citet{Malyshkin} leads to a much faster dynamo than the unlimited-Braginskii-MHD dynamo it is intended to model (e.g., compare the blue-dashed and purple-solid lines). The reason for this difference is that, while the two simulations exhibit similar stretching motions, the resistive dissipation is effectively much higher in the unlimited Braginskii system due to the increased fraction of field-mixing motions in the rate-of-strain tensor (relative to the field-stretching motions, which are viscously suppressed). This increased magnetic dissipation results in a slower dynamo (see \S\,\ref{sec:kazantsev} for more on this subject). This indicates that the details of the sub-parallel-viscous range are important for the overall operation of the dynamo, and that the closure advocated by \citet{Malyshkin} is inappropriate for the case of large isotropic Reynolds number. 

In the right panel of figure \ref{fig:energy_unlim}, the Braginskii viscosity is fixed at $\mu^{-1}_{\rm B} = 20$ while the isotropic viscosity is varied. Somewhat counter-intuitively, the growth rate appears to decrease as the isotropic viscosity is decreased. A plausible explanation of this result is that a small isotropic viscosity can allow a cascade of perpendicular (or `interchange'-like) motion to small-scales. These motions, while not capable of growing the magnetic field, can bring field lines closer together and thereby accelerate the resistive reconnection of the field.  The deleterious effect of these mixing motions is a central issue in the $\mathrm{Pm} < 1$ dynamo \citep{Vincenzi2002,Boldyrev2004,Iskakov2007,Scheko_NJP2007}, where mixing from all kinetic energy scales promotes resistive annihilation, while only motions larger than the resistive scale can aid in amplification of the magnetic energy via stretching. This idea is further developed in \S\,\ref{sec:kazantsev}. Another anti-dynamo factor is that the rate of strain of the smaller-scale motions allowed by the decreased isotropic viscosity could act to cancel partially the parallel rate of strain driven by the large scales, modifying the dynamo growth rate without significantly increasing resistive mixing. This effect has been recently seen in Braginskii-MHD simulations of the magnetorotational instability~\citep{Kempski_2019}. This partial cancellation, related to the concept of `magneto-immutability', is investigated further in appendix \ref{sec:magnetoimmutability}. It will be shown that, while both effects are present in our simulations, the latter is of only minor consequence, while the former -- that of the efficiency of sub-parallel-viscous mixing motions -- has a significant impact on whether or not unlimited Braginksii MHD can exhibit a dynamo.

\subsubsection{Power spectra of the velocity and magnetic fields}

%
%
\begin{figure}
    \centering
    \includegraphics[scale=0.79]{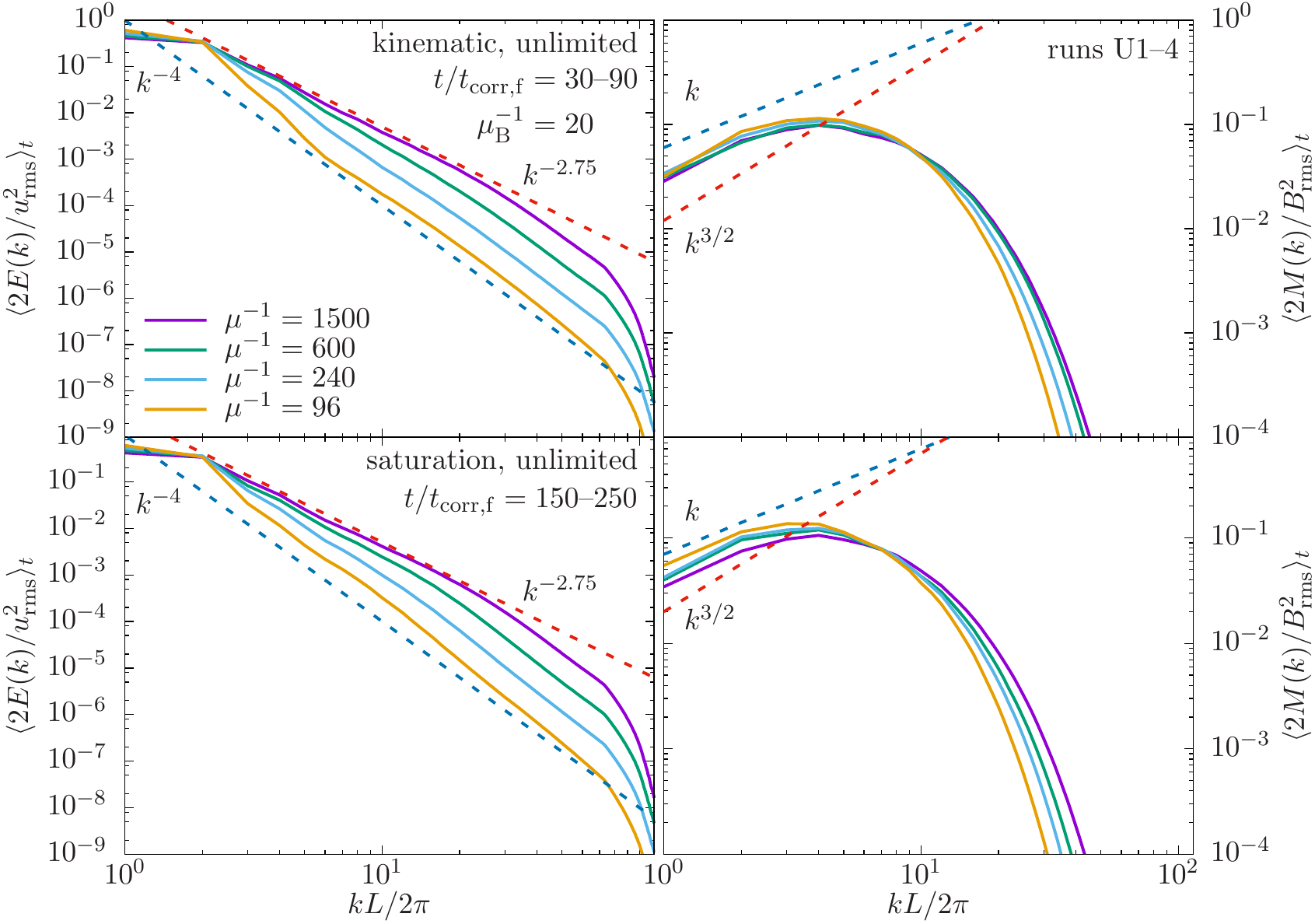}
    \caption{Time-averaged wavenumber spectra of kinetic energy (left) and magnetic energy (right) from simulations with unlimited pressure anisotropy for various values of isotropic viscosity $\mu$ in the kinematic stage (top) and the saturated state (bottom).}
    \label{fig:unlim_spec}
\end{figure}

A theme of the last two sub-sections is that, in Braginskii MHD, the perpendicular component of the fluid motions is allowed to cascade to sub-parallel-viscous scales. The left panel of figure \ref{fig:unlim_spec} provides quantitative evidence that this is the case, with less isotropic viscosity allowing a shallower spectrum and thus stronger small-scale velocity fluctuations. At the smallest nonzero value of isotropic viscosity (purple line), the unlimited runs exhibit a kinetic-energy spectrum $E(k)\sim k^{-2.75}$, which appears to be asymptotic in $\mu \rightarrow 0$ at fixed $\mu_\mathrm{B}^{-1} = 20$ (based on preliminary studies at even higher Re). While steep, this power law still implies a local rate of strain that increases with $k$. Eventually the spectrum experiences a break at $kL/2\upi \approx 30$, whereupon it exhibits a $k^{-4}$ power law down to the grid. This break can be taken as the effective \emph{perpendicular} viscous scale, the slope beyond which is sufficiently steep (spectral index $<{-3}$) for the rate of strain to decrease with $k$, i.e., for the fastest eddies to occur at the largest scales. The run with the next smallest value of isotropic viscosity (green line) also exhibits a similar spectrum, but the spectral break occurs around $kL/2\upi \approx 8$. For $\mu^{-1}=96$ (orange line), almost the entire kinetic-energy spectrum is proportional to $k^{-4}$.

Note that a small kink exists at the de-aliased grid-scale wavenumber $kL/2\upi = 224/3 \approx 75$ in all of the kinetic-energy spectra, regardless of the value of the isotropic viscosity. This is a result of small-scale energy injection by the unregulated mirror and firehose instabilities that are present in unlimited Braginskii MHD, the latter of which exhibits an ultraviolet catastrophe when $\mu = \eta = 0$ (see appendix \ref{ap:linear}). In our simulations, this small-scale energy injection is balanced by dissipation via the isotropic viscosity (this is shown later in figure \ref{fig:trans_unlim}).

The kinetic-energy spectra shown in figure \ref{fig:unlim_spec} appear to be independent of whether the dynamo is in the kinematic stage (top row) or in saturation (bottom row) -- a notable difference from the high-Re MHD dynamo, in which the kinetic-energy spectrum steepens from the Kolmogorovian $k^{-5/3}$ to $k^{-7/3}$ in the saturated state (at least at these limited resolutions; cf.~\citealt{Scheko_sim}). The accompanying magnetic-energy spectra, shown in the right panels of figure \ref{fig:unlim_spec}, also show little evolution from the kinematic stage to saturation (consistent with the visualization of the magnetic-field evolution in figure \ref{fig:printout_unlim}). In both the kinematic and saturated stages, these magnetic spectra are shallower than the Kazantsev $3/2$ scaling, and are instead closer to the scaling seen in the saturated state of the high-Re MHD simulation (cf.~green dotted line in the right panel of figure~\ref{fig:lim_spec}).

\subsubsection{Scale-by-scale energy transfer}

%
%
\begin{figure}
    \centering
    \includegraphics[scale=0.9]{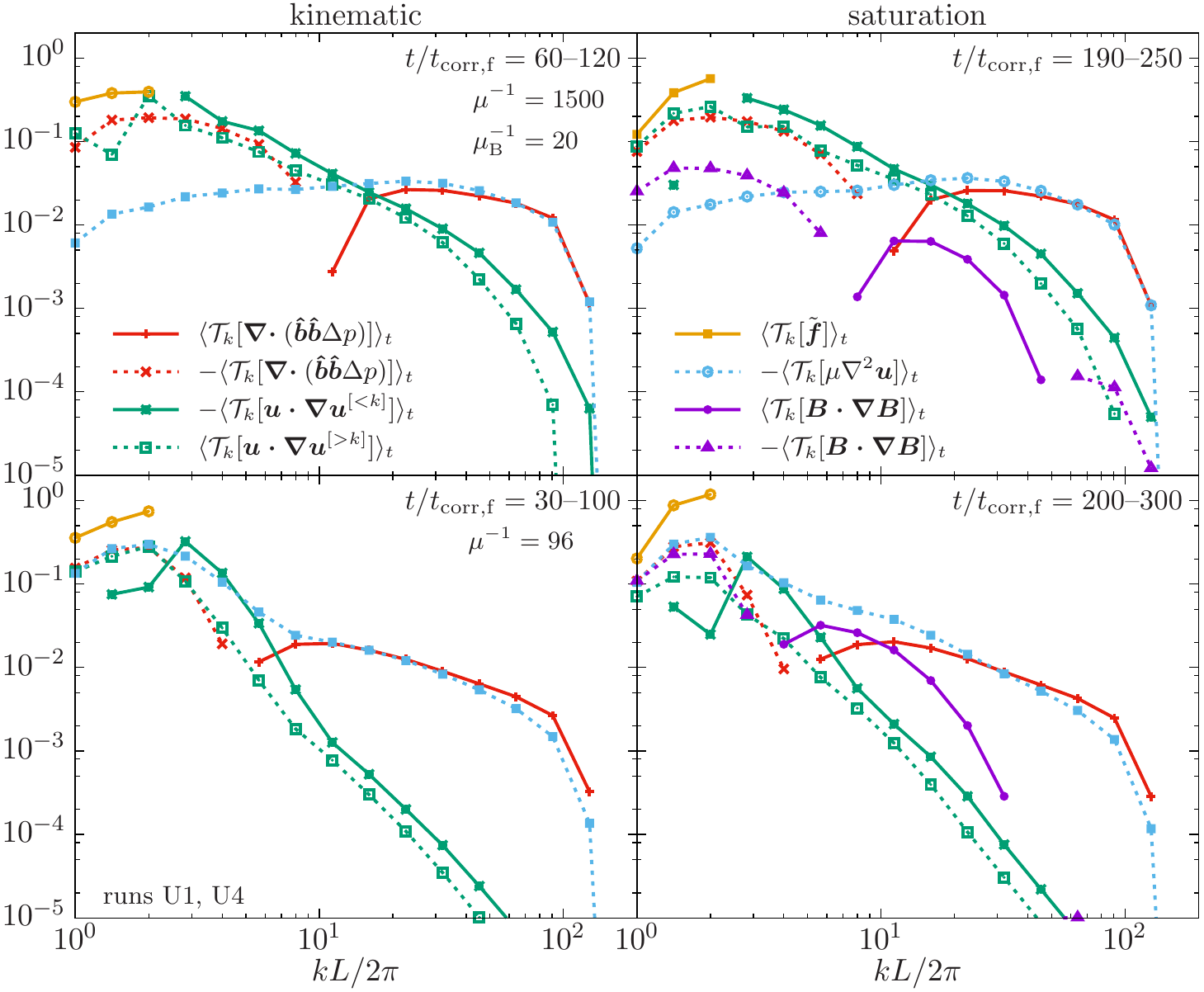}
    \caption{Shell-filtered kinetic-energy transfer functions $\mc{T}_k$ [see \eqref{eq:shell_trans}] for the unlimited Braginskii-MHD system in the (left) kinematic and (right) saturated stages. Top (bottom) row corresponds to $\mu^{-1} = 1500$ ($\mu^{-1} = 96$). Solid (dotted) lines denote energy flowing into (out of) the shell centred at mode $k$.}
    \label{fig:trans_unlim}
\end{figure}

The shell-filtered energy-transfer functions are again used here to probe the origin of the kinetic energy spectra found in the previous subsection and are shown in figure \ref{fig:trans_unlim} for the unlimited Braginskii-MHD simulations with $\mu^{-1} = 1500$ (top) and $\mu^{-1}= 96$ (bottom). For both systems, the $k^{-4}$ spectrum of sub-perpendicular-viscous motions (i.e., motions occurring at scales smaller than the viscous cutoff caused by the isotropic viscosity) is seen to be the result of a balance between the pressure-anisotropy stress (red lines) and the isotropic viscosity (blue lines). Similar behaviour was measured and explained in \citet{Scheko_sim} by balancing the viscous dissipation $\mu \nabla^2 \bb{u}$ and the magnetic tension $(\bb{B}\bcdot \grad \bb{B})$, resulting in sub-viscous motions that, while initially small, grew along with the magnetic energy. In our simulations of unlimited Braginskii-MHD, the same scenario applies, but now the pressure anisotropy stress $\grad \bcdot (\eb \eb \rmDelta p)$ takes on the role of the magnetic tension $\grad \bcdot (\eb \eb B^2)$. In order for the balance $\grad \bcdot (\eb\eb\Deltap) \sim \visc \nabla^2 \bb{u}$ to be valid, it must be the case that $|\bb{u}\bcdot \grad \bb{u}| \ll |\mu \nabla^2 \bb{u}|\sim|\grad \bcdot (\eb \eb \rmDelta p) |$, inequalities that are satisfied in the unlimited Braginskii-MHD simulation beyond the spectral break. Similar to the role of magnetic tension on small scales in the isotropic MHD dynamo, the Braginskii viscosity in this range \emph{gives} energy to the velocity field instead of dissipating it, a feature that is discussed further in appendix \ref{sec:m-i_null} in the context of magneto-immutability. Both the sub-viscous motions in the $\mr{Pm} \gg 1$ MHD dynamo (\S\,\ref{sec:limited}) and sub-perpendicular-viscous motions in unlimited Braginskii-MHD dynamo are passive and do not influence the evolution of either the turbulent motions above the viscous scale or the magnetic field.

The observed $k^{-2.75}$ slope in the kinetic-energy spectrum for the simulation with $\mu^{-1}= 1500$ can be seen as a balance between the Braginskii viscosity (red line) and the hydrodynamic nonlinearity (solid and dotted green lines). In this range of $k$, the Braginskii viscosity removes kinetic energy at every scale, resulting in a spectrum that is steeper than $-5/3$. Much like the spectral index of $-7/3$ reported for the kinetic-energy spectra of saturated MHD dynamo \citep{Scheko_sim}, the exact nature of this index is difficult to pin down, although the difference in the two indices (2.75 vs.~$7/3\simeq 2.33$) does represent one of the minor distinctions between the unlimited Braginskii-MHD and the saturated MHD dynamo. Because the unlimited Braginskii viscosity renders the turbulence anisotropic with respect to the magnetic field regardless of the strength of the latter, the unlimited Braginskii-MHD dynamo increasingly resembles the MHD dynamo in the saturated state as the parallel viscosity is increased. For a system with a sufficiently large parallel viscosity and small perpendicular viscosity (resulting in strongly anisotropic turbulence), one only needs a small amount of magnetic tension to anisotropize the velocity field further and push the dynamo into its saturated state. This is clearly seen in figure~\ref{fig:trans_unlim}, in which the magnetic tension is shown to be sub-dominant to the Braginskii stress for the $\mathrm{Re} \gg 1$ unlimited Braginskii system, even in the saturated state. It is thus no surprise then that the spectrum of the underlying fluid motions remains relatively constant throughout the entire dynamo process when the pressure anisotropy is unlimited. The reduced magnetic tension required to saturate the dynamo also implies that the unlimited Braginskii system saturates with a smaller magnetic to kinetic energy ratio than the MHD dynamo, a feature that can be seen by comparing the purple, green, blue, and orange solid lines in figure \ref{fig:energy_unlim} that show how the evolution of the magnetic energy changes with decreasing $\mu_{\rm B}$. However, the reduction in this ratio is minor (order unity) unless the parallel-viscous scale implied by $\Reprl$,  {\it viz.}~$\ell_{\mu_\parallel} \sim \ell_0 \, \Reprl^{-3/4}$, resides near the outer scale. In the next two sections, we show that the field-biased structure of the flow, the characteristic scales of the magnetic field, and the statistics of the magnetic-field curvature are all relatively constant as well.

\subsubsection{Structure functions of the velocity field}

%
%
\begin{figure}
    \centering
    \includegraphics[scale=0.9]{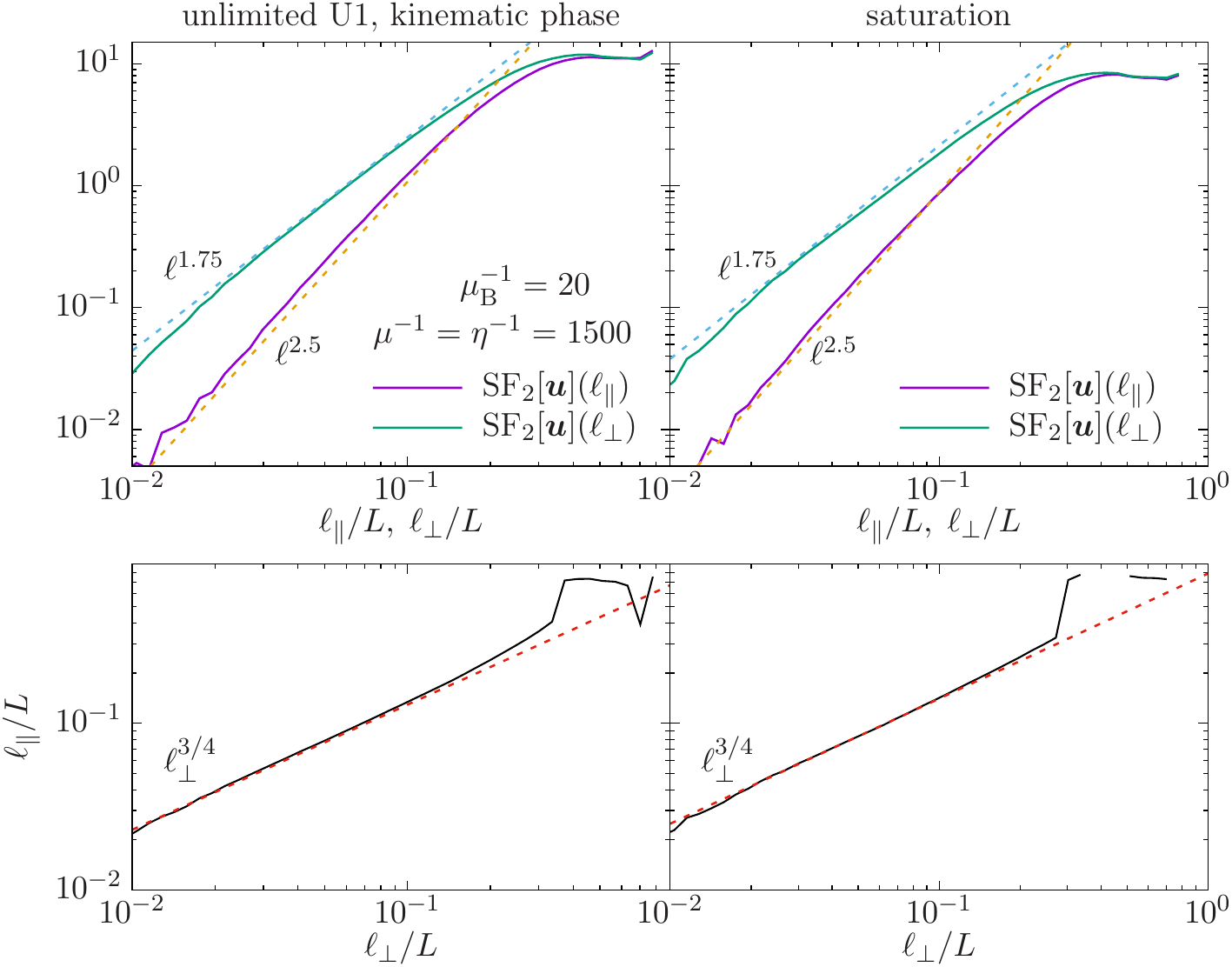}
    \caption{Perpendicular and parallel three-point, second-order structure functions (top) and corresponding spectral anisotropy (bottom) from the unlimited Braginskii-MHD run U1 in the kinematic (left) and saturated (right) stages.}
    \label{fig:struct_unlim}
\end{figure}

We have shown that an unlimited anisotropic viscous stress strongly biases the properties of the flow with respect to the magnetic-field direction. To quantify this bias further, we calculate the structure functions \eqref{eqn:strf} and \eqref{eqn:strf_prl_prp} of the velocity field in the unlimited $\mu^{-1}_\mr{B}=20$ Braginskii-MHD simulation (run U1) during the kinematic phase and in saturation. The result is shown in the top panels of figure \ref{fig:struct_unlim}. We find $\mathrm{SF}_2 \propto \ell_\perp^{1.75}$ for the perpendicular structure function (corresponding to the $-2.75$ slope seen in figure \ref{fig:unlim_spec}) and a much steeper $\mathrm{SF}_2 \propto \ell_\parallel^{2.5}$ for the parallel structure function. The steepness of the parallel structure function confirms that small-scale stretching motions play no dynamical role in this Braginskii-MHD dynamo, and so the largest scales are those primarily responsible for growing the magnetic field. The corresponding spectral anisotropy in both the kinematic and saturated stages, shown in the bottom panels of figure \ref{fig:struct_unlim}, scales roughly as $\ell_\parallel \sim \ell_\perp^{3/4}$, the same as in the saturated states of the limited Braginskii-MHD and isotropic-MHD dynamos (cf.~figure \ref{fig:struct_lim}). Most notably, none of these properties change from the kinematic stage to saturation.

\subsubsection{Characteristic scales of the magnetic field}

%
%
\begin{figure}
    \centering
    \includegraphics[scale=0.8]{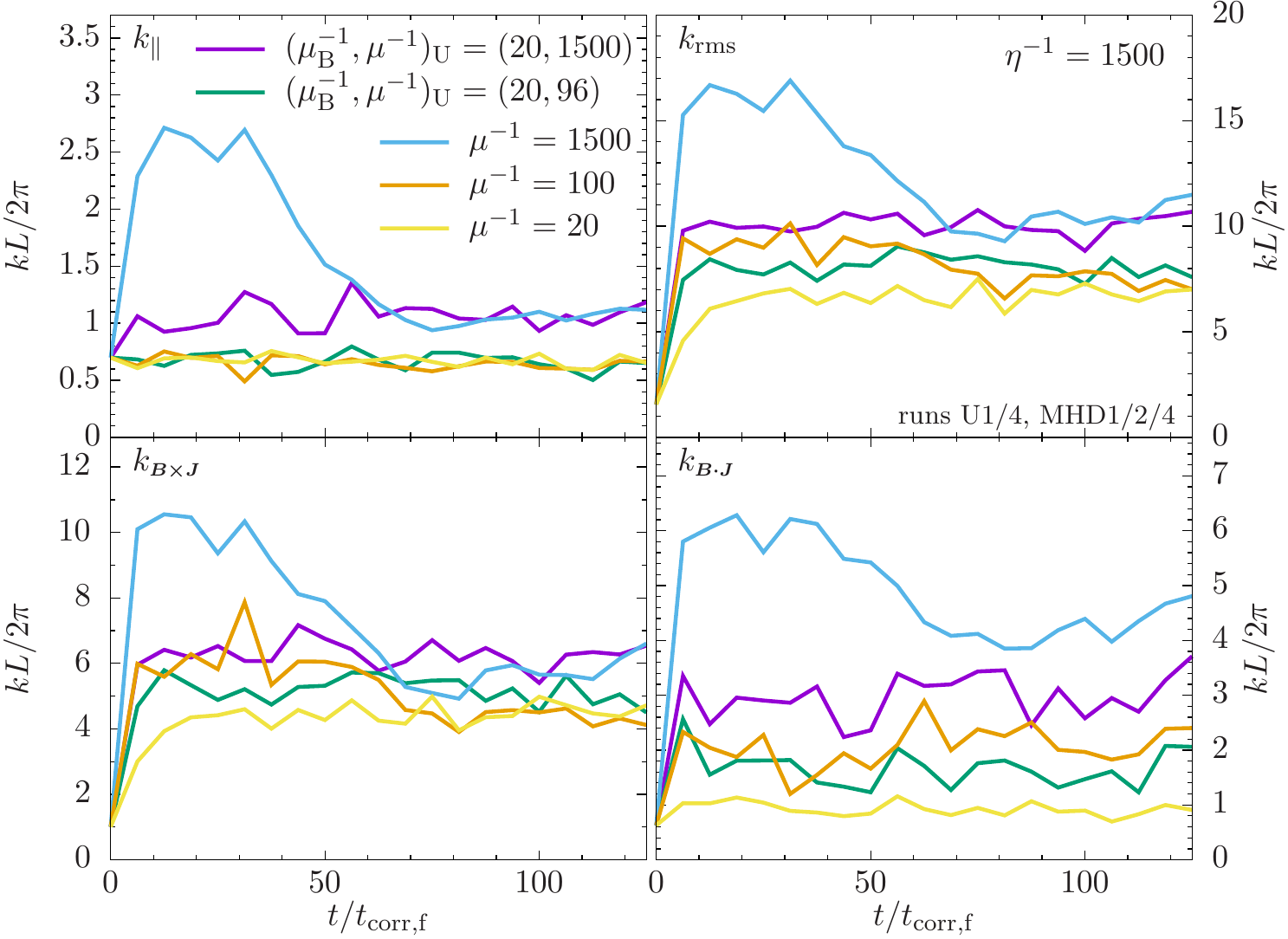}
    \caption{Evolution of the characteristic wavenumbers \eqref{char-wavenumbers} in the unlimited Braginskii-MHD runs U1 and U4 (purple and green lines, respectively) and in runs MHD1, 2, 4 (yellow, orange and blue lines, respectively).}
    \label{fig:wavenumbers_unlim}
\end{figure}

The geometry of the magnetic field in the unlimited Braginskii-MHD runs does not change much either as the dynamo saturates. This is diagnosed quantitatively by measuring the characteristic wavenumbers \eqref{char-wavenumbers} of the magnetic field and the distributions of the field-line curvature. Figure \ref{fig:wavenumbers_unlim} shows the evolution of the former from runs U1, U4, and MHD1, 2 and 4. The wavenumbers in both unlimited Braginskii-MHD runs (U1 and U4, denoted by the purple and green lines) are roughly constant in time, with those from run U1 holding values similar to those found in the saturated state of the $\mu^{-1}=1500$ MHD run (blue lines) and those from run U4  matching closely its MHD counterpart ($\mu^{-1} = 100$, orange line). This is to be contrasted with the high-$\mathrm{Re}$ MHD simulation (light blue line) whose characteristic scales of the magnetic field increase as the system saturates, which is due to selective decay of small-scale modes of the magnetic field as a result of the anisotropization of the turbulence by the magnetic tension. 

A somewhat surprising result for the Braginskii-MHD system is that $k_\parallel$ is larger in run U1 than in the isotropic-MHD simulations with either $\mu^{-1}= \mu_\mr{B}^{-1} = 20$ or $\mu^{-1} = \mu_\mr{B}^{-1}/5 = 100$. This is likely because anisotropic viscosity allows a cascade of perpendicular energy to the perpendicular viscous scale (which is set by $\mu$), thereby allowing small-scale mixing motions that can bring field lines closer together and promote resistive annihilation. This causes characteristically shorter fold lengths. As the isotropic viscosity for the unlimited Braginskii simulations is increased, the saturated value of $k_\parallel$ approaches that found in the small-Re MHD runs (as demonstrated by U4). 

Figure \ref{fig:wavenumbers_unlim} also suggests that Braginskii viscosity does not promote magnetic fields with larger-scale structure than those produced in isotropic MHD. It has been conjectured by \citet{Malyshkin} that anisotropic viscosity might cause the turbulent dynamo to inverse-cascade the saturated magnetic fields from resistive scales to the larger viscous scales (i.e., those independent of ${\rm Rm}$) by allowing small-scale perpendicular motions that might unfold the field. If this conjecture turned out to be true, then it might explain the relatively large scale of the observed magnetic fields in the weakly collisional ICM. At least at our modest resolutions, no additional unfolding seems to take place, as the magnetic fields generated in the Braginskii simulations exhibit similar $k_{\bs{B}\bstimes \bs{J}}$ and $k_\mr{rms}$ to those found in the saturated state of the isotropic-MHD runs. Efforts to extend our work to larger Rm, and thus larger scale separation, could clarify what sets the peak of the magnetic-energy spectrum in the Braginskii-MHD dynamo.

%
%
\begin{figure}
    \centering
    \includegraphics[scale=0.85]{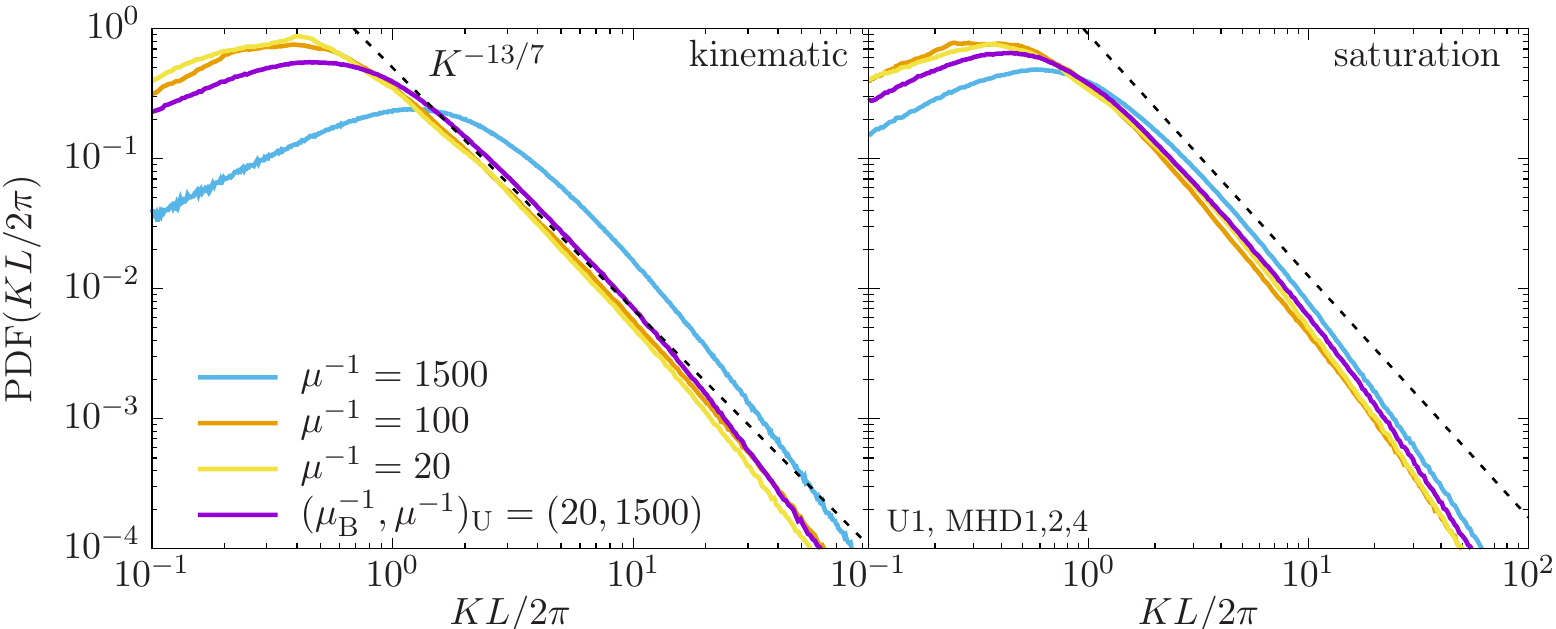}
    \caption{PDF of the magnetic-field curvature $K \doteq |\eb \bcdot \grad \eb|$ in run U1 (purple line) and in runs MHD1, 2 and 4 (yellow, orange and blue lines, respectively) in the kinematic stage and in saturation.}
    \label{fig:curvature}
\end{figure}

The PDF of the field-line curvature $K \doteq |\eb\bcdot \grad \eb|$ is displayed in figure \ref{fig:curvature} for the same runs as in figure \ref{fig:wavenumbers_unlim}. The asymptotic form of the curvature PDF predicted by \citet{Scheko_theory2} ({\em viz.}, a peak concentrated near the viscous scale and a power-law tail of $K^{-13/7}$ -- see \S\,\ref{sec:limscales}) is manifest in all of our simulations. As with the other quantities that we have analyzed, the curvature PDFs from the unlimited Braginskii run in both its kinematic and saturated stages match closely those measured in the saturated state of isotropic MHD. The form of the PDF changes somewhat for the $\mathrm{Pm} = 1$ case (blue line), in which the peak moves to smaller scales in accordance with the higher Reynolds number. By contrast, the unlimited Braginskii-MHD simulation exhibits slightly more curvature than comparable MHD simulations (runs MHD1 and MHD2; compare the purple line with the yellow and orange lines). This small increase likely occurs for the same reasons that the characteristic wavenumber $k_\parallel$ increases in the unlimited Braginskii case (see preceding paragraph): small-scale motions that mix field lines and promote resistive reconnection of the magnetic field are allowed by the Braginskii viscosity (but not by the isotropic viscosity).

\section{Anisotropization of the rate of strain and its consequences}\label{sec:anisotropization}

As was motivated at the start of \S\,\ref{sec:unlimited} and supported by the accompanying figures, the `kinematic' stage of the  unlimited Braginskii-MHD dynamo has many characteristics in common with the saturated state of the ${\rm Pm}\gtrsim{1}$ isotropic-MHD dynamo. This is because the Braginskii viscous stress has a mathematical form very similar to that of the magnetic tension, which, in saturation, biases the fluid velocity with respect to the magnetic-field direction to reduce the parallel rate of strain $\ROS$. In this section, we explore this bias further. Namely, we focus on how the rate-of-strain tensor remains field-anisotropic throughout the evolution of the Braginskii-MHD dynamo and on the consequences of this anisotropy. We also provide further evidence for, and formulate a theoretical dynamo model that is based on, the observed similarity between the unlimited-Braginskii and saturated-MHD dynamos.

\subsection{Constraints on stretching versus mixing}\label{sec:constraints_discussion}

The starting point for this discussion is an assessment of the roles that field-line stretching and mixing play in the efficacy of the dynamo -- roles that have already been glimpsed in the preceding sections. These two effects are quantified, respectively, by the parallel rate of strain, $\nabla_\parallel u_\parallel \doteq\ROS $, and by the perpendicular variations of the perpendicular velocity, $\grad_\perp \bb{u}_\perp \doteq (\mathsfbi{I}-\eb\eb)\bcdot \grad \bb{u}\bcdot(\mathsfbi{I}-\eb\eb)$. The key property of the fluctuation dynamo, exemplified by the \citet{zeldovich57} anti-dynamo theorem, is that amplification of the magnetic energy by field-line stretching can overcome resistive diffusion only in three dimensions. This is because a folded magnetic field undergoing stretching in a two-dimensional incompressible flow necessarily involves bringing together oppositely directed magnetic fields in the compression direction of the flow (see figure 10 of \citealt{Scheko_sim}). This causes very rapid annihilation of the field by resistivity. In three dimensions, this dynamo-adverse configuration may be avoided by orienting the directions of any magnetic-field reversals (which are parallel to $\bb{B}\btimes \bb{J}$) to be perpendicular to both the stretching and compression directions of the flow -- the `null' direction -- thus ensuring that current sheets in the folded field are not subject to further compression and thus to enhanced resistive diffusion. Of course, this special configuration cannot be maintained everywhere in a chaotic flow, and so the triumph of field-line stretching over field-line mixing and enhanced diffusion is a statistical one: in the now-classic \citet{Zeldovich} model of the kinematic fluctuation dynamo in a random, linear velocity field, only an exponentially decreasing fraction of initial wavenumbers $\bb{k}(t=0)$ characterizing the magnetic-field geometry -- those corresponding to field reversals occurring only in the null direction --  contribute to the growing field at any given time.

In the Braginskii-MHD dynamo, the relative sizes of $|\nabla_\parallel u_\parallel|$ and $|\grad_\perp \bb{u}_\perp|$ are controlled both by the dynamical influence of the magnetic field on the flow through the Lorentz force (in the saturated state) and by the disparate values of the parallel ($\visc_\mr{B}$) and perpendicular ($\visc$) viscosities (throughout the entire evolution, including in the `kinematic' regime). The latter truncate the various components of the rate-of-strain tensor at disparate scales, thereby forcing an imbalance in the maximal rates of strain along and across the magnetic field and placing constraints on the alignment between the magnetic folds and the stretching direction of the flow. The principal theme of the following sub-sections is that it is this imbalance, controlled by the anisotropic viscosity, that determines whether or not the Braginskii-MHD dynamo is viable. In effect, the magnetic Prandtl number $\mu/\eta$, which governs the relative rates of viscous-scale stretching and resistive-scale diffusion in MHD, is replaced as the key parameter in Braginskii MHD by the pair of parameters $\mu/\eta$ and $\mu/\mu_\mr{B}$. The survival of the Braginskii-MHD dynamo depends on both of these ratios, the latter (when small) constraining the geometry of the flow and thus its ability to amplify magnetic fields.

%
%
\begin{figure}
    \centering
    \includegraphics[scale=0.8]{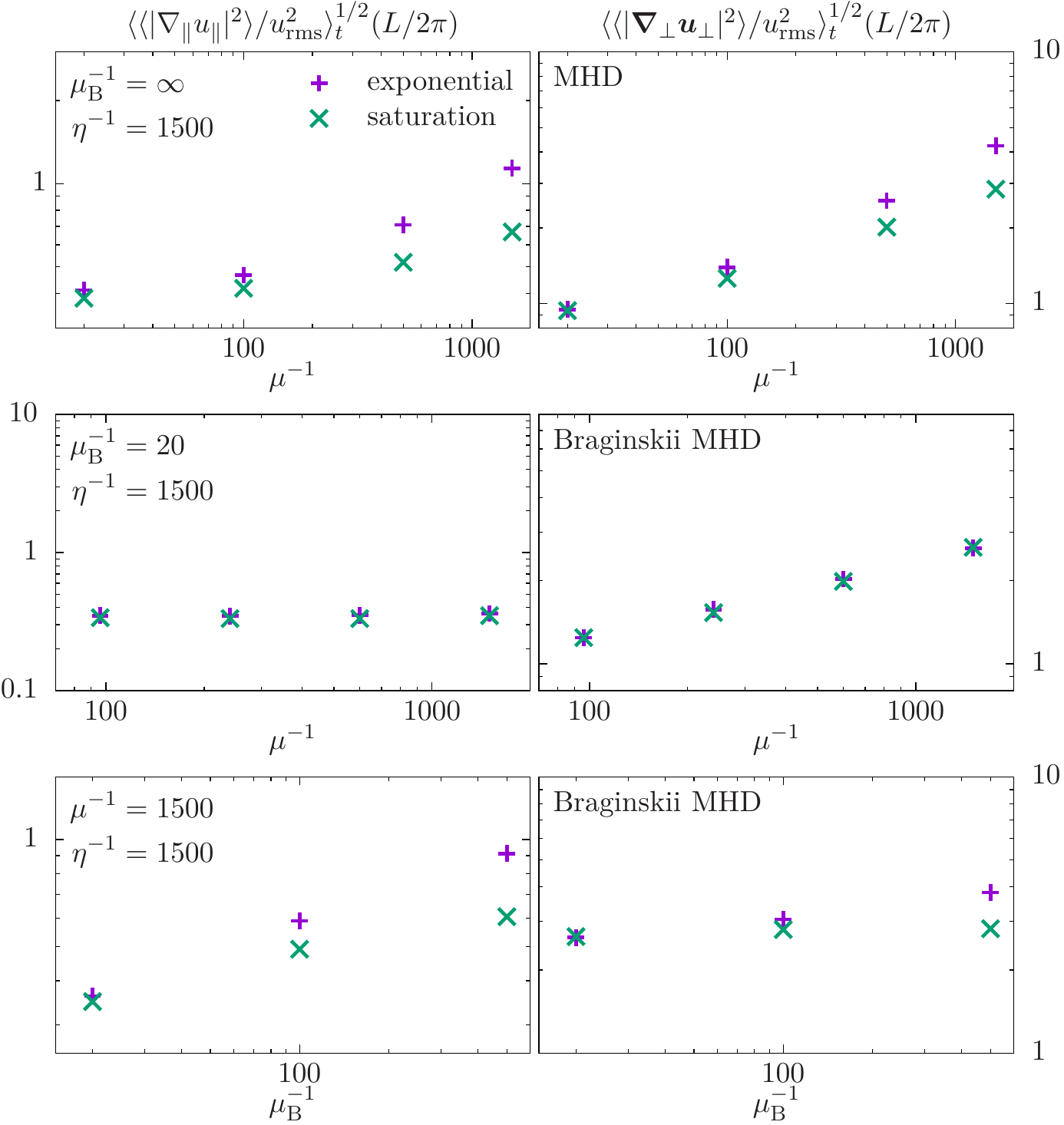}
    \caption{Time- and box-averaged values of the parallel ($\nabla_\parallel u_\parallel \doteq \ROS$, left) and perpendicular \dblbrck{$\grad_\perp \bb{u}_\perp \doteq ( \mathsfbi{I} - \eb\eb) \bcdot \grad \bb{u} \bcdot ( \mathsfbi{I} - \eb\eb) $, right} rates of strain for MHD (top) and unlimited Braginskii MHD with varying isotropic viscosity $\visc$ (centre) and varying parallel viscosity $\visc_\mr{B}$ (bottom). Purple pluses (green crosses) are time-averages taken in the exponential-growth phase (saturated state). All simulations have $\eta^{-1}= 1500$.}
    \label{fig:ROS_all}
\end{figure}

\subsection{Alignment of the rate-of-strain tensor with the magnetic-field direction}\label{sec:alignment}

It is therefore of interest to use our numerical simulations to measure directly the alignment of the rate-of-strain tensor with the magnetic-field direction during the various stages of the dynamo and across a range of parallel and isotropic viscosities.

Figure \ref{fig:ROS_all} displays the time-averaged values of $\nabla_\parallel u_\parallel$ (left) and $\grad_\perp \bb{u}_\perp$ (right) for various simulations using isotropic MHD (top) and unlimited Braginskii MHD (middle and bottom). For the high-$\mathrm{Re}$ MHD simulations, the saturated state is characterized by a reduction in the magnitude of the parallel rate of strain $|\nabla_\parallel u_\parallel|$ (upper left panel for $\visc^{-1}$ = 500, 1500) as the magnetic tension force becomes dynamically important. The mixing motions (top row, right panel) are partially suppressed as well. By contrast, the parallel rate of strain in the unlimited Braginskii run with $\mu^{-1}_{\rm B}=20$ is nearly constant in time and independent of isotropic viscosity (middle row, left panel). The perpendicular rate of strain is nearly unchanged in the saturated state as well, but varies with $\mu^{-1}$ in a predictable way: smaller viscosity allows smaller-scale mixing motions, as in isotropic MHD. When the parallel viscosity is decreased at fixed $\mu$ (bottom panels), the evolution of both the parallel and perpendicular rates of strain in the Braginskii-MHD system (rightmost points  in bottom row) converges to those seen in the analogous isotropic MHD system (rightmost points in in top row), as expected. Interestingly, the overall levels of the perpendicular mixing motions in the unlimited-Braginskii MHD simulations (middle row, right panel) are comparable to those found in the isotropic-MHD systems \emph{in the saturated state} (top row, right panel), indicating that these motions are partially suppressed relative to those in the kinematic stage of analogous isotropic MHD systems, even though the parallel viscosity does not affect them directly. It will be shown in \S\,\ref{sec:kazantsev} that, in order for the dynamo to be viable, the relative size of the perpendicular (mixing) motions cannot greatly exceed that of the parallel (stretching) ones.

An alternative way to quantify the anisotropization of the velocity caused by the Braginskii viscosity is to examine the relationship between the magnetic-field direction $\eb$ and the eigenvectors of the symmetrized rate-of-strain tensor $\mathsfbi{S} \doteq (\grad \bb{u} + \grad \bb{u}^\mathrm{T})/2$.
As we are considering only incompressible turbulence, the trace of the tensor $\mathsfbi{S}$, and thus the sum of its eigenvalues $\lambda_i$, are zero. We order these eigenvalues so that $\lambda_1 > \lambda_2 > \lambda_3$. Because $\mathsfbi{S}$ is real and symmetric, its eigenvalues are real, and its eigenvectors $\boldsymbol{\hat{e}}_i$ are orthogonal, hence $|\eone\bcdot \eb|^2 + |\etwo\bcdot \eb|^2 + |\ethree\bcdot \eb|^2 = 1$. The eigenvectors $\eig_1$ and $\eig_3$ correspond to the directions of field-line stretching and compression, while $\eig_2$ points in the so-called `null' direction (which can be either stretching or compressing). The alignment angles $\theta_i$ are defined by $|\eig_i \bcdot \eb| = \cos \theta_i$. For isotropic hydrodynamic turbulence, $\lambda_1 \sim -\lambda_3 \sim (3\textrm{--}5) \lambda_2 > 0$~\citep{ashurst1987,tsinober1992}.

%
%
\begin{figure}
    \centering
    \includegraphics[scale=0.8]{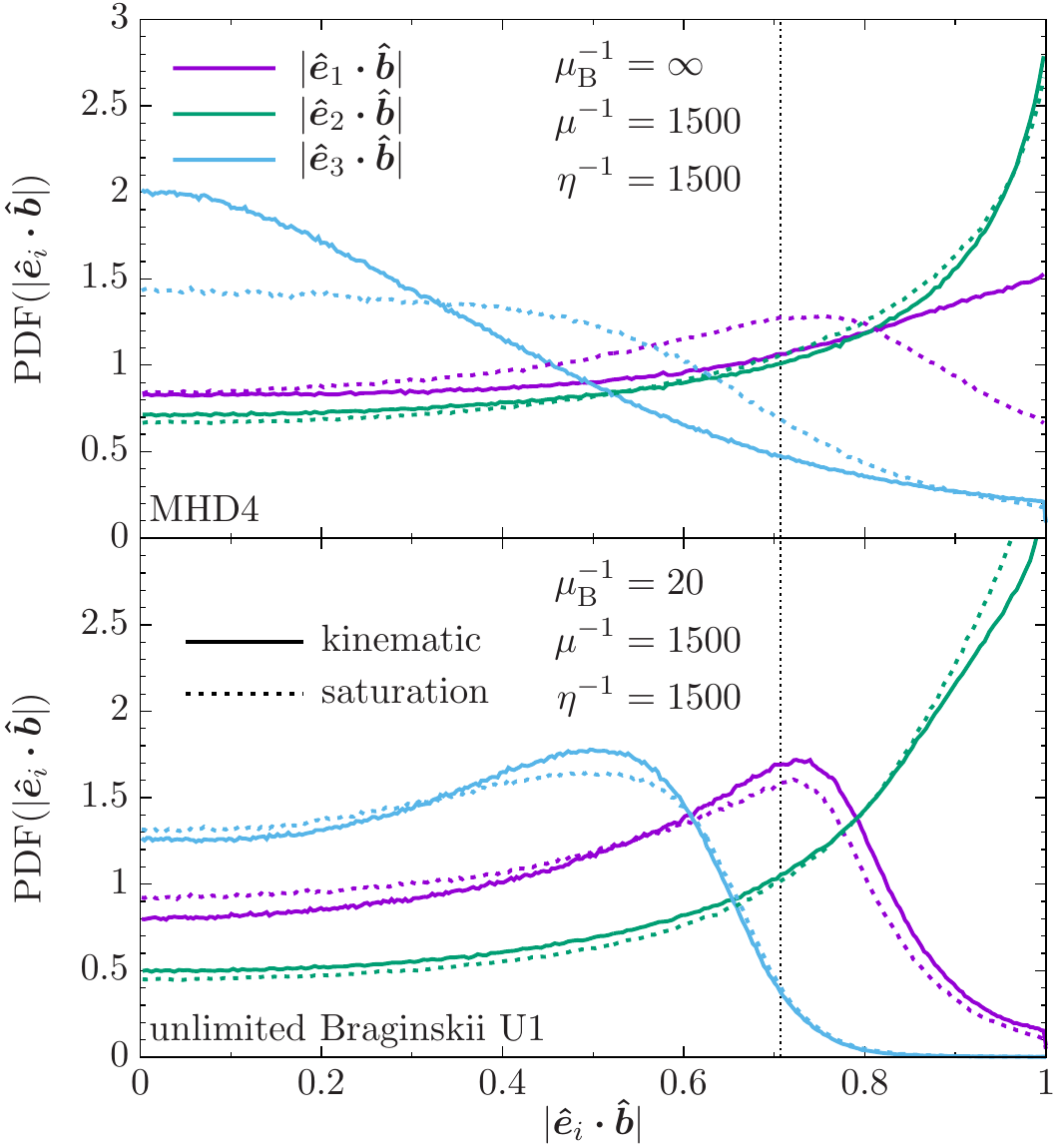}
    \caption{PDF of the rate-of-strain eigenvector alignment cosines $\cos \theta_i = |\eig_i \bcdot \eb|$ for simulations of isotropic MHD (top) and unlimited Braginskii MHD (bottom) in the kinematic stage (solid line) and saturated state (dotted line). Eigenvalues are ordered so that $\lambda_1 > \lambda_2 > \lambda_3$. The vertical dotted line marks $\theta_i = 45^\circ$.}
    \label{fig:angle_some}
\end{figure}

%
%
\begin{figure}
    \centering
    \includegraphics[width=\textwidth,interpolate=false]{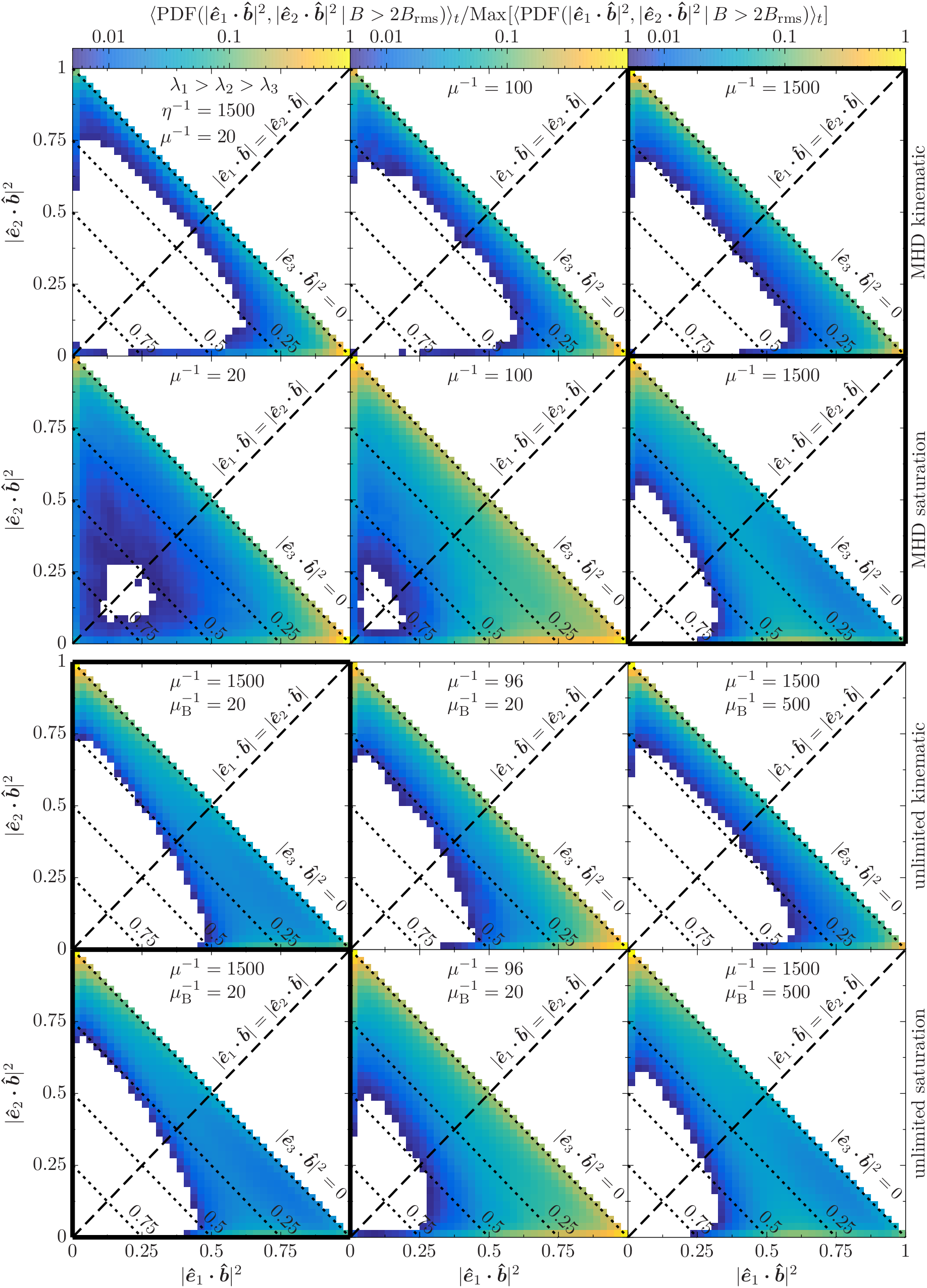}
    \caption{Time-averaged joint PDFs of the cosines of the angles between the magnetic field and the first two ordered eigenvectors of the rate-of-strain tensor for the MHD simulations MHD1/2/4, and for the unlimited Braginskii-MHD simulations U1, U4, and U1b. The PDFs are conditioned on high magnetic energy ($B > 2 B_\mathrm{rms}$) to probe the strongest dynamo-generated fields. Dashed lines correspond to $|\eone\bcdot \eb| = |\etwo\bcdot \eb|$; dotted lines correspond to solutions of  $|\eone\bcdot \eb|^2 + |\etwo\bcdot \eb|^2+ |\ethree\bcdot \eb|^2 =1$ for fixed $|\eig_3\bcdot \eb|$ (i.e.~$|\eig_3\bcdot \eb|$ is fixed along these lines). The simulations shown in figure \ref{fig:angle_some} are marked by the thicker boxes.}
    \label{fig:angle2d_all}
\end{figure}

Figure \ref{fig:angle_some} displays the PDFs of the alignment cosines $|\eig_i \bcdot \eb|$ for ${\rm Re}\gg{1}$, $\mathrm{Pm} =1$ MHD (run MHD4; top) and unlimited Braginskii MHD (run U1; bottom) in the kinematic stage (solid lines) and saturated state (dashed lines).\footnote{\. Similar PDFs were calculated recently for the isotropic-MHD dynamo by~\citet{Seta2020}, whose results agree broadly with those presented here in figures~\ref{fig:angle_some} and~\ref{fig:angle2d_all} for isotropic MHD.} For the MHD run in the kinematic stage, the most probable values of both $\eig_1 \bcdot \eb$ and $\eig_2 \bcdot \eb$  are 1, while for $\eig_3\bcdot \eb$, it is 0. Thus, the magnetic field aligns itself either with $\eig_1$ or $\eig_2$. In saturation, the PDF of $\eig_2\bcdot \eb$ remains largely unchanged, while the part of the field that is not aligned with $\eig_2$  aligns itself predominantly between $\eig_1$ and $\eig_3$.  
In contrast, the unlimited Braginskii-MHD run mimics the statistics of the saturated MHD dynamo throughout its evolution, with the alignment angles hardly changing at all between the exponential growth regime and saturation.

These alignments can also be seen in figure \ref{fig:angle2d_all}, which shows time-averaged joint PDFs of $|\eone\bcdot \eb|^2$ and $|\etwo\bcdot \eb|^2$ both for MHD simulations (top two rows) and for unlimited Braginskii MHD ones (bottom two rows) in the kinematic stage (first row of each group) and in the saturated state (second row of each group). To highlight the alignment of dynamically important fields, these PDFs are conditioned on regions of high field strength, $B > 2 B_\mathrm{rms}$. As the alignment of $\eb$ with $\eig_3$ can be related to the other two alignments via the identity $|\eone\bcdot \eb|^2 + |\etwo\bcdot \eb|^2 + |\ethree\bcdot \eb|^2 = 1$, these PDFs contain all the information about the field-alignment statistics (e.g., any probability density appearing in the bottom left corner of these plots  denotes magnetic fields primarily aligned with $\eig_3$). It is clear that, in the MHD simulations, the magnetic field tends to align itself principally with either $\eig_1$ or $\eig_2$ in the kinematic stage, rather than with some combination of all three eigenvectors. The statistics in saturation vary significantly for different amounts of viscosity; for $\mathrm{Re}\sim 1$, the dynamo saturates by having the magnetic field align itself uniformly between $\eig_1$ and $\eig_2$ along the $|\eig_3 \bcdot \eb| = 0$ contour, while for $\mathrm{Re} \gg 1$, it aligns itself between $\eig_1$ and $\eig_3$ (namely, $\theta_1 \approx 40^\circ$ and $\theta_3 \approx 50^\circ$). The simulation with intermediate Reynolds number (centre panel) contains regions where both of these situations occur. 

For unlimited Braginskii MHD with large parallel viscosity $\visc_\mr{B}$ (left panels of bottom group in figure \ref{fig:angle2d_all}), the statistics of the field alignment hardly change between the kinematic stage and saturation, and both are similar to those found in the high-Re MHD dynamo ($\mu^{-1}=1500$) in its saturated state (top group, bottom right panel). (The only notable difference is that the enhancement in probability density localized near $|\eig_1 \bcdot \eb|^2 \approx 0.6$ in the latter is slightly less pronounced in the Braginskii-MHD case.) This again emphasizes that the role of the parallel Braginskii viscosity is to render the statistics of the magnetic field nearly identical to those found in the saturated state of the MHD dynamo.\footnote{\,Increasing the isotropic viscosity (centre panels of bottom two rows of figure \ref{fig:angle2d_all}) leads to an increase of field alignment with $\eig_1$ both in the kinematic stage and in saturation, which may be explained by the elimination of small-scale fluctuations that tend to \emph{decrease} the parallel rate of strain when the isotropic viscosity is increased (see appendix \ref{sec:m-i_null} for more).} 

%
%
\begin{figure}
    \centering
    \includegraphics[scale=0.79]{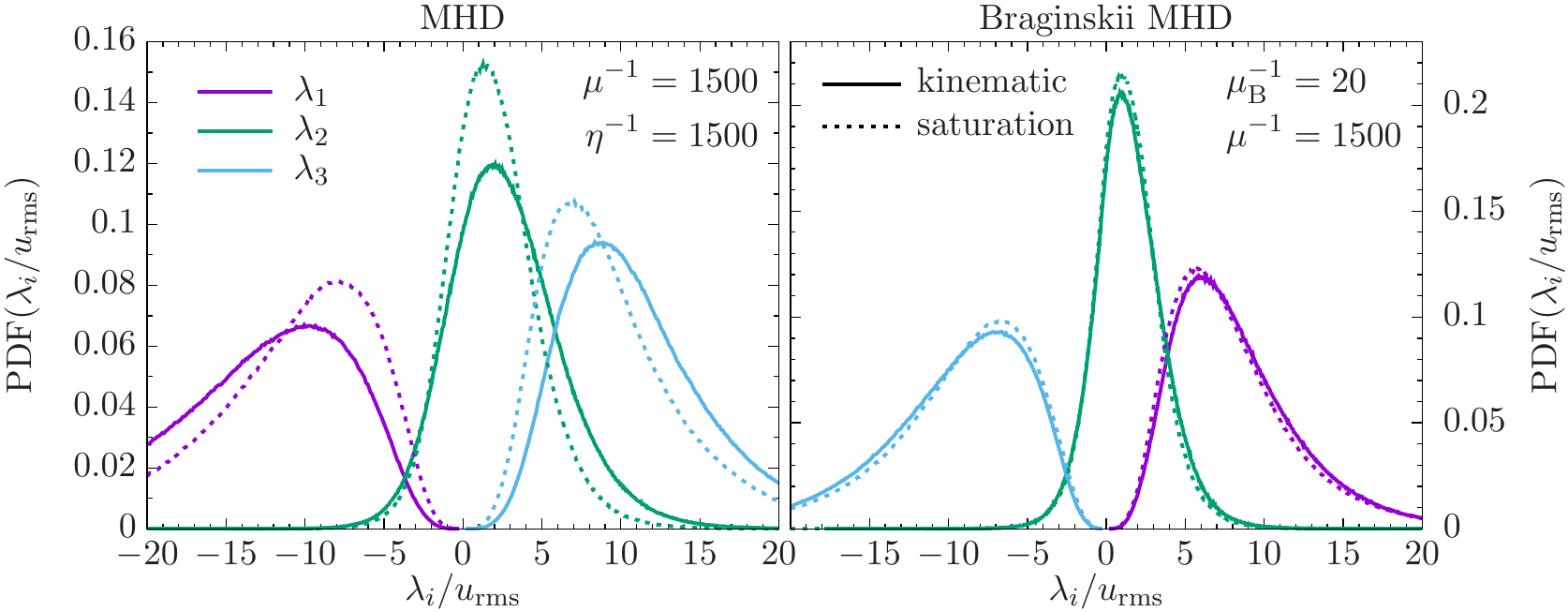}
    \caption{PDF of the rate-of-strain eigenvalues for simulations of isotropic MHD (left) and unlimited Braginskii MHD (right)  in the kinematic stage (solid lines) and saturated state (dotted lines). Eigenvalues are ordered so that $\lambda_1 > \lambda_2 > \lambda_3$. }
    \label{fig:eigen_some}
\end{figure}

Finally, figure \ref{fig:eigen_some} displays the PDFs of the eigenvalues $\lambda_1 > \lambda_2 > \lambda_3$ of the rate-of-strain tensor $\mathsfbi{S}$ for ${\rm Re}\gg{1}$ MHD (run MHD4; left) and for unlimited Braginskii MHD (run U1; right) in the kinematic (solid line) and saturated states (dashed line). The statistics of the rate-of-strain tensor change noticeably between the two stages in the isotropic-MHD simulation, while the change is minimal (if any) in the unlimited Braginskii-MHD system. Once again, the unlimited Braginskii-MHD dynamo behaves throughout its evolution much like the isotropic-MHD dynamo in its saturated state.


\subsection{Modified Kazantsev--Kraichnan model for the Braginskii-MHD dynamo}\label{sec:kazantsev}

In light of the main conclusion of \S\S\,\ref{sec:unlimited}--\ref{sec:alignment} -- that the turbulence statistics during the `kinematic' phase of the unlimited Braginskii-MHD dynamo are strikingly similar to those found in the saturated state of the isotropic-MHD dynamo -- we propose that the `kinematic' stage of the unlimited Braginskii-MHD dynamo can be described by an analytic model originally developed to describe the saturated state of the MHD dynamo by \citet{Scheko_saturated,Scheko_sim}. This model uses a modified form of the Kazantsev--Kraichnan \citep{Kazantsev,kraichnan} velocity correlation tensor
\begin{equation}\label{eqn:kazantsev}
    \ea{u^i(t,\bb{x})u^j(t',\bb{x}')} = \delta(t-t')\kappa^{ij}(\bb{x}-\bb{x}'), 
\end{equation}
where an overline denotes the ensemble average, that allows anisotropic statistics with respect to the magnetic-field direction $\eb$: the Fourier transform of $\kappa^{ij}(\bb{x}-\bb{x}')$ is
\begin{equation}\label{eqn:kappa_tensor}
    \kappa^{ij}(\bb{k}) = \kappa^{\rm (i)}(k,|\xi|) \bigl( \delta^{ij} - \hat{k}_i \hat{k}_j \bigr) + \kappa^{\rm (a)}(k,|\xi|) \bigl( \nrndb^i \nrndb^j + \xi^2 \hat{k}_i \hat{k}_j - \xi \nrndb^i \hat{k}_j - \xi \hat{k}_i \nrndb^j \bigr),
\end{equation}
where $\hat{\bb{k}} \doteq \bb{k}/k$, $\xi \doteq \hat{\bb{k}}\bcdot\eb$, and $\nrndb^i$ denotes the $i$th component of the magnetic-field unit vector $\eb$. The $k$- and $|\xi|$-dependent amplitudes $\kappa^{\rm (i)}$ and $\kappa^{\rm (a)}$ quantify the sizes of the isotropic and anisotropic components of the correlation tensor. Note that $k_i\kappa^{ij} = k_j \kappa^{ij} = 0$ as a result of incompressibility. When this model was put forward, the idea was that the dynamically important magnetic field in saturation exerts a feedback on the velocity by means of the Lorentz force, biasing the velocity's statistics with respect to the local field direction. Here, we propose that this model might accurately describe the impact of the Braginskii viscosity on the turbulent flow during the {\em kinematic} stage.

An outline and derivation of the model is given in \S\,\ref{sec:kazantsev_formulation} and appendix~\ref{ap:kazantsev}. The model is then solved in the general case of finite dynamo growth rate and non-negligible resistivity in \S\,\ref{sec:kazantsev_solution}. In \S\S\,\ref{sec:anti-Stokes}--\ref{sec:stokes}, we use this solution and the results of numerical simulation to show that the dynamo fails in the limit $\Reprl/\mathrm{Re} \rightarrow 0$.

\subsubsection{Formulation of the model}\label{sec:kazantsev_formulation}

We first obtain equations for the evolution of magnetic-field fluctuations, the wavevector of these fluctuations, and the scale-dependent magnetic-field direction as functions of the fluctuating velocity field. These are used to determine the statistics of the magnetic-field strength, structure, and direction, from which we then derive an evolution equation for the magnetic-energy spectrum.

To simplify the calculation, we focus only on the straight regions of the resulting folded magnetic field, so that spatial variations in $\eb$ are limited to changes of sign. Because $\eb$ arises in the momentum equation as part of the dyadic $\eb\eb$, these changes of sign cancel each other out, and so $\eb$ can be taken to depend only on time. This approximation is not as drastic as it may appear, as the bends of the magnetic folds occupy a small fraction of the total volume \citep{Scheko_theory2,Scheko_sim}. The evolution equations then follow straightforwardly from the non-ideal induction equation after adopting the \ansatz
\begin{equation}
    \bb{B}(t,\bb{x}) = \eb(t) \int \od^3\bb{k}' \, {B}(t,\bb{k}') \rme^{\imag \bs{x}\bscdot {\bs{k}}(t,\bs{k}')} ,
\end{equation}
where $\bb{k}(t,\bb{k}')$ is the wavevector that evolves in time from the initial value $\bb{k}'$. Assuming statistical homogeneity and setting $\bb{x} = 0$ results in the closed set of equations
\begin{subequations}\label{eqn:kaz_base}
\begin{gather}
    \partial_t {B} = \nrndb^i \nrndb^m (\partial_m u^i) {B} - \eta {k}^2 {B}, \\
    \partial_t {k}_m = - (\partial_m u^i) {k}_{i}, \\
\partial_t \nrndb^i = \nrndb^m (\partial_m u^i) - \nrndb^l \nrndb^m (\partial_m u^l) \nrndb^i,
    \end{gather}
\end{subequations}
where we have used the Einstein summation convention for repeated indices. The upper and lower indices are used for convenience of keeping track of the summations. The joint probability density function of $B$, $\bb{k}$ and $\eb$ must then have the form
\begin{equation}
    \mc{P}({B},{\bb{k}},\eb) = \delta(|\eb|^2 -1)\delta(\eb\bcdot{\bb{k}})(4\upi^2 {k})^{-1}P({B},{k}),
\end{equation}
and a closed equation for $\mathcal{P}(B,k)$
can be derived using (\ref{eqn:kazantsev}). Several quantities of interest may then be calculated by taking the appropriate moments of $\mc{P}(B,k)$, e.g., the magnetic-energy spectrum  
\begin{equation}\label{eqn:Mk}
    M(k) \doteq \frac{1}{2} \int_0^\infty \od {B}\, {B}^2 P({B},{k}) .
\end{equation}
A detailed derivation of the evolution equation for $\mc{P}({B},k)$ is given in appendix \ref{ap:kazantsev}. 

For $k\gg k_\mu$ (i.e, at sub-viscous scales), the evolution equation for the magnetic-energy spectrum obtained via \eqref{eqn:Mk} is
\begin{equation}\label{eqn:mod_kazantsev}
    \pD{t}{M} = \frac{\gamma_\perp}{8} \pD{k}{} \biggl[ \bigl( 1 + 2\sigma_\parallel\bigr) k^2 \pD{k}{M} - \bigl( 1 + 4\sigma_\perp + 10\sigma_\parallel \bigr) kM \biggr] + 2 \bigl( \sigma_\perp + \sigma_\parallel\bigr) \gamma_\perp M - 2\eta k^2 M,
\end{equation}
where
\begin{subequations}\label{eqn:gamma_delta_sing}
\begin{equation}\label{eqn:gamma_delta}
    \gamma_\perp = \int\frac{\rmd^3\bb{k}}{(2\upi)^3} \, k^2_\perp \kappa_\perp(\bb{k}) , \quad
    \sigma_\perp = \frac{1}{\gamma_\perp} \int\frac{\rmd^3\bb{k}}{(2\upi)^3} \, k^2_\parallel \kappa_\perp(\bb{k}) , \quad
    \sigma_\parallel = \frac{1}{\gamma_\perp} \int\frac{\rmd^3\bb{k}}{(2\upi)^3} \, k^2_\parallel \kappa_\parallel(\bb{k}) . \tag{\theequation {\it a,b,c}}
\end{equation}
\end{subequations}
Here, $k_\parallel = k\xi$, $k_\perp = k(1-\xi^2)^{1/2}$, and 
\begin{align}
    \kappa_\perp(\bb{k}) &= \frac{1}{2} \bigl( \delta^{ij} - \nrndb^i\nrndb^j\bigr) \kappa^{ij}(\bb{k}) \nonumber\\*
    \mbox{} &= \frac{1}{2} \Bigl[ \bigl(1+\xi^2)\kappa^{\rm (i)}(k,|\xi|) + \xi^2\bigl(1-\xi^2\bigr) \kappa^{\rm (a)}(k,|\xi|) \Bigr] ,\label{eqn:kappa_prp} \\*
    \kappa_\parallel(\bb{k}) &= \frac{1}{2} \nrndb^i \nrndb^j \kappa^{ij}(\bb{k}) \nonumber\\*
    \mbox{} &= \frac{1}{2} \bigl(1-\xi^2\bigr) \Bigl[ \kappa^{\rm (i)}(k,|\xi|) + \bigl(1-\xi^2\bigr) \kappa^{\rm (a)}(k,|\xi|) \Bigr] \label{eqn:kappa_prl} 
\end{align}
are the correlation spectra of velocities respectively perpendicular and parallel to the local magnetic field, and $\kappa^{ij}$ is defined by \eqref{eqn:kazantsev} and \eqref{eqn:kappa_tensor}. The quantity $\gamma_\perp$ measures the perpendicular variation of the perpendicular velocities, and thus gives the mixing rate of the magnetic field. It is at this rate that interchange-like motions shuffle and bring direction-reversing magnetic fields close enough together for them to diffuse resistively. The quantities $\sigma_\perp$ and $\sigma_\parallel$ measure the relative strengths compared to this mixing rate of the field-aligned stretching rates of the perpendicular and parallel velocities. For reference, the isotropic case with $\kappa^{\rm (i)} = \kappa^{\rm (i)}(k)$ and $\kappa^{\rm (a)} = 0$ has $\sigma_\perp = 2/3$, $\sigma_\parallel = 1/6$, and $\gamma_\perp = (6/5)\overline{\gamma}$, where
\begin{equation}
    \overline{\gamma} = \frac{1}{3}\left[\int_{k_0}^\infty \od k \, k^2 E(k)\right]^{1/2},
\end{equation}
$k_0 \doteq \ell_0^{-1} = 2 \upi / L$ is the wavenumber of the outer scale,
 $E(k) = \int \od \Omega_\bb{k}\, k^2\delta_{ij}I^{ij}(\bb{k})$ is the kinetic-energy spectrum, and $\od \Omega_\bb{k}$ is an element of solid angle in $k$ space.

\subsubsection{Solution of the model and comparison with simulation results}\label{sec:kazantsev_solution}
 
Equation \eqref{eqn:mod_kazantsev} is solved by assuming  $M(k,t) \propto e^{\gamma t}$, where $\gamma$ is the magnetic-energy growth rate.  Requiring $M(k) \rightarrow 0$ as $k \rightarrow \infty$ leads to the solution
\begin{equation}\label{eqn:kazantsev_solution}
 M(k) = k^s \rme^{\gamma t}K_{\imag \varpi}(k/k_\eta),
\end{equation}
where $K_{\imag \varpi}$ is the Macdonald function, $k_\eta = [(1+2\sigma_\parallel)\gamma_\perp/ 16 \eta]^{1/2}$ is the (inverse) resistive scale,
\begin{equation}\label{eq:spectral_index}
    s = \frac{2(\sigma_\perp + 2 \sigma_\parallel)}{1+2 \sigma_\parallel}
\end{equation}
is the spectral index, and
\begin{equation} \label{eqn:kazantsev_varpi}
 \varpi = \frac{1}{1 + 2\sigma_\parallel}\left[16(\sigma_\perp + \sigma_\parallel)(1+2\sigma_\parallel) + 8(\gamma/\gamma_\perp)(1+2\sigma_\parallel) - (1+2\sigma_\perp + 6 \sigma_\parallel)^2 \right]^{-1/2}.
\end{equation}
In order for this solution to be uniquely determined, the four coefficients $\gamma$, $\gamma_\perp$, $\sigma_\perp$, and $\sigma_\parallel$ must be determined. One constraint is obtained by imposing a zero-flux boundary condition at small $k$, i.e., the term in square brackets in \eqref{eqn:mod_kazantsev} is set to zero:
\begin{equation}\label{eqn:kaz_zeroflux}
 \frac{k_\ast}{k_\eta}K'_{\imag \varpi}\biggl(\frac{k_\ast}{k_\eta}\biggr) -  \frac{1+2 \sigma_\perp + 6\sigma_\parallel}{1 + 2\sigma_\parallel}K_{\imag \varpi}\biggl(\frac{k_\ast}{k_\eta}\biggr)=0,
\end{equation}
where $k_\ast$ is the infrared cutoff of the magnetic-energy spectrum. In practice, we let $k_\ast = k_0$.  Three more constraints must then be provided.\footnote{\,
An analytic expression for the eigenvalue $\gamma$ may be obtained in the $\eta\rightarrow +0$ limit, for which $k_\ast/k_\eta \rightarrow 0$, $\varpi = 0$, leading to $\gamma = [\gamma_\perp/(8+16\sigma_\parallel)][16(\sigma_\perp + \sigma_\parallel)(1+2\sigma_\parallel) - (1+2\sigma_\perp + 6 \sigma_\parallel)^2]$.  
This is the growth rate originally obtained by~\citet{Scheko_saturated}. The isotropic case with $\sigma_\perp = 4\sigma_\parallel = 2/3$ and $\gamma_\perp = (6/5)\overline{\gamma}$ (\S\,\ref{sec:kazantsev_formulation}) corresponds to the classic \citet{Kazantsev} and \citet{Kulsrud1992} magnetic-energy spectrum: $M(k) \approx k^{3/2}\, \rme^{(3/4)\overline{\gamma} t} K_0(k\sqrt{10\eta/\overline{\gamma}})$.}

%
%
\begin{figure}
    \centering
    \includegraphics[scale=0.90]{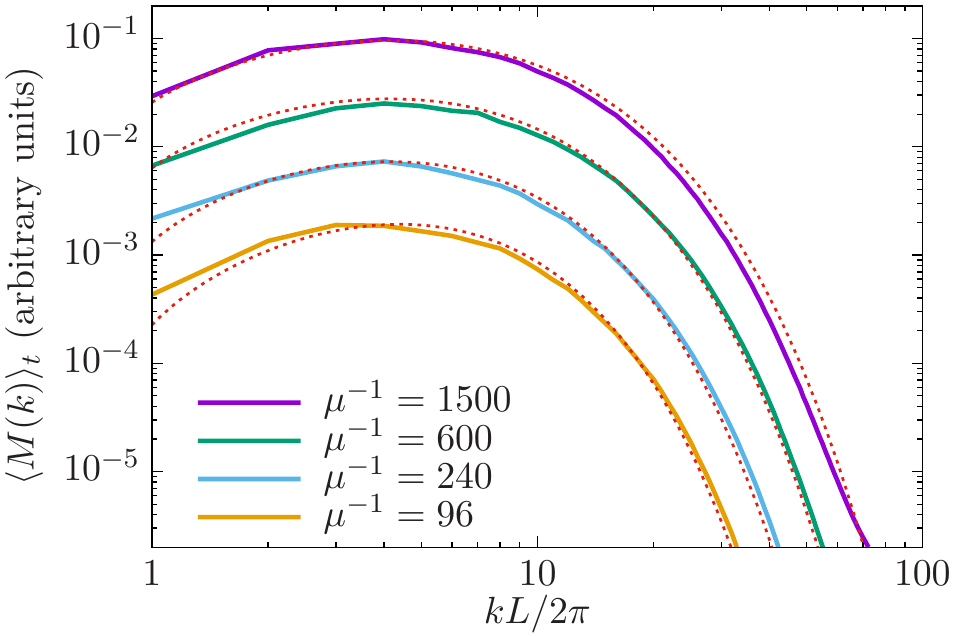}
    \caption{Time-averaged magnetic energy spectra from simulations of unlimited Braginskii MHD (solid line), along with predicted spectra calculated from solutions of~\eqref{eq:spectral_index} and~\eqref{eqn:kaz_zeroflux} (dotted lines) using the assumption $\sigma_\perp = 4 \sigma_\parallel$. Curves with varying $\visc$ have been shifted vertically for clarity.}
    \label{fig:kaz_spectra}
\end{figure}

We can gauge the validity of this model (and, therefore, of the idea of the local anisotropization) by comparing the magnetic-energy spectra obtained from our simulations to those predicted by \eqref{eqn:kazantsev_solution} with \eqref{eqn:kaz_zeroflux}. To do so, we specify the three remaining constraints as follows. First, we model the mixing rate by
\begin{equation}\label{eqn:kaz_mod_gammaperp}
\gamma_\perp = c \, (\epsilon/\visc)^{1/2},
\end{equation}
where $\epsilon$ is the energy injection rate and $c$ is an adjustable parameter (we use $c=0.3$ for all calculations, as did~\citealt{Scheko_sim}). Secondly, we assume $\sigma_\perp = 4\sigma_\parallel$, as in the isotropic case. We justify this simple closure by noting that the structure functions in figure~\ref{fig:struct_unlim} indicate that the parallel scales of velocity variations are predominantly on the outer scale (i.e.~$k_\parallel \sim k_0$) and that, at these scales, we indeed measure $\sigma_\perp \approx 4 \sigma_\parallel$ \dblbrck{using \eqref{eqn:kazantsev_param} below; see figure \ref{fig:kazantsev}}. Finally, the magnetic-energy growth rate $\gamma$ is obtained by time-averaging the measured growth rate during the exponential-growth phase of the simulated dynamo. With this information, equation \eqref{eqn:kaz_zeroflux} can be solved to obtain $\sigma_\parallel$ (and thus $\sigma_\perp$), which then  completely determines the magnetic-energy spectra. 

The results of this calculation and a comparison of the model and simulated spectra are shown in figure~\ref{fig:kaz_spectra}. Surprisingly good quantitative agreement is found between the time-averaged spectra obtained from the simulations and those predicted by~\eqref{eqn:kazantsev_solution}.

\subsubsection{Is unlimited Braginskii-MHD dynamo possible when ${\rm Re}\rightarrow\infty$?}\label{sec:anti-Stokes}

Numerical investigation of solutions to~\eqref{eqn:kaz_zeroflux} has shown that, in the limit $\mr{Rm} \rightarrow \infty$, the viability of the dynamo -- that is, whether its growth rate is positive or negative -- depends only on the \emph{relative} size of the parallel (stretching) motions compared to the perpendicular (mixing) ones, rather than on their absolute magnitude. In particular, as $\gamma_\perp$ increases and the mixing motions become more important (thereby \emph{decreasing} $\sigma_\parallel$ and $\sigma_\perp$), the dynamo growth rate will decrease if the parallel motions do not increase commensurately. 
Physically, this is due to an enhancement of small-scale mixing, which tends to bring field lines closer together and promotes their resistive annihilation. This suggests that {\em there is no unlimited Braginskii-MHD dynamo in the limit $\mu\rightarrow 0$}, a somewhat paradoxical result whose physical origin is further discussed at the end of \S\,\ref{sec:stokes}. 
This scenario can be inferred from figure \ref{fig:energy_unlim}, where the unlimited Braginskii-MHD system with the largest isotropic viscosity (and thus the weakest mixing motions) has the fastest growing dynamo. It is also consistent with the observation that the structure functions in figure~\ref{fig:struct_unlim} reveal that parallel variations of the velocity field reside mainly at the outer scale, while its perpendicular variations tend to reside at the viscous cutoff.  This indicates that $\sigma_\parallel$, $\sigma_\perp \rightarrow 0$ as $\mathrm{Re} \rightarrow \infty$ if $\Reprl$ is held fixed. 

\begin{figure}
    \centering
    \includegraphics[width=\textwidth]{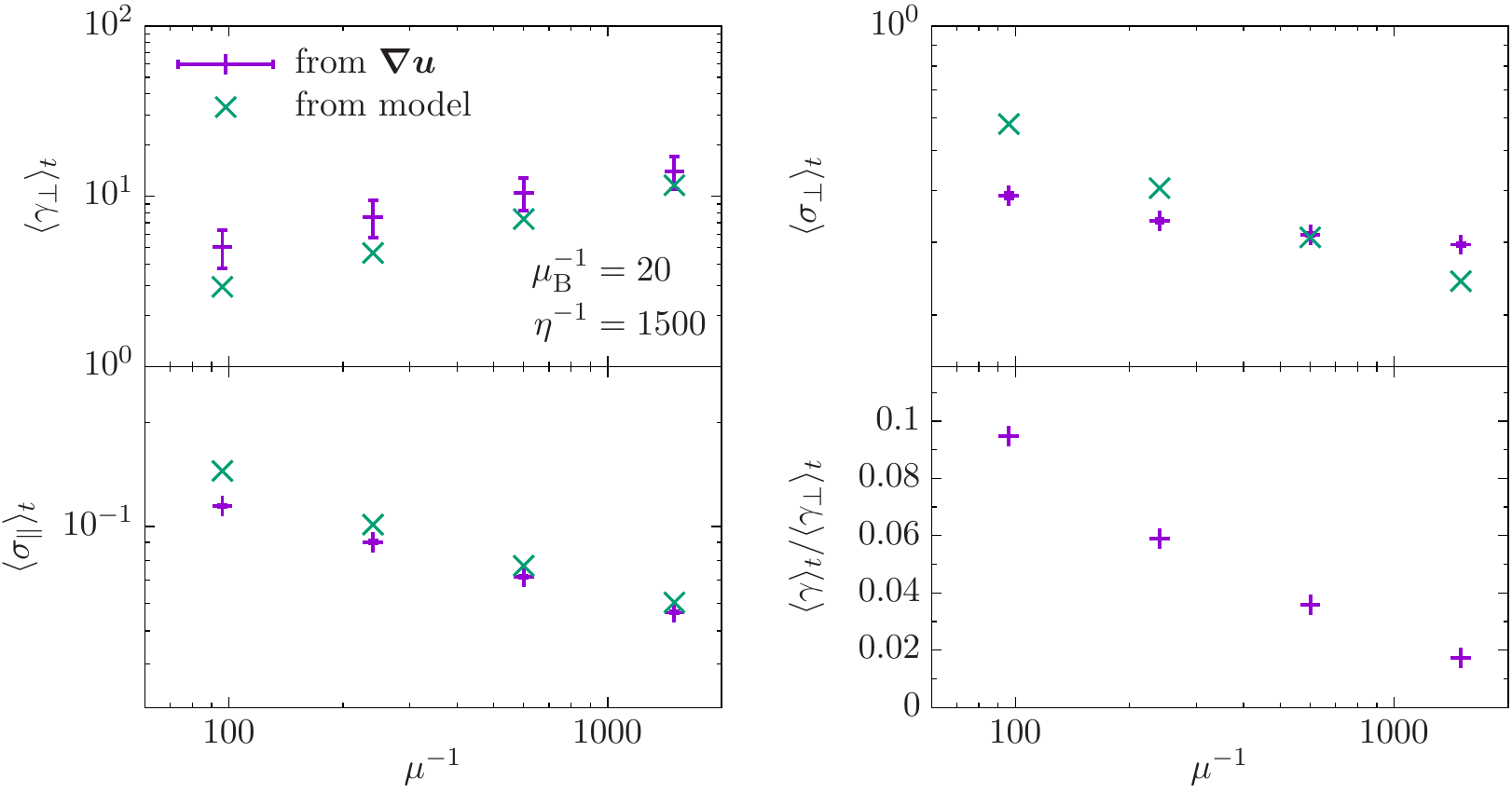}
    \caption{Modified Kazantsev--Kraichnan model parameters $\gamma_\perp$ (top left), $\sigma_\perp$ (top right) and $\sigma_\parallel$ (bottom left), calculated using the velocity gradient from simulations of unlimited Braginskii MHD according to~\eqref{eqn:kazantsev_param}, assuming $C_{\perp\perp}= 1$ (purple crosses), as well as from~\eqref{eqn:kaz_zeroflux} assuming $\sigma_\perp = 4 \sigma_\parallel$ (green x's). The former values are time-averaged over the exponential-growth phase. \emph{Bottom right:} magnetic energy growth rate $\gamma$ normalized by $\gamma_\perp$ for simulations of unlimited Braginskii MHD, time-averaged over the exponential-growth phase. 
    \label{fig:kazantsev}}
\end{figure}

This asymptotic behaviour is inferred by determining the dependence of $\sigma_\perp$, $\sigma_\parallel$ on $\visc^{-1}$ from our simulations and then extrapolating the trends. We do so using two separate methods, which fit the model outlined above to our simulation data in fundamentally different ways and are compared in figure \ref{fig:kazantsev}. The first method is to use the model solution~\eqref{eqn:kazantsev_solution} and the constraints given by~\eqref{eqn:kaz_zeroflux}, \eqref{eqn:kaz_mod_gammaperp}, and $\sigma_\perp = 4 \sigma_\parallel$. The final undetermined variable, the magnetic-energy growth rate $\gamma$, is obtained directly by measuring the growth rate from simulation data. The second method is to extract $\gamma_\perp$, $\sigma_\parallel$, and $\sigma_\perp$ directly from the simulations by modifying equations \eqref{eqn:gamma_delta_sing} to account for the finite correlation time $\tau_\mathrm{c}$ of the velocity field. In this case, the Fourier-space velocity correlation tensor $I^{ij}(\bb{k}) \doteq \int \od^3 \bb{k}\, \rme^{\imag \bb{k}\bcdot(\bb{x}-\bb{x}')}\ea{ u^i(\bb{x})u^{j}(\bb{x}')}$ is related to $\kappa^{ij}(\bb{k})$ via $I^{ij}(\bb{k}) = \tau^{-1}_\mathrm{c} \kappa^{ij}(\bb{k})$ \dblbrck{cf.~\eqref{eqn:kazantsev}}. We then take the correlation time of a given type of motions to be proportional to the `turnover' time of these motions; e.g., the correlation time of the perpendicular variations of velocity perpendicular to the magnetic field is $\tau_{\rm c,\perp\perp} = C_{\perp\perp} \gamma^{-1}_\perp$, where $C_{\perp\perp}$ is an undetermined coefficient. In principle, the correlation time may depend on both the alignment of the motion with respect to the magnetic field and the alignment of its variation, so we must also introduce $\tau_{\rm c,\parallel\perp} = C_{\parallel \perp} (\sigma_\perp \gamma_\perp)^{-1}$ and $\tau_{\rm c,\parallel\parallel} = C_{\parallel \parallel} (\sigma_\parallel \gamma_\perp)^{-1}$, where $C_{\parallel\perp}$ and $C_{\parallel\parallel}$ are further undetermined constraints. Then equations~\eqref{eqn:gamma_delta_sing} become
\begin{subequations}\label{eqn:kazantsev_param}
\begin{align}\label{eqn:kazantsev_gamma}
       \gamma_\perp &= \left[C_{\perp\perp}\int\frac{\rmd^3\bb{k}}{(2\upi)^3} \, k^2_\perp  I_\perp\right]^{1/2}  = \left[C_{\perp\perp} \langle | (\mathsfbi{I} - \eb\eb)\bcdot \grad\bb{u}\bcdot (\mathsfbi{I} - \eb\eb)|^2 \rangle \right]^{1/2} , \\
    \sigma_\perp &= \frac{1}{\gamma_\perp}\left[C_{\parallel\perp}     \int\frac{\rmd^3\bb{k}}{(2\upi)^3} \,k^2_\parallel  I_\perp\right]^{1/2} = \frac{1}{\gamma_\perp}\left[C_{\parallel \perp}\langle | \eb\bcdot \grad\bb{u}\bcdot (\mathsfbi{I} - \eb\eb)|^2 \rangle \right]^{1/2}, \label{eqn:kazantsev_sprp}\\
    \sigma_\parallel &= \frac{1}{\gamma_\perp} \left[C_{\parallel\parallel}\int\frac{\rmd^3\bb{k}}{(2\upi)^3} \, k^2_\parallel   I_\parallel\right]^{1/2} = \frac{1}{\gamma_\perp}\left[C_{\parallel \parallel}\langle | \ROS |^2 \rangle \right]^{1/2}, \label{eqn:kazantsev_sprl}
\end{align}
\end{subequations}
where $I_\perp$ and $I_\parallel$ are defined in a similar way to \eqref{eqn:kappa_prp} and \eqref{eqn:kappa_prl}, respectively. In order to eliminate some of the undetermined coefficients, we make the assumptions $C_{\parallel\parallel} = (1/6)C_{\perp\perp}$,   $C_{\parallel\perp} = (2/3) C_{\perp\perp}$, chosen to ensure that $\sigma_\perp = 2/3$ and $\sigma_\parallel = 1/6$ in the isotropic case $I^{\mr{(a)}} =0$.

Figure \ref{fig:kazantsev} shows the results of these measurements. The assumption  $c = 0.3$ in~\eqref{eqn:kaz_mod_gammaperp} appears to agree well with $\gamma_\perp$ 
estimated directly from the simulation data. The predicted value of $\sigma_\parallel$ also closely matches the estimated value obtained from the simulated $\grad \bb{u}$. Finally, and most importantly for this discussion, the scaled magnetic-energy growth rate decreases along with the isotropic viscosity $\visc$, indicating that there exists a critical value of $\visc$ at which the dynamo ceases to operate.\footnote{\,The value of $\sigma_\perp$ obtained from the model does not do as good of a job matching the estimated value from \eqref{eqn:kazantsev_sprp}. One possible reason is that \eqref{eqn:kazantsev_sprp} does not capture the correct correlation time for this motion. Alternatively, the assumption $\sigma_\perp = 4 \sigma_\parallel$ may not hold in general. The latter is not, however, essential to our argument:  if it is found in the future
that $\sigma_\perp$ and $\sigma_\parallel$ do, in fact, scale differently,  our qualitative conclusion still holds so long as $\sigma_\parallel$ and $\sigma_\perp$ continually decrease with $\mathrm{Re}$. Our limited empirical evidence at this point suggests that the latter statement is true.}

%
%

\subsection{Stokes flow: Is unlimited Braginskii-MHD dynamo possible when $\Reprl < 1?$}\label{sec:stokes}

It was predicted in \S\,\ref{sec:kazantsev_solution} that the unlimited Braginskii-MHD dynamo ceases to operate in the limit $\mathrm{Re} \rightarrow \infty$, $\Reprl = \mathrm{const}$. In this section, we ask whether the same is true in the opposite, `Stokes-flow' limit of $\mathrm{Re} = \mathrm{const}$, $\Reprl \rightarrow 0$. We will show that it is: at fixed $\mathrm{Re}$, the dynamo does cease to operate at sufficiently small $\Reprl$  (although it may be reactivated by decreasing $\mathrm{Re}$ to values comparable to $\Reprl$).

In the isotropic-MHD case, the correlation time of a white-noise-forced flow vanishes in the limit ${\rm Re}\rightarrow{0}$ (provided the forcing amplitude is scaled with viscosity) and the dynamo is well described by the Kazantsev--Kraichnan model. Provided there is sufficient scale separation between the forcing and resistive scales, the dynamo will continue to operate during this stage regardless of $\mathrm{Re}$ \citep{Scheko_theory}. To investigate whether a similar result holds for the Braginskii system, we perform a set of low-$\mathrm{Re}$ isotropic-MHD and low-$\Reprl$ unlimited Braginskii-MHD simulations at reduced resolution ($N_\mathrm{cell} = 112^3$). For each run, the forcing amplitude $\varepsilon$ is adjusted so that $u_\mathrm{rms}\sim 1$ in steady state. Details of these runs are given in the last block of table \ref{tab:runs}.

%
%
\begin{figure}
    \centering
    \includegraphics[scale=0.76]{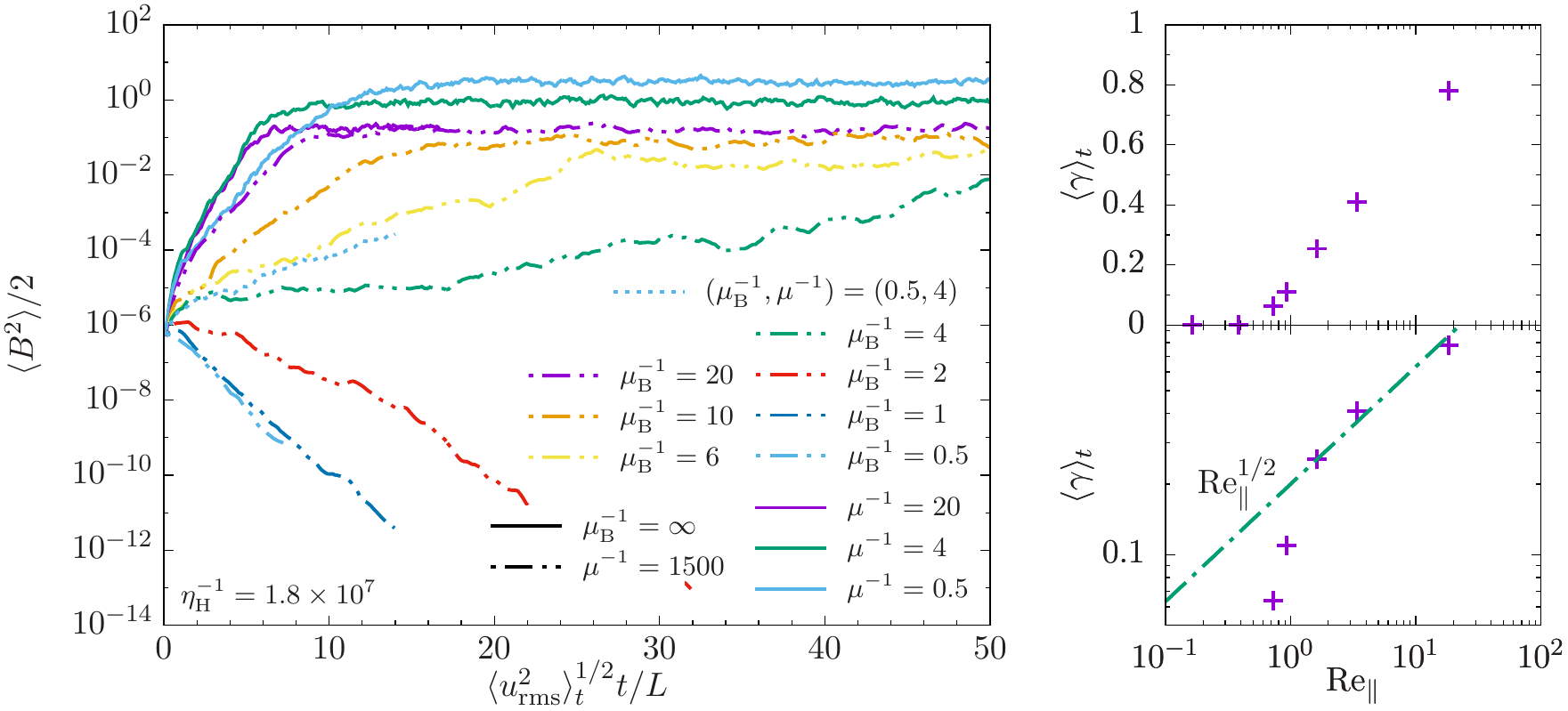}
    \caption{\emph{Left:} Evolution of the magnetic energy for various $\mathrm{Re} < 1$ (isotropic MHD, solid lines) and $\Reprl < 1$, $\mathrm{Re} \gg 1$ (unlimited Braginskii-MHD, dotted and dash-double-dotted lines) simulations; $\mu^{-1}_{\rm B}$ is varied from $20$ down to $0.5$ for simulations of unlimited Braginskii MHD. For $\visc^{-1}_\mathrm{B} \lesssim 4$ and $\visc^{-1} = 1500$, the unlimited Braginskii-MHD simulations do not exhibit a dynamo. \emph{Right:} Time-averaged growth rate $\gamma = \od \,\ln B^2_\mathrm{rms}/\od t$ of the magnetic energy during the kinematic stage as a function of  parallel Reynolds number for unlimited Braginskii MHD in the Stokes regime on a linear (top) and logarithmic (bottom) scale. }
    \label{fig:stokes_en_growth}
\end{figure}

The evolution of the magnetic energy for simulations in this Stokes-flow regime is plotted in figure \ref{fig:stokes_en_growth}. As predicted by the Kazantsev--Kraichnan model, the behaviour of the isotropic MHD simulations changes little, provided $u_\mathrm{rms}$ is kept constant across all runs. They all have magnetic-energy growth and reach saturation. The unlimited Braginskii-MHD dynamo, on the other hand, operates only beyond a certain critical value of $\Reprl\gtrsim{1}$, with an abrupt cut-off in the growth rate for smaller $\Reprl$. Well above this cut-off, the scaling of the growth rate still follows the expected \citet{Kolmogorov1941} scaling $\Reprl^{1/2}$ (figure \ref{fig:stokes_en_growth}, bottom right). These results, as well as those from the analytic model solved in \S\,\ref{sec:kazantsev_solution}, indicate that the dynamo is only viable for moderate values of the ratio $\Reprl/\mathrm{Re}$: too small a value results in a dynamo with too much mixing (relative to stretching) and thus excessive resistive dissipation. 

%
%
\begin{figure}
    \centering
    \includegraphics[scale=0.7]{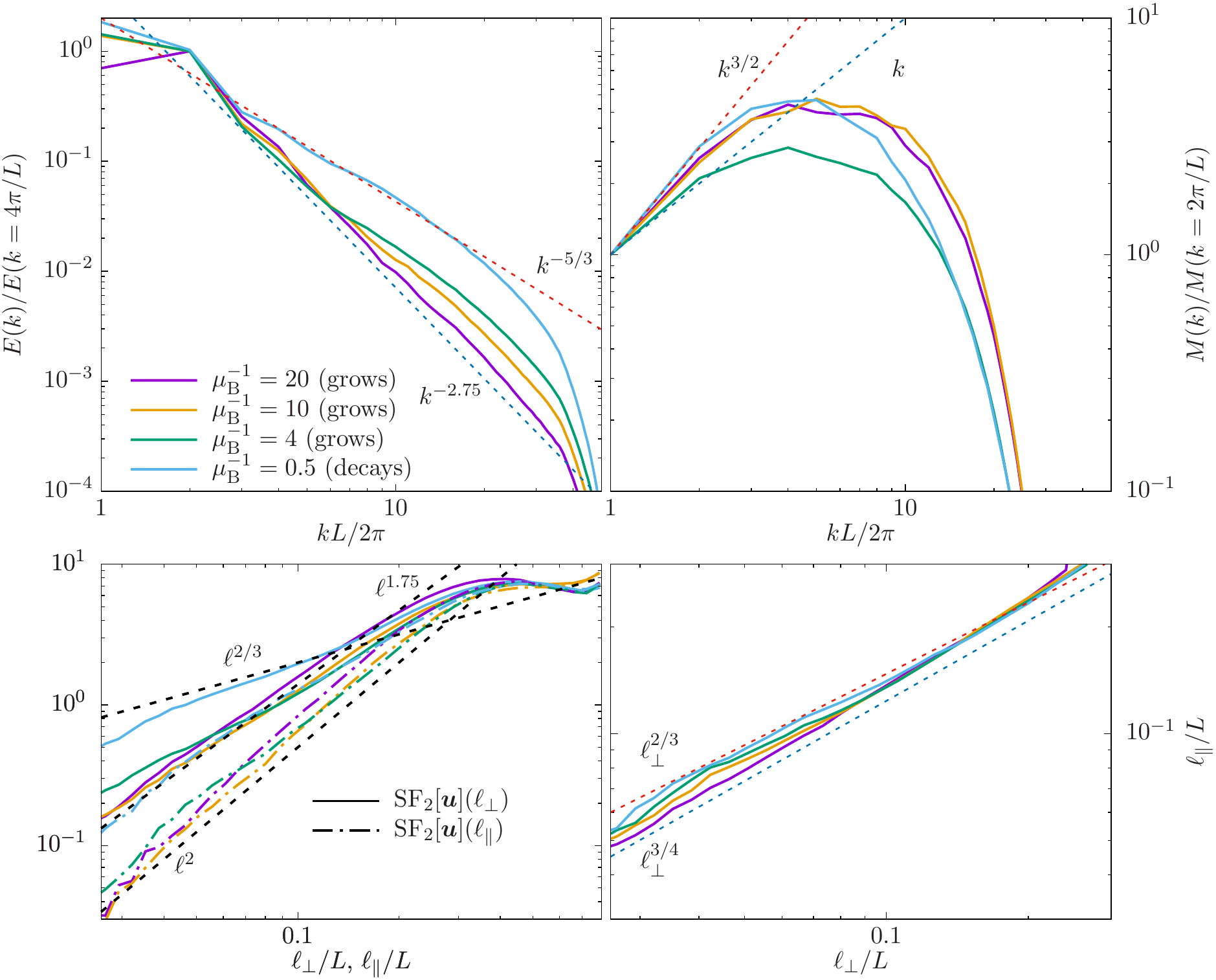}
    \caption{\emph{Top:}  Kinetic (left) and  magnetic (right) energy spectra for various unlimited Braginskii-MHD simulations with small and large parallel Reynolds number $\Reprl$. \emph{Bottom:}  Perpendicular and parallel structure functions (left) and  spectral anisotropy scalings (right) for the same runs \dblbrck{see \eqref{eqn:scale_aniso}}.}
    \label{fig:stokes_spec}
\end{figure}

The extinction of dynamo in both limits $\Reprl\rightarrow 0$, $\mathrm{Re}= \mathrm{const}$ and $\Reprl =\mathrm{const}$, $\mathrm{Re}\rightarrow \infty$ can be thought of as the result of a strong anisotropization of the underlying turbulence, leading to motions that are more two-dimensional than three-dimensional (see \S\,\ref{sec:constraints_discussion}). Interestingly, simulations that have $\Reprl$, $\mathrm{Re} < 1$ yet $\Reprl/\mathrm{Re}\sim 1$ still experience a viable dynamo; compare the dotted and dot-dashed light blue lines in figure \ref{fig:stokes_en_growth}. This shows that it is the \emph{relative size of the stretching to mixing time scales}, rather than their absolute values, that determine whether there is dynamo (cf.~\S\,\ref{sec:kazantsev}). Having an isotropic viscosity comparable to an anisotropic viscosity allows the momentum to diffuse more isotropically, undermining the anisotropic viscosity's tendency to make the flow more two-dimensional.
Indeed, the case with $\mathrm{Re} \sim \Reprl$ is more similar to the isotropic $\mathrm{Pm} \gtrsim 1$ dynamo, where the contributions of both stretching and mixing come from the same range of scales.

%
%
\begin{figure}
    \centering
    \includegraphics[width=\textwidth]{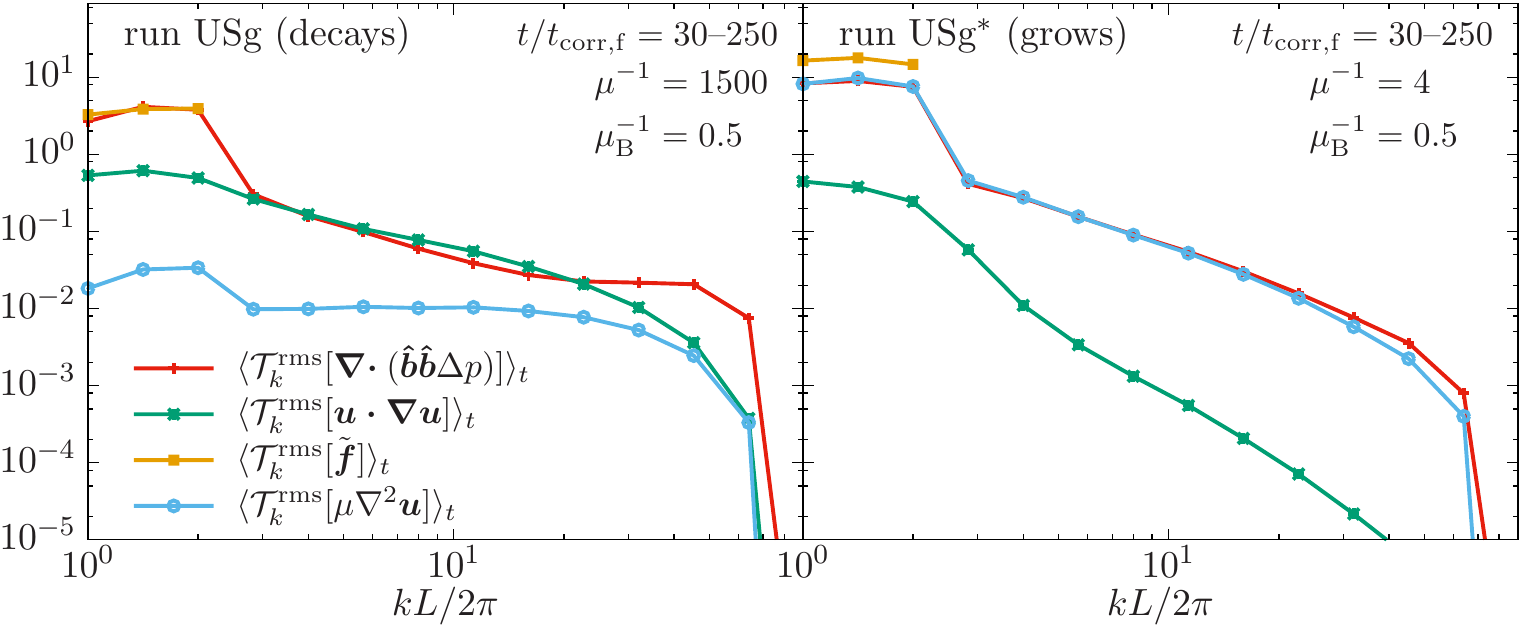}
    \caption{Root-mean-square shell-filtered kinetic-energy transfer functions $\mc{T}_k$ \dblbrck{see \eqref{eq:shell_trans_rms}} for the unlimited Braginskii-MHD system in the Stokes-flow regime employing $\mu_\mathrm{B}^{-1} = 0.5$ and $\mu^{-1} = 1500$ ($\mu^{-1} = 4$) in the left (right)  frame. The runs chosen here are one that corresponds to magnetic-field decay (USg, left) and one that is a viable dynamo (USg$^*$, right).}
    \label{fig:stokes_trans}
\end{figure}

\subsubsection{Mixing through a critically balanced pressure-anisotropy cascade}

Despite not supporting dynamo action for $\Reprl\lesssim{1}$, the unlimited Braginskii-MHD system still features a cascade of turbulent energy to small scales. The kinetic and magnetic energy spectra of various unlimited Braginskii-MHD simulations in the Stokes-flow regime are displayed in the top panels of figure \ref{fig:stokes_spec}. As $\Reprl$ is decreased, the kinetic-energy spectrum beyond the parallel viscous scale approaches a power law with the exponent $-5/3$, a slope seen most clearly in the $\mu_\mathrm{B} = 0.5$ run (which is a failed dynamo; see the light-blue dot-dot-dashed line in figure \ref{fig:stokes_en_growth}). For that run, the scaling of the wavevector anisotropy (bottom-right panel) is measured \dblbrck{via \eqref{eqn:scale_aniso}} to be $\ell_\parallel \propto \ell_\perp^{2/3}$. These scaling exponents, $-5/3$ and $2/3$, match those predicted by the \citet{Goldreich1995} theory for MHD turbulence in the presence of a strong guide field, where the linear (Alfv\'{e}n-wave) and nonlinear (turbulence) timescales are taken to be comparable (`critical balance'). Here, though, there is no guide field, and the Alfv\'{e}n-wave timescale is much too long at these low magnetic-field strengths to matter dynamically. So why then do these scalings arise? If the observed scalings are representative of a critically balanced state, just what time scales is it a critical balance between?

To answer these questions, we compute the shell-filtered kinetic-energy transfer functions \dblbrck{see \eqref{eq:shell_trans} and \eqref{eq:shell_trans_rms}} for the unlimited Braginskii-MHD system in the Stokes-flow regime. These are shown in figure~\ref{fig:stokes_trans}.\footnote{\,The root-mean-square transfer functions rather than the mean transfer functions are shown because they provide a clearer picture of the magnitude of non-linearities without capturing their sign.} In the failed Braginskii-MHD dynamo that nevertheless exhibits a robust turbulent cascade (left panels), we find from the root-mean-square transfer functions (bottom row) that in the inertial range there is a scale-by-scale balance between the hydrodynamic nonlinearity (green line) and the (unlimited) Braginskii viscous stress (red line). Thus, we can picture a cascade of MHD turbulence consisting of Alfv\'en-like waves propagating along the \emph{local} magnetic-field direction, with `critical balance' occurring between the linear frequency associated with the pressure anisotropy, ${\sim}k_\parallel\sqrt{\Deltap}$, and the nonlinear frequency associated with the eddy-turnover time, ${\sim}k_\perp u_\perp$. Effectively, the unchecked viscous stress supplants the (negligible) Maxwell stress in setting the characteristic linear frequency of the plasma.\footnote{\,Note that, in our dynamo runs, $\rmDelta p$ is positive throughout most of the volume and thus far away from the firehose instability threshold. As a result, the `interruption' effect proposed by~\citet{Squire2016}, which occurs at the firehose threshold, is not active.} Adopting the usual assumption of a conservative cascade with $k_\perp u^3_\perp \sim \const$, this balance gives $k_\parallel\sqrt{\Deltap} \sim k_\perp u_\perp \sim k_\perp^{2/3}$. Noting that the pressure anisotropy in this simulation is dominated by its mean value (and thus is largely scale-independent), the scaling $k_\parallel \sim k^{2/3}_\perp$  results (as is indeed seen in figure \ref{fig:stokes_spec} for this run). By contrast, the transfer functions in the complementary run with stronger isotropic viscosity (run Usg$^\ast$, which {\em is} a dynamo -- see the light-blue dotted line in figure \ref{fig:stokes_en_growth}) are qualitatively similar to those displayed in the lower panels ($\mu^{-1}_{\rm B} = 20$, $\mu^{-1} = 96$) of figure \ref{fig:trans_unlim}.

\section{Summary and Discussion}\label{sec:discussion}

In this work, we have studied the fluctuation dynamo in a weakly collisional, magnetized plasma using incompressible Braginskii-MHD simulations and semi-analytic modeling. The pressure anisotropy, and thus the parallel viscous stress, was either hard-wall limited to lie within the firehose and mirror instability thresholds or allowed to venture freely beyond those thresholds. While the latter option has traditionally been considered unphysical due to the plethora of evidence -- both theoretical and observational -- that kinetic-instability thresholds are well respected in collisionless, magnetized plasmas, its study is important for (at least) four reasons. First, the unlimited case teaches us about the fundamental dynamical properties of the pressure anisotropy stress, which has received relatively little attention in the past (particularly in the case of dynamo). Second, it offers an additional point of comparison to dynamo behaviour in isotropic MHD and limited Braginskii MHD; in particular, our finding of its similarity to the saturated state of the MHD dynamo affords a better understanding of that more traditional case. Third, many aspects of its evolution are remarkably similar to those found in the more {\em ab initio} hybrid-kinetic simulations of plasma dynamo presented in \citetalias{StOnge_2017}; this is fortunate, as the unlimited Braginskii-MHD simulations provide a better controlled and more economical test bed with which to diagnose the field and flow statistics in this regime.  Finally, we have argued that a significant period in the dynamo amplification of the intracluster magnetic field occurs at a time when the plasma $\beta$ is too large for kinetic instabilities to regulate the pressure anisotropy efficiently enough to pin it tightly near its ${\sim}1/\beta$ stability boundaries. During this phase, a constant collision frequency that partially restrains the pressure anisotropy (as in the unlimited simulations presented above) may in fact be the more realistic `closure'. Indeed, much of the evolution of the collisionless dynamo found in \citetalias{StOnge_2017} using a hybrid-kinetic approach occurred during such a phase, with a suppressed parallel rate of strain, an anisotropization of the flow velocity, an imperfect regulation of the pressure anisotropy, and a Kolmogorov-like cascade of perpendicular kinetic energy -- all of which are manifest in our unlimited Braginskii-MHD runs.

Our main conclusions are as follows:
\begin{enumerate}
    \item \label{bragcon1} The chaotic flow driven by large-scale forcing produces highly intermittent and structured magnetic fields, which are organized into folds and grow exponentially until the Lorentz tension force is strong enough to back-react dynamically on the velocity field. This folded structure, a hallmark of the ${\rm Pm}\gtrsim{1}$ fluctuation dynamo, persists into the saturated state. These results hold regardless of whether the plasma is described by isotropic MHD or Braginskii MHD with either limited or unlimited pressure anisotropy, so long as the ratio of the isotropic and Braginskii viscosities (or the ratio of the parallel and perpendicular Reynolds numbers; see \eqref{bragcon5} below) is not too small. 
    \item \label{bragcon2} Hard-wall limiters on the parallel viscosity, intended to mimic the rapid regulation of pressure anisotropy by kinetic Larmor-scale instabilities otherwise not properly captured in Braginskii MHD, effectively reduce the Braginskii-MHD dynamo to its ${\rm Re}\gg{1}$, ${\rm Pm}\sim{1}$ MHD counterpart. With the exception of some minor differences, such as a slight suppression of $\ROS$ in firehose/mirror-stable regions (figure \ref{fig:pdfROS}), the two are nearly indistinguishable. This conclusion is broadly consistent with the findings of \citet{SantosLima}, who used the CGL equations with limiters. Regions of the plasma lying near or beyond the mirror-instability threshold are subject to a magnetic tension that is effectively enhanced by a factor of $3/2$ and thus are qualitatively unaffected by the positive, limited viscous stress. The regions of the plasma lying near or beyond the firehose instability threshold, at which the effective tension from the Maxwell and viscous stresses is zero, appear to be unimportant to the dynamics of the dynamo, most likely because these regions have a smaller volume-filling factor and contain comparatively lower magnetic energy.
    \item \label{bragcon3} When the dynamical feedback of parallel-viscous dissipation on the field-stretching motions is allowed (the unlimited Braginskii-MHD model), the dynamo takes on a different character. Not only is it slower, but many characteristics of the flow and magnetic field change very little from the kinematic stage to the saturated state. Further, most of these characteristics bear a striking resemblance to those found in the saturated, rather than kinematic, state of the ${\rm Re}\gg{1}$, ${\rm Pm}\gtrsim{1}$ isotropic-MHD dynamo. These include: the magnetic-energy spectrum (figure \ref{fig:unlim_spec}), the characteristic scales of the folded magnetic-field geometry (figure \ref{fig:wavenumbers_unlim}), the PDF of the magnetic-field-line curvature (figure \ref{fig:curvature}), and the PDF of the alignment angles between the magnetic field and the rate-of-strain tensor eigenvectors (figures \ref{fig:angle2d_all} and \ref{fig:eigen_some}). In addition, a scale-dependent anisotropy of the velocity field was found in the saturated state of the limited Braginskii-MHD dynamo and throughout the entire evolution of the unlimited Braginskii-MHD dynamo (specifically, $\ell_\parallel\propto\ell^{3/4}_\perp$; see figures \ref{fig:struct_lim} and \ref{fig:struct_unlim}); a similar anisotropy was found in the saturated state of the isotropic-MHD dynamo.
    \item \label{bragcon4} Motivated by this resemblance between the saturated MHD state and `kinematic' unlimited Braginskii-MHD dynamos and by the structural similarity of the magnetic-tension and Braginskii-viscous stresses, we have constructed a theory for the unlimited Braginskii-MHD dynamo analogous to that developed by \citet{Scheko_saturated} to model the saturated sate of the ${\rm Pm}\gg{1}$ MHD small-scale dynamo (\S\,\ref{sec:kazantsev}). This theory introduces a field-biased rate-of-strain tensor into the Kazantsev--Kraichnan model of the fluctuation dynamo that captures the partial two-dimensionalization of the velocity-gradient statistics with respect to the local magnetic-field direction caused by the anisotropic viscosity. (In \citealt{Scheko_saturated}, this partial two-dimensionalization is instead caused by the dynamically important Lorentz force in the saturated state.) This model predicts magnetic-energy spectra in remarkably good quantitative agreement with those found in our simulations (arguably, more quantitative an agreement than we had the right to expect).
    \item \label{bragcon5} Another prediction of our modified Kazantsev--Kraichnan model is that enhanced small-scale mixing and local two-dimensionalization of the flow shut down the Braginskii-MHD dynamo as the isotropic viscosity $\visc \rightarrow 0$ (at fixed Braginskii viscosity). We have confirmed this behaviour numerically in the complementary limit of $\mu_{\rm B}\rightarrow\infty$ at constant $\mu$ -- the `Braginskii Stokes-flow' regime --  where we see that the unlimited Braginskii-MHD dynamo fails at sufficiently small parallel Reynolds numbers (at fixed $u_{\rm rms}$). Isotropic viscosity comparable to the Braginskii viscosity saves the dynamo in this situation by diffusively bleeding momentum into the field-parallel direction; indeed, we find that the Stokes-flow dynamo works in isotropic MHD regardless of the value of ${\rm Re}$ so long as there is sufficient scale separation between the forcing and resistive scales. Combined with conclusion \eqref{bragcon4} above, this implies that the Braginskii-MHD dynamo is only viable for moderate values of the ratio $\Reprl/{\rm Re}$: too small a value results in too much field-line mixing by the perpendicular flows and gradients relative to field-line stretching and thus to excessive resistive dissipation of the magnetic field. This principle is quantified in our modified Kazantsev--Kraichnan model by the quantities $\sigma_\parallel$ and $\sigma_\perp$ \dblbrck{see \S\,\ref{sec:kazantsev} and, in particular, \eqref{eqn:gamma_delta_sing}}.
\end{enumerate}

Going forward from here, three avenues for progress are open. First, there are striking similarities in the morphology and statistics of the flows and fields and the evolution of the pressure anisotropy PDF between the hybrid-kinetic runs reported by \citetalias{StOnge_2017} and the unlimited-Braginskii runs presented herein. This suggests that some version of the unlimited Braginskii-MHD equations might be used to describe the `magnetized, kinetic regime' of the plasma dynamo (i.e., one in which the effective collision frequency needed to constrain the pressure anisotropy within the firehose/mirror stability thresholds $\nu_{\rm eff} \sim \beta|\ROS| \gtrsim\Omegai$). Secondly, it should be assessed how accurate are the pressure-anisotropy limiters for describing the dynamo evolution in the `magnetized, fluid regime' ($\nu_{\rm eff}\lesssim \Omegai$), in which the pressure anisotropy can (in principle) be well regulated by the instabilities. If this regime is realized and the scalings conjectured in \S\,\ref{sec:Reeff} hold, the interpolation between these two regimes implies a parallel Reynolds number increasing with magnetic-field strength and thus the possibility of an explosive dynamo. This idea will be the subject of a separate publication.

Finally, we have not been able to assess definitively whether Braginskii viscosity breaks the tendency for the small-scale dynamo to saturate with the majority of its magnetic energy residing near resistive scales. On the one hand, it is interesting that the magnetic spectrum found in the unlimited Braginskii-MHD dynamo shows little tendency to concentrate power on resistive scales, and even less tendency to evolve in time in going from the kinematic stage to the saturation (see figure \ref{fig:unlim_spec}).
On the other hand, because computational expense limits the maximum achievable scale separation between the viscous and resistive scales in our simulations, we cannot yet establish whether the peak of the spectrum is independent of ${\rm Rm}$, or whether the conjecture by \citet{Malyshkin} -- that the interchange motions that are undamped by Braginskii viscosity unwrap the folded magnetic fields and thus promote their inverse cascade to larger scales -- can be realized. Future numerical work, both fluid and kinetic, should maximize scale separation with the goal of definitively evaluating the ability of the fluctuation dynamo to generate saturated magnetic fields with large-scale coherence in weakly collisional plasmas such as the intracluster medium.

\vspace{1ex}
Support for D.A.S.~and M.W.K.~was provided by US DOE Contract DE-AC02-09-CH11466 and an Alfred P.~Sloan Research Fellowship in Physics. Support for J.S.~was provided by Rutherford Discovery Fellowship RDF-U001804 and Marsden Fund grant UOO1727, which are  managed through the Royal Society Te Ap\=arangi. A.A.S.~was supported in part by the UK STFC Consolidated grant ST/N000919/1 and by UK EPSRC grants EP/M022331/1 and EP/R034737/1. The authors thank Fran\c{c}ois Rincon, Philipp Kempski, Eliot Quataert, and Valentin Skoutnev for valuable conversations; the anonymous referees for the time and care they spent in reading our manuscript; and Steve Cowley, who with Jason Maron initiated a study of this problem in 2001 \citep{MaronCowley01}, for his insight and encouragement. A.A.S.~thanks Marek~Strumik, whose (unpublished) investigations of the dynamical effects of the pressure-anisotropy stress informed some of our early thinking on this problem. The completion of this work was facilitated by the generous hospitality of the Wolfgang Pauli Institute in Vienna and of the Kavli Institute for Theoretical Physics in Santa Barbara (supported by the National Science Foundation under Grant No.~NSF PHY-1748958).

\appendix

\section{\label{ap:linear} Linear stability analysis of the unlimited Braginskii-MHD equations}

In this appendix, we present the linear theory for the unlimited, incompressible, Braginskii-MHD system of equations \eqref{brag_MHD}. We demonstrate that, while these equations exhibit a parallel firehose instability, they do not correctly capture the mirror instability. 

We assume an unforced, homogeneous, stationary equilibrium state with a uniform pressure anisotropy $\Deltap \ne 0$ and $\bb{B}_0 = B_0\ez$, subject to small-amplitude perturbations (subscripted with a `1') whose amplitudes are of the form $\exp(\gamma t + \imag\bb{k}\bcdot\bb{x})$.
 While this equilibrium is formally incompatible with the Braginskii closure $\Deltap = 3\mu_\mr{B} \ROS$, we consider perturbations with sufficiently high frequencies ($\omega \gg | \grad \bb{u}|$) and small scales ($k \gg k_\mu$) that $\Deltap$ may be considered constant in space and time (a similar analysis using a $\Deltap$ obtained from the drift kinetic equation was carried out by \citealt{Scheko_2005}).

Without loss of generality, let $\bb{k} = k_\perp \ex + k_\parallel \ez$. The equations, written to first order in the perturbation amplitudes and assuming Laplacian diffusion, are
\begin{align}
( \gamma + \mu k^2 ) \bb{u}_1 &= - \imag\bb{k} p_1  +  \imag k_\parallel  B_0^2 \bb{b}_1 + \imag k_\parallel \Deltap \left(\bb{b}_1 - 2b_{1z}\ez\right) - \mu_\mr{B} k^2_\parallel u_{1z}\ez,\label{eqn:linmom}
\\ \bb{b}_1 &= \frac{\imag k_\parallel }{\gamma + k^2 \eta} \bb{u}_1,  \label{eqn:A2}
\end{align}
where $\bb{b}_1 \doteq \bb{B}_1/B_0$. If one were to consider other types of diffusion, $\mu$ and $\eta$ may be simply redefined (e.g.~$\mu \rightarrow \mu_\mr{H}k^2$ for hyper-diffusion). Incompressibility ($\bb{k}\bcdot\bb{u}_1$ = 0) is used to determine $p_1$:
\begin{equation}\label{eqn:p1}
p_1 = ( \imag k_\parallel \mu_\mr{B} u_{1z} - 2b_{1z}\Deltap  )\frac{k_\parallel^2}{k^2} ,
\end{equation}
where we have used the solenoidality constraint on the perturbed magnetic field, \emph{viz}.,  $\bb{k}\bcdot\bb{b}_1=0$. Substituting \eqref{eqn:p1} into \eqref{eqn:linmom} leads to 
\begin{align} \label{eqn:A4}
(\gamma + \mu k^2) \bb{u}_1 &=  \imag k_\parallel B_0^2 \bb{b}_1 + \imag k_\parallel \Deltap \Bigl[\bb{b}_1 - 2b_{1z}\ez\bcdot(\msb{I}-\bb{\hat{k}\hat{k}})\Bigr] - \mu_\mr{B} k_\parallel^2 u_{1z}\ez \bcdot(\msb{I}-\bb{\hat{k}\hat{k}}).
\end{align}
Combining \eqref{eqn:A2} and \eqref{eqn:A4} yields the matrix equation
\begin{align}\label{eqn:dr_matrix}
\bigl( \gamma + \eta k^2 \bigr) &\bigl[ (\gamma + \mu k^2 )\msb{I} + \mu_\mr{B}k_\parallel^2  (\msb{I}-\bb{\hat{k}\hat{k}})\bcdot\ez\ez  \bigr]\bcdot \bb{b}_1  
\nonumber\\* \mbox{} &= - k_\parallel^2 B_0^2 \bb{b}_1 -  k_\parallel^2 \Deltap \, \bigl[\msb{I} - 2(\msb{I}-\bb{\hat{k}\hat{k}})\bcdot \ez\ez\bigr]\bcdot \bb{b}_1.
\end{align}
Solenoidality requires that $b_{1z} = -(k_\perp/k_\parallel) b_{1x}$, splitting the dispersion relation into two separate branches:
\begin{align}
\bigl(\gamma + \eta k^2 \bigr) \bigl(\gamma + \mu k^2\bigr)  &=  - k_\parallel^2  \bigl(B_0^2 + \Deltap \bigr) , \label{ap:branch1} \\*
\bigl(\gamma + \eta k^2 \bigr) \left(\gamma + \mu k^2 + \mu_\mr{B}  \frac{k_\parallel^2k_\perp^2}{k^2} \right)  &=  - k_\parallel^2 \left(B_0^2 +   \Deltap \frac{k_\parallel^2 -k_\perp^2}{k^2} \right). \label{ap:branch2}
\end{align}
The first branch, corresponding to \eqref{ap:branch1}, involves magnetic-field perturbations that are perpendicular to the background magnetic field, with $\eb_1 \doteq \bb{b}_1/|\bb{b}_1| = (1,0,0)$ or $\eb_1 = (0,1,0)$ (i.e.~$\delta\bb{B}_\perp$). The second branch, corresponding to \eqref{ap:branch2}, involves both transverse and longitudinal (i.e.~$\delta B_\parallel$) magnetic-field fluctuations, so that $\eb_1 = (-k_\parallel/k_\perp, 0 ,1)$.

The solution of \eqref{ap:branch1} for the first branch is
\begin{equation}
    \gamma =  -D_+ \pm \sqrt{ D_-^2 - k_\parallel^2(B_0^2+\Deltap)},
\end{equation}
where $D_\pm \doteq (\mu \pm \eta) k^2/2$. The dissipationless limit $\mu = \eta =0$ returns the firehose stability threshold $\Deltap > -B_0^2$ \dblbrck{see \eqref{stability}}. Instability occurs once the destabilizing pressure anisotropy nullifies the otherwise restoring magnetic-tension force that is responsible for the shear-Alfv\'en wave.

Equation \eqref{ap:branch2} for the second branch is most clearly analyzed in the limit of $\mu_\mr{B} k_\parallel^2 \gg \gamma$ and $\mu = \eta = 0$, which gives
\begin{align}\label{eqn:mirror_stab}
 \gamma  \approx \frac{1}{\mu_\mr{B}} \frac{k^2}{k_\perp^2} \biggl( - B_0^2 + \Deltap\frac{k_\perp^2-k_\parallel^2}{k^2} \biggr).
\end{align}
The stability criterion is  $\Deltap(1-2k_\parallel^2/k^2) < B_0^2$, beyond which the mode exhibits a positive but \emph{scale-independent} growth rate in the limit $k_\parallel / k \ll 1$, in which the magnetic-field perturbation is dominated by $\delta B_\parallel$. This illustrates the inability of the Braginskii-MHD system to capture correctly the mirror instability, which in a kinetic calculation exhibits a growth rate that increases as $|k_\parallel|$ (at least until finite ion-Larmor-radius effects intercede and reduce growth) and has a stability threshold given by $\Deltap < B_0^2/2$~\citep{Hasegawa69}. Instead, the Braginskii mirror instability grows at a rate proportional to $\Deltap/\mu_\mathrm{B}$, i.e., comparable to the rate of strain of the field-stretching motions that are responsible for driving the pressure anisotropy in the first place (which makes its calculation above formally invalid, since constancy of the background rate of strain was assumed in it).


\section{Magneto-immutability in unlimited Braginskii-MHD dynamo}\label{sec:magnetoimmutability}

The minimization of $\ROS$ seen in our unlimited Braginskii-MHD simulations is reminiscent of what was found in the recent study of guide-field Braginskii-MHD turbulence by \citet{Jono_magnetoimmutability}, in which the flow self-organizes to minimize changes in magnetic-field strength and the consequent production of pressure anisotropy -- an effect named `magneto-immutability' by those authors. Given the discussion in the main text concerning the anisotropization of the velocity field by the unlimited Braginskii viscosity, it is worth asking whether there is any physical relationship between the results in \S\,\ref{sec:anisotropization} and those presented by \citet{Jono_magnetoimmutability}. We also ask whether certain minor differences between the limited-Braginskii-MHD dynamo and its isotropic-MHD counterpart in their saturated states may be attributable to magnetoimmutability effects.

%
%
\begin{figure}
    \centering
    \includegraphics[scale=0.73]{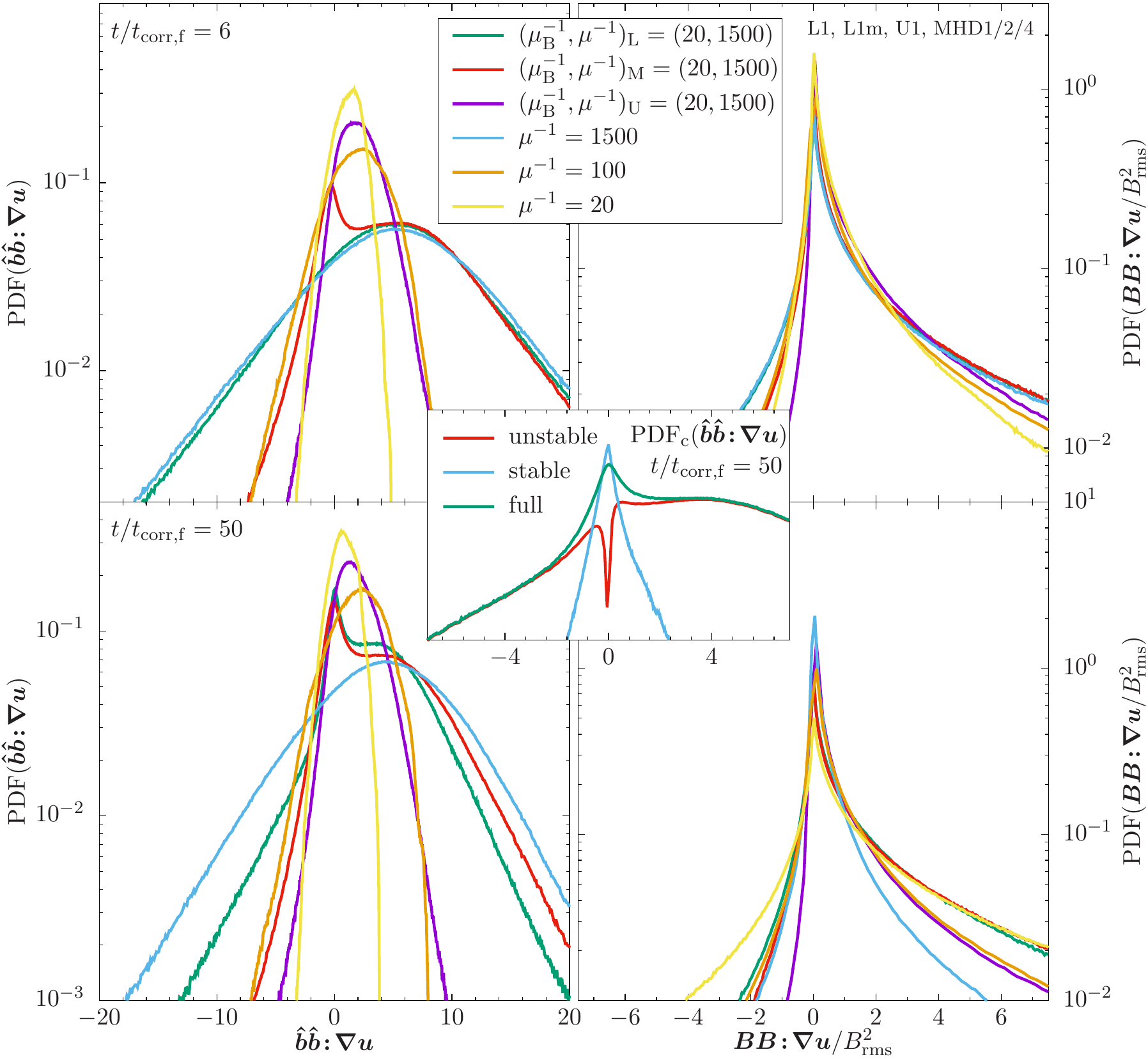}
    \caption{\label{fig:pdfROS} Probability distribution function of the parallel rate of strain (left) and the energy-weighted parallel rate of strain (right) for unlimited Braginskii-MHD (purple line; run U1) and limited Braginskii-MHD employing limiters on the firehose and mirror side (green line; run L1) and on the mirror side only (red line; run L1m). Also included for reference are three simulations of isotropic MHD (blue/orange/ yellow lines; runs MHD1/2/4). The PDFs are shown after six correlation times of the random forcing, corresponding to the diffusion-free regime where the resistive scale has not been reached (top) and after fifty correlation times, corresponding to the end of the exponential-growth phase and onset of saturation (bottom). 
    \emph{Inset:} PDF of the parallel rate of strain for a limited Braginskii-MHD simulation, conditioned on the stability criteria given by equation \eqref{stability} at $t/\tcorrf=50$. }
\end{figure}

\subsection{Statistics of $\ROS$ and $\bb{BB\!:\!\nabla u}$}
\label{sec:m-i_stats}

To address these questions, we follow \citet{Jono_magnetoimmutability} in calculating the PDFs of both the parallel rate of strain $\ROS$ and the energy-weighted parallel rate of strain $\boldsymbol{BB\!:\!\nabla u}/B^2_\mathrm{rms}$. The former quantity gives insight into how magneto-immutability affects the statistics of the parallel rate of strain, while the latter can be used to see how magneto-immutability affects the growth of the magnetic field on average. The results are displayed in figure \ref{fig:pdfROS}, whose upper panels show these PDFs after six correlation times of the random forcing (corresponding to diffusion-free stage; see \S\,\ref{sec:overview}) and the bottom panels show the same PDFs after $50$ correlation times (corresponding to the end of the exponential-growth stage and the onset of saturation). 

At the start of the kinematic stage, the PDF from the limited Braginskii-MHD run is almost identical to that from the $\mathrm{Pm} = 1$, $\mathrm{Re} \gg 1$ MHD run (MHD4). By contrast, the unlimited Braginskii-MHD dynamo resembles more closely the ${\rm Pm}\gg{1}$ MHD dynamo (MHD1), suggesting a strong tendency towards a magneto-immutable state in which $\ROS$ is dynamically regulated (although the former's PDF is slightly broader, because Braginskii viscosity is overall less efficient at damping motions than is an isotropic viscosity of similar magnitude.) As the limited Braginskii-MHD dynamo evolves (green line in figure \ref{fig:pdfROS}), the PDF of the parallel rate of strain develops a large notch centred around $\ROS = 0$, indicating a relative preference in Braginskii MHD for motions with $\ROS \approx 0$. This is consistent with the finding that even systems with hard-wall pressure-anisotropy limiters show a tendency towards magneto-immutability~\citep{Jono_magnetoimmutability}. The notch in the PDF, clearly seen at $t/\tcorrf = 50$ in the green and red curves, can be attributed to the unlimited Braginskii viscosity acting on the mirror- and firehose-stable regions of the plasma. This becomes clear if one conditions the PDF to examine regions that lie either within or outside of the stability region as defined by \eqref{stability}; this conditional PDF is displayed in the inset at the centre of figure \ref{fig:pdfROS}. As portions of plasma enter the region of stability \eqref{stability}, unlimited viscous forces quickly damp the parallel rate of strain, condensing the PDF in that region near zero and forming a sharp peak that contrasts with the otherwise wide PDF in the unstable regions where the Braginskii viscosity is greatly reduced. This has the effect of rendering the PDF of $\ROS$ in the limited Braginskii simulation near saturation much thinner than its MHD counterpart (cf.~green and blue lines in the bottom panels).

The simulations of Braginskii-MHD dynamo performed here, in conjunction with the PDFs displayed in figure~\ref{fig:pdfROS}, can also be used to validate certain conjectures on the behaviour of the dynamo.

First, it was envisioned by~\citet{Melville} that microscale instabilities could produce an environment  more favourable to motions that amplify the magnetic field than to those that damp it, which would result in a dynamo potentially faster than its isotropic-MHD counterpart. This proposition is tested here in the extreme limit where hard-wall pressure-anisotropy limiters are employed only on the mirror side (run L1m), resulting in a situation where only motions that decrease the magnetic-field strength are strongly damped. One might expect that this case would lead to the fastest possible fluctuation dynamo for a given set of diffusion coefficients. Indeed, the left panels of figure \ref{fig:pdfROS} show that negative rates of strain are suppressed compared to the fully limited simulation (compare the red and green lines), suggesting the potential for faster dynamo growth in the case without a firehose limiter. However, these negative rates of strain only amount to modest changes to low-probability regions in the PDF of the \emph{energy-weighted} rate of strain, whose average drives magnetic-energy growth. The evolution of the box-averaged magnetic energy in this run is not substantially different than in those employing limiters on both regions of instability (see figure \ref{fig:energy_lim}).

Secondly, it was argued by \citet{Malyshkin} that, for an isotropically tangled magnetic field on sub-viscous scales, the anisotropic viscous stress on the average resembles an isotropic viscous stress with effective viscosity $\mu_\mr{B}/5$. To test this hypothesis, the PDFs of $\ROS$ and $\boldsymbol{BB\!:\!\nabla u}/B^2_\mathrm{rms}$ from an isotropic-MHD simulation (run MHD2, $\mu=1/100$) are included in figure \ref{fig:pdfROS} as the orange lines. These PDFs resemble closely those found in the unlimited Braginskii-MHD case with $\mu_{\rm B}=1/20$ (purple line), as argued. However, the magnetic-energy growth rate is greatly reduced in the unlimited Braginskii system (compare the blue dashed and purple solid lines in the left panel of figure \ref{fig:energy_unlim}). As explained in \S\,\ref{sec:unlim_evolution}, this is because sub-parallel-viscous mixing motions allowed by the Braginskii viscosity promote resistive dissipation of the magnetic field, a fact not considered in the work of~\citet{Malyshkin}. Thus, while an effective isotropic viscosity of $1/5$ its anisotropic counterpart may indeed give an idea of the overall levels of $\ROS$, its use does not accurately capture all the features of the Braginskii-MHD dynamo.

%
%
\begin{figure}
    \centering
    \includegraphics[scale=0.85]{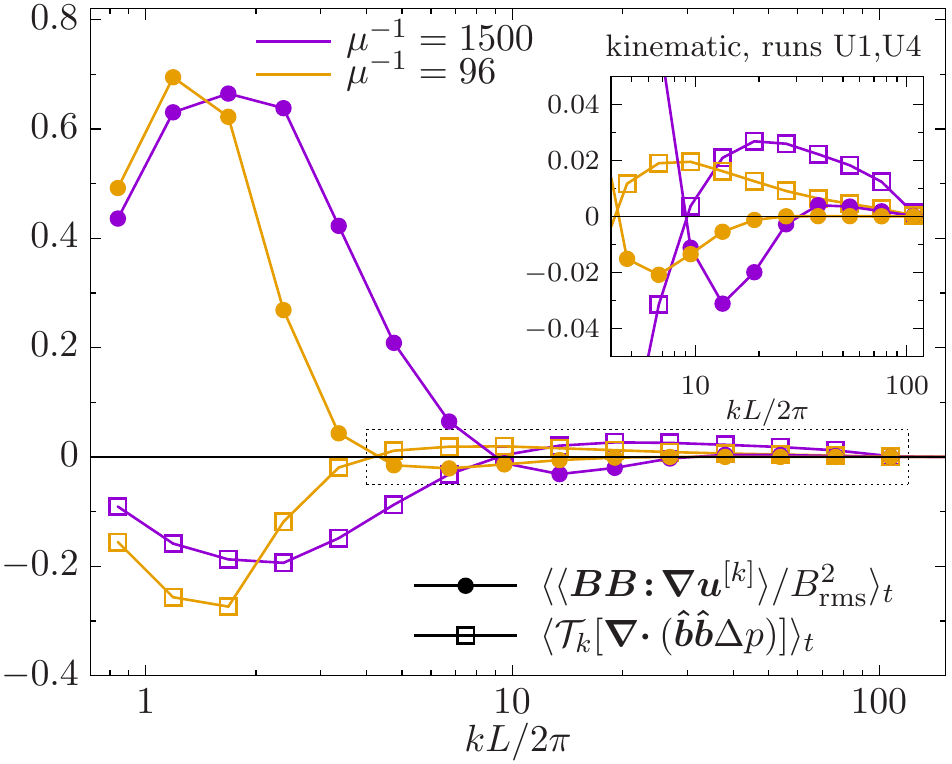}
    \caption{\label{fig:unlim_trans} Time-averaged $k$-shell-filtered energy transfer function of the pressure anisotropy $\mc{T}_k[\grad \bcdot (\eb \eb \rmDelta p)]$ \dblbrck{see \eqref{eq:shell_trans}} and the shell-filtered energy-weighted rate of strain $\langle\boldsymbol{BB\!:\!\nabla u}^{[k]} \rangle/B_\mathrm{rms}^2$ \dblbrck{see \eqref{eq:shell_u}} for runs U1 (purple) and U4 (orange). Bins have size $\sqrt{2}k$, where $k$ is the centre of the bin. These quantities have been averaged over 60 correlation times during the kinematic stage of the unlimited Braginskii-MHD dynamo. \emph{Inset:} Enlargement of the region enclosed by the dotted rectangle. Note that regions of kinetic-energy injection coincide with regions of negative $\bb{BB\!:\!\grad u}$; small-scale fluctuations attempt to nullify the net-positive parallel rate of strain and the consequent growth of magnetic energy driven by the large scales.}
\end{figure}

\subsection{Nullification of $\ROS$ by small-scale modes}
\label{sec:m-i_null}

Further evidence of magneto-immutability can be seen in the $k$-space transfer of kinetic energy by the Braginskii viscous stress, which is shown in figure \ref{fig:unlim_trans} for the kinematic stage of runs U1 (purple lines) and U4 (orange lines). While some of the energy extracted from the large-scale motions is damped away -- note that $\mc{T}_k[\grad \bcdot (\eb \eb \rmDelta p)] < 0$ at large scales -- a small portion of it is transferred to small-scale fluctuations.\footnote{\,The net effect of this term is damping, because
$3\mu_\mr{B}\langle\bb{u}\bcdot \{\grad \bcdot [\eb\eb(\ROS)] \}\rangle = - 3\mu_\mr{B} \langle|\ROS|^2\rangle$,
with the surface term vanishing in a triply periodic box.} These small-scale fluctuations attempt to nullify the net positive parallel rate of strain by introducing \emph{negative} rates of strain, as highlighted in the inset of figure \ref{fig:unlim_trans}. In both runs, these motions attempt to counteract the growth of magnetic energy driven by the large scales. In a system where the collisional relaxation of pressure anisotropy is governed by a constant $\mu_{\rm B}$ (as in our unlimited runs), the only means of regulating the pressure anisotropy and thereby avoiding strong damping of the large-scale motions is to re-organize the fields and flows to control $\ROS$ dynamically.\footnote{\,Similar dynamical regulation is done by the nonlinear parallel firehose instability \citep{Scheko_2008,Rosin_2011}, where the contribution to $\ROS$ from the small-scale firehose fluctuations partially offsets the contribution to $\ROS$ from the large-scale motions to maintain the plasma at marginal stability. Such microphysical control of $\ROS$ was invoked in \cite{Kunz_2011} to argue that the associated parallel-viscous heating rate, $Q^+ \sim \mu_{\rm B} |\ROS|^2$, was also regulated (see their footnote 8). The resulting $Q^+ \propto \mu^{-1}_{\rm B} B^4 \propto T^{-5/2} B^4$ causes the plasma to be thermally stable to isobaric perturbations when the cooling rate of the plasma is due to thermal Bremsstrahlung (as it is in the bulk ICM).} In kinetic systems, the pressure anisotropy, and thus the parallel viscous stress, can be reduced instead by anomalously increasing the collision frequency through the pitch-angle scattering of particles off firehose/mirror fluctuations. Indeed, such an enhanced collisionality has been directly measured in hybrid-kinetic simulations of the firehose and mirror instabilities \citep{Kunz_kin, Melville}, and is what underpins the very idea of hall-wall limiters, in which $\mu_{\rm B}$ is effectively modified to maintain a kinetically stable plasma. No such collisional regulation can occur in our unlimited Braginskii-MHD runs, and so the motions responsible for driving $p_\perp\ne p_\parallel$ must adjust. Interestingly, the high-Re isotropic MHD simulation in the saturated state also exhibits small-scale structures that attempt to nullify $\ROS$ (see figure~\ref{fig:trans_mhd_lim} and \S\,\ref{sec:limited}), a feature not previously noted in the literature. Apparently, the unlimited Braginskii-MHD dynamo takes similar steps as the saturated MHD dynamo to reduce the overall stretching rate of the magnetic field.

%
%
\begin{figure}
    \centering
    \includegraphics[scale=0.67]{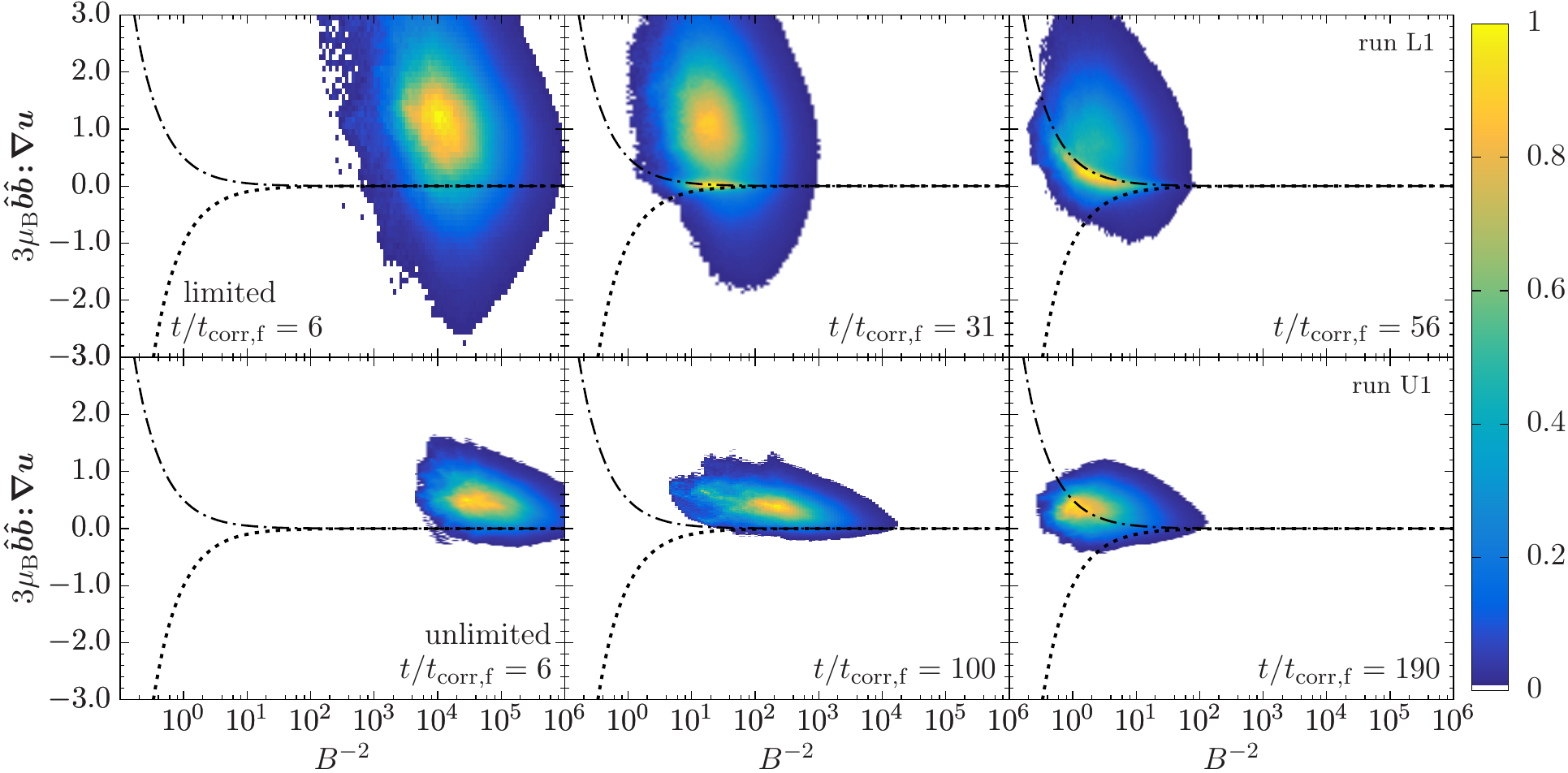}
    \caption{ \label{fig:brazil} PDF of the parallel rate of strain with respect to $B^{-2}$ for the hard-wall limited run L1 (top) and unlimited Braginskii-MHD run U1 (bottom) at successive times. Dashed-dotted (dotted) lines trace the mirror (firehose) instability thresholds given by  \eqref{stability}.}
\end{figure}

\subsection{Regulation of pressure anisotropy}
\label{sec:m-i_brazil}
 
Signatures of this adjustment of $\ROS$ can be seen in figure \ref{fig:brazil}, which displays the PDF of the parallel rate of strain with respect to the reciprocal of the magnetic energy $B^{-2}$ (which serves as a proxy for the plasma $\beta$ seen in similar histograms of pressure anisotropy; see, e.g., \citealp{Bale} and \citetalias{StOnge_2017}). Dash-dotted (dashed) lines trace the mirror (firehose) instability thresholds given by \eqref{stability}, with the stable region lying to the left of these lines. In the limited case, the bulk of the plasma stays beyond the mirror-instability threshold, also migrating toward larger magnetic energies as the dynamo progresses. Because the viscous stress is limited, its ability to damp out the motions driving the PDF in the direction of mirror instability is attenuated. As a result, the portions of the plasma lying beyond the stability boundaries are subject to a greatly reduced viscous stress that is comparable to the (small) Maxwell stress. However, as the magnetic energy grows, the limited viscous stress is able to regain its dynamical importance. More and more plasma exhibits pressure anisotropy in the stable region (as also seen in figure \ref{fig:pdfROS}) and evolves in a way that appears to respect the mirror boundary. No such behaviour is seen in the unlimited run, although the bulk of the plasma does seem to sit on the mirror threshold $\Deltap \approx B^2_0/2$ once saturation is reached.\footnote{\,This result is not too surprising: $u^2 \sim B^2$ implies $\Reprl \times\visc_\mathrm{B} u/\ell_0 \sim B^2$, by definition of the parallel Reynolds number. Thus, for a $\Reprl \sim 1$ flow, the pressure anisotropy implied by the parallel rate of strain in saturation is of the same order as the magnetic energy.} Instead, there is a dynamical feedback on the motions with $\ROS > 0$ that limits their ability to amplify the magnetic field and drive $p_\perp > p_\parallel$.

Figure \ref{fig:brazil} is to be compared to the analogous figure for pressure anisotropy versus plasma $\beta$ in the magnetized kinetic regime computed using a collisionless hybrid-kinetic code (figure 2 of \citetalias{StOnge_2017}). The behaviour throughout much of the kinetic calculation bears a stronger resemblance to that found in the \emph{unlimited} Braginskii run than in the limited run. As explained in \S\,\ref{sec:comparison}, the effective collision frequencies required at such large plasma $\beta$ to pin the pressure anisotropy instantaneously at its stability thresholds are difficult to realize in a simulation with limited scale separation. Indeed, the simulations of \citetalias{StOnge_2017} have a box-averaged anomalous collision frequency that is much smaller than $\beta|\ROS|$ throughout the kinematic state of evolution, which puts further in doubt whether hard-wall limiters can truly serve as a realistic model for the action of mirror and firehose instabilities in kinetic plasmas.

\section{\label{ap:kazantsev}Derivation of anisotropic Kazantsev--Kraichnan model}

For completeness, we provide here the full derivation of \eqref{eqn:mod_kazantsev}, the anisotropic Kazantsev--Kraichnan model presented and utilized in \S\,\ref{sec:kazantsev}. This derivation was omitted in \citet{Scheko_saturated} for lack of space and never came to be published anywhere, but we feel that it is important to spell out all the steps and assumptions that went into it in order for the reader to be able to judge the level of plausibility of our arguments in \S\,\ref{sec:kazantsev}.

We start with the evolution equations
\begin{subequations}\label{eqn:kaz_eqns}
\begin{align}
    \partial_t \rndB &= \rndb^i\rndb^m \partial_m \rndu^i \rndB - \eta \rndk^2 \rndB, \\*
    \partial_t \rndk_m &= - \partial_m \rndu^i \rndk_{i}, \\*
\partial_t \rndb^i &= \rndb^m \partial_m \rndu^i - \rndb^l \rndb^m \partial_m \rndu^l \rndb^i,
\end{align}
\end{subequations}
where $\nrndb^i$ denotes the $i$th component of the magnetic-field unit vector $\eb$ and tildes are used in this section to denote random variables [cf.~\eqref{eqn:kaz_base}]. We assume that $\tilde{\bb{u}}$ is approximately linear in space,\footnote{\,Such a Taylor expansion of the flow is only a good approximation when both the anisotropic (i.e., Braginskii) and isotropic magnetic Prandtl numbers are large.} white in time, and anisotropic with respect to the local magnetic-field direction:
\begin{equation}\label{eqn:ap:velo}
    \rndu^i(t,\bb{x}) = \rnds^i_m (t) x^m,
\end{equation}   
where the rate-of-strain tensor $\rnds^i_m \doteq \partial_m \rndu^i$ has the two-time correlation function
\begin{equation}\label{eqn:ttcorr}
    \ea{\rnds^i_m(t)\rnds^j_n(t')} = \tilde{\Gamma}^{ij}_{mn}(t)\delta(t-t') ,
\end{equation}
where an overline denotes the ensemble average and
\begin{align}\label{eqn:Gammaijmn}
    \tilde{\Gamma}^{ij}_{mn}(t) &= \kappa_2 \Bigl[\delta^{ij}\delta_{mn} + a(\delta^i_m \delta^j_n + \delta^i_n \delta^j_m) + \chi_1 \delta^{ij}\rndb^m\rndb^n + \chi_2 \rndb^i\rndb^j \delta_{mn} + \chi_3 \rndb^i\rndb^j \rndb^m \rndb^n \nonumber\\*
        \mbox{} &\quad +\chi_4(\delta^i_m \rndb^n \rndb^j + \delta^i_n \rndb^m \rndb^j + \rndb^i\rndb^m \delta^j_n + \rndb^i\rndb^n \delta^j_m)\Bigr]
\end{align}
\noindent is the general fourth-rank tensor that is anisotropic with respect to the magnetic-field direction and symmetric under interchange of $i$, $m$ and $j$, $n$. Here, $\kappa_2$ is the second-order coefficient in the Taylor expansion of the field-anisotropic velocity correlation tensor defined by \eqref{eqn:kazantsev}:
\begin{align}
    \tilde{\kappa}^{ij}(\bb{y}) &= \kappa_0 \delta^{ij} - \frac{1}{2} \kappa_2 \Bigl[ y^2 \delta^{ij} + 2a y^i y^j + \chi_1 \delta^{ij} (y_m\rndb^m)^2 + \chi_2 \rndb^i \rndb^j y^2 + \chi_3 \rndb^i \rndb^j (y_m\rndb^m)^2
    \nonumber\\*
    \mbox{} &+ 2\chi_4 (y_m\rndb^m) \bigl( \rndb^j y^i + \rndb^i y^j \bigr) \Bigr] + \dots ,
\end{align}
and the constants $\chi_i$ and $a$ parametrize the rate of strain. Two of them can be fixed by assuming an incompressible flow: $\tilde{\Gamma}^{ij}_{in}=0$ and $\tilde{\Gamma}^{ij}_{mj} = 0$, so
\begin{equation}\label{eqn:kappa_incomp}
    a = -\frac{1+\chi_4}{1+d} , \quad \chi_1+\chi_2+\chi_3=-(d+2)\chi_4,
\end{equation}
where $d$ is the dimensionality of the system. The isotropic case is recovered when $\chi_i = 0$ for all $i$. The tensors $\tilde{\Gamma}^{ij}_{mn}$ and $\tilde{\kappa}^{ij}(\bb{k})$ \dblbrck{see \eqref{eqn:kappa_tensor}} are related by 
\begin{equation}\label{eqn:kaz_gamma}
 \tilde{\Gamma}^{ij}_{mn}(t) = \int \frac{\od^d \bb{k}}{(2\pi)^d} \, k_mk_n \tilde{\kappa}^{ij}(\bb{k}).
\end{equation}
Formally, $\tilde{\Gamma}^{ij}_{mn}(t)$ depends on the random quantity $\rndb_i$, and so the averaging performed in~\eqref{eqn:ttcorr} is understood to be a time average taken over times longer than the velocity correlation time (which is zero), but shorter than the time over which $\eb\eb$ changes significantly.  This is a reasonable simplification if $\eb$ aligns itself along the direction of the maximal Lyapunov exponent of the flow, which occurs exponentially fast in time~\citep{Goldhirsch}. Therefore, in the (Lagrangian) frame of the flow, the vector $\eb$ changes slowly in time.

We aim to obtain an evolution equation for the magnetic-energy spectrum $M(k)$. This can be done by first deriving an evolution equation for the joint probability density function of the magnetic field $\bb{B}$ and its wavenumber $\bb{k}$,
\begin{equation}\label{eqn:PtBkb}
   \mc{P}(t;B,\bb{k},\eb) = \ea{\tilde{\mc{P}}} \doteq \ea{\delta(B-\rndB(t))\delta(k_m - \rndk_m(t))\delta(\nrndb^i - \rndb^i(t))},
\end{equation} 
and then taking the appropriate moments; to wit,
\begin{equation}
    M(t,k) = \int \od \Omega_\bb{k}\, k^2 \int \od^3 \eb \int\od B\, B^2  \mc{P}(t; B,\bb{k},\eb),
\end{equation}
where $\od \Omega_\bb{k}$ is an element of solid angle in wavenumber space.
Taking the time derivative of \eqref{eqn:PtBkb} and using \eqref{eqn:kaz_eqns}, we get

\begin{align}\label{eqn:dPdt}
\pD{t}{\mc{P}} &= \ea{\pD{t}{\tilde{\mc{P}}}} = \ea{\left[-\pD{t}{\rndB(t)} \pD{B}{} - \pD{t}{\rndk_m(t)} \pD{k_m}{} - \pD{t}{\rndb^i(t)} \pD{\nrndb^i}{} \right]\tilde{\mc{P}}}
\nonumber\\*
\mbox{} &= -\ea{\biggl[ \rndb^i(t) \rndb^m(t) \rnds^i_m(t) \rndB(t) \pD{B}{}  - \eta \rndk^2(t)\rndB(t) \pD{B}{}  -   \rnds^i_m (t)\rndk_i(t) \pD{k_m}{}}
\nonumber\\*
\mbox{} & \quad
\ea{\mbox{} + \rnds^i_m(t) \rndb^m(t) \pD{\nrndb^i}{} - \rnds^l_m(t) \rndb^l(t) \rndb^m(t) \rndb^i(t) \pD{\nrndb^i}{}\biggr]\tilde{\mc{P}}}
\nonumber\\*
\mbox{} &= - \hat{L}_i^m \ea{\rnds^i_m(t) \tilde{\mc{P}}} + \eta k^2 \pD{B}{}B \mc{P},
\end{align}
where
\begin{equation}
\hat{L}^m_i \doteq \pD{B}{} B \nrndb^i \nrndb^m  - \pD{k_m}{} k_i + \pD{\nrndb^i}{} b^m - \pD{\nrndb^l}{} \nrndb^l \nrndb^i \nrndb^m.
\end{equation}
To arrive at the final line of \eqref{eqn:dPdt}, the identity $a\,\delta(a-b) = b\,\delta(a-b)$ was used. Note that everything in the square brackets in the final line is non-random. To perform the remaining ensemble average, we make use of the Furutsu--Novikov formula \citep{furutsu,Novikov}, which generalizes Gaussian splitting to functions:
\begin{equation}\label{eq:furutsu-novikov}
\ea{\rnds_m^i(t) \tilde{\mc{P}}(t)} = \int_0^t \od t' \ea{\rnds^i_m(t) \rnds^j_n(t')} \, \ea{\frac{\delta \tilde{\mc{P}}(t) }{\delta \rnds^j_n(t')}} .
\end{equation}
The functional derivative with respect to $\rnds^j_n(t')$ can be calculated by formally integrating $\partial \tilde{\mc{P}}(t)/\partial t$ with respect to time:
\begin{align}
\frac{\delta \tilde{\mc{P}}(t)}{\delta \rnds^j_n(t')} &=- \hat{L}^m_i\int_0^{t} \od t'' \, \frac{\delta \rnds^i_m(t'')}{\delta \rnds^j_n(t')} \tilde{\mc{P}}(t'') - \int_0^{t} \od t'' \,  \left(\hat{L}^m_i \rnds^i_m(t'') - \eta k^2 \pD{B}{} B\right)\!\cancel{\frac{\delta\tilde{\mc{P}}(t'')}{\delta \rnds^j_n(t')} }
\nonumber\\* 
\mbox{} &= -\hat{L}^n_j\int_0^{t} \od t'' \,\delta(t'-t'')\tilde{\mc{P}}(t'') = -\hat{L}^n_j\tilde{\mc{P}}(t') .
\end{align}
The last term in first line is dropped because it disappears when $t=t'$, owing to causality (i.e.~$\tilde{\mc{P}}$ cannot depend on future values of $\rnds$).
Using this result in \eqref{eq:furutsu-novikov} alongside \eqref{eqn:ttcorr}, we find
\begin{equation}
\ea{\rnds_m^i(t) \tilde{\mc{P}}(t)} = -\frac{1}{2} \hat{L}^n_j \tilde{\Gamma}_{mn}^{ij}(t)  \ea{\tilde{\mc{P}}(t)} \approx-\frac{1}{2} \hat{L}^n_j  \ea{\tilde{\Gamma}_{mn}^{ij}(t) \tilde{\mc{P}}(t)} = -\frac{1}{2} \hat{L}^n_j \Gamma_{mn}^{ij}  \ea{\tilde{\mc{P}}(t)} ,
\end{equation}
where $\Gamma^{ij}_{mn}$ is defined similarly to~\eqref{eqn:Gammaijmn} but with the substitution $\rndb \rightarrow \nrndb$.
The second (approximate) equality relies on the assumption of a slowly varying $\eb\eb$, as was used to obtain~\eqref{eqn:ttcorr}. Recall that, in the striped volume-filling region of a magnetic fold, only the \emph{sign} of $\eb$ changes in space, and so the dyad $\eb\eb$, a quadratic quantity, changes on a much longer timescale than either the magnetic energy or the  turbulent velocity. Once $\tilde{\Gamma}_{mn}^{ij}$ is brought within the average, it becomes $\Gamma_{mn}^{ij}$ via the delta functions in the definition of~$\tilde{\mc{P}}$.
The result of these manipulations is a closed equation for the joint probability density function:
\begin{equation}\label{eqn:dPdt_averaged}
    \pD{t}{\mc{P}} = \frac{1}{2} \hat{L}^m_i \hat{L}^n_j {\Gamma}^{ij}_{mn} \mc{P} + \eta k^2 \pD{B}{} B\mc{P} .
\end{equation}

Equation~\eqref{eqn:dPdt_averaged} can be greatly simplified by noting that $\mc{P}(B,\bb{k},\eb)$ must have the following factorization:
\begin{equation}\label{eqn:Pfactor}
    \mc{P}(B,\bb{k},\eb) = \delta(|\eb|^2 -1) \delta(\eb\bcdot \bb{k})   \Punorm(B,k) .
\end{equation}
The two delta functions result, respectively, from $\eb$ being a unit vector and from the solenoidality constraint $\eb\bcdot\bb{k} = 0$. The remaining factor in \eqref{eqn:Pfactor}, $\Punorm(B,k)$, is a result of the statistics being homogeneous and the relative alignment of $\eb$ and $\bb{k}$ being fixed.\footnote{\,$\Punorm(B,k)$ receives proper normalization below in \eqref{eqn:Pnorm}, after which its ornamental hat is dropped.} In order to express~\eqref{eqn:dPdt_averaged} in terms of $\Punorm$, we use the chain rule to write
\begin{equation}
\hat{L}_j^n = -k_j \frac{\partial}{\partial k_n} + \left(\nrndb^n \frac{\partial}{\partial \nrndb^j} -\nrndb^j \nrndb^n \nrndb^q \frac{\partial}{\partial \nrndb^q} \right) + \nrndb^j \nrndb^n \left(\frac{\partial}{\partial B}B - d - 2\right),
\end{equation}
where $d$ is the dimensionality of the system, and then calculate the combination $\hat{L}^n_j \Gamma^{ij}_{mn}\mc{P}(B,\bb{k},\eb)$. 
For any function $f$,
\begin{align}
\hat{L}^n_j \delta(\eb\bcdot \bb{k})f &=\delta(\eb\bcdot \bb{k})\hat{L}^n_j  f  - k_j \nrndb^n \delta'(\eb\bcdot\bb{k}) f+ (\nrndb^n k_j - \nrndb^j \nrndb^n \eb \bcdot \bb{k})\delta'(\eb\bcdot\bb{k})f
\nonumber\\*
\mbox{} &= \delta(\eb\bcdot\bb{k}) \hat{L}_j^n f- \nrndb^j \nrndb^n (\eb\bcdot\bb{k})\delta'(\eb\bcdot \bb{k})f
\nonumber\\*
\mbox{} &= \delta(\eb \bcdot \bb{k})(\hat{L}^n_j + \nrndb^n \nrndb^j)f,\label{eqn:Lop1}
\end{align}
where we have used $x \delta'(x) = -\delta(x)$ to obtain the final equality. Similarly,
\begin{equation}
\hat{L}^n_j \delta(|\eb|^2 -1)f = \delta(|\eb|^2-1)(\hat{L}^n_j + 2 \nrndb^n \nrndb^j)f.\label{eqn:Lop2}
\end{equation}
Combining \eqref{eqn:Lop1} and \eqref{eqn:Lop2} leads to
\begin{align}
\hat{L}^n_j\Gamma^{ij}_{mn}\mc{P}  &= \delta(\eb\bcdot\bb{k})\delta(|\eb|^2-1)(\hat{L}^n_j + 3\nrndb^n\nrndb^j)\Gamma^{ij}_{mn}\Punorm(B,k)
\nonumber\\*
\mbox{} &= \delta(\eb\bcdot\bb{k})\delta(|\eb|^2-1)\bigg[-\frac{k_jk_n}{k^2} \Gamma_{mn}^{ij}k \pD{k}{} + \nrndb^j \nrndb^n \Gamma^{ij}_{mn}\left(\pD{B}{} B - d +1\right) 
\nonumber\\*
\mbox{}&\quad + \nrndb^n \pD{\nrndb^j}{\Gamma^{ij}_{mn}} - \nrndb^j\nrndb^n\nrndb^q \pD{\nrndb^q}{\Gamma^{ij}_{mn}} \bigg]\Punorm(B,k). \label{eqn:Lop3}
\end{align}
Taking into account the solenoidality constraint $\eb\bcdot \bb{k} = 0$ imposed by the prefactor in \eqref{eqn:Lop3}, the following combinations in \eqref{eqn:Lop3} may be calculated:
\begin{subequations}
\begin{align}
\frac{k_jk_n}{k^2}\Gamma^{ij}_{mn} &=\kappa_2 \left[a\delta^i_m + (a+1) \frac{k_ik_m}{k^2} + \chi_4 \nrndb^i \nrndb^m \right], 
\\*
\nrndb^j \nrndb^n \Gamma^{ij}_{mn} &= \kappa_2 \bigl[(a+\chi_4)\delta^i_m + (1+a+\chi_1+\chi_2+\chi_3 +3\chi_4)\nrndb^i\nrndb^m \bigr], 
\\*
\nrndb^n \frac{\partial}{\partial \nrndb^j}\Gamma^{ij}_{mn} &= \kappa_2 \bigl[\delta^i_m (\chi_1 + (d+2)\chi_4) 
\nonumber \\*
\mbox{} &\quad + \nrndb^i\nrndb^m(\chi_1+ (d+1)\chi_2 + (d+3)\chi_3 + (d+4)\chi_4) \bigr], 
\\*
\nrndb^j \nrndb^n \nrndb^q \frac{\partial}{\partial \nrndb^q} \Gamma^{ij}_{mn} &= \kappa_2 \bigl[2\chi_4 \delta^i_m + 2(\chi_1+\chi_2 + 2\chi_3 + 3\chi_4)\nrndb^i\nrndb^m \bigr].
\end{align}
\end{subequations}
Using these formulae in \eqref{eqn:Lop3} gives
\begin{align}
\hat{L}^n_j\Gamma^{ij}_{mn}\mc{P}  &= \kappa_2 \delta(\eb\bcdot\bb{k})\delta(|\eb|^2-1)\bigg\{  -\bigg[a\delta^i_m + (a+1)\frac{k_ik_m}{k^2} + \chi_4 \nrndb^i \nrndb^m\bigg] k \frac{\partial}{\partial k}
\nonumber\\*
\mbox{} &\quad + \bigl[(a+\chi_4)\delta^i_m + (1+a+\chi_1+\chi_2+\chi_3+3\chi_4)\nrndb^i\nrndb^m \bigr]\frac{\partial}{\partial B}B 
\nonumber\\*
\mbox{} &\quad + \bigl[-a(d-1) + \chi_1 + \chi_4\bigr]\delta^i_m 
\nonumber\\
\mbox{} &\quad + \bigl[-(1+a)(d-1) -d \chi_1 - (2d-1)\chi_4\bigr] \nrndb^i \nrndb^m \bigg\}\Punorm(B,k).
\end{align}
Further applying the operator $\hat{L}_i^m$ and expending much effort along the same lines gives us the  expression for the first term in \eqref{eqn:dPdt_averaged}:
\begin{align}\label{eqn:Lop4}
\hat{L}_i^m\hat{L}^n_j&\Gamma^{ij}_{mn}\mc{P} \\
&= \bigg\{(2a+1)k^2 \frac{\partial^2}{\partial k^2} + (1+2a+\chi_1 + \chi_2 + \chi_3 + 4\chi_4) \frac{\partial}{\partial B}B\frac{\partial}{\partial B}B 
\nonumber\\*
\mbox{} &\quad -2(a+\chi_4) \frac{\partial}{\partial B}B k \frac{\partial}{\partial k} + \bigl[d+(3d-1)a-\chi_1 - \chi_4\bigr]k \frac{\partial}{\partial k}
\nonumber\\*
\mbox{} &\quad - (1+3a+\chi_1 + 3\chi_4)(d-1)\frac{\partial}{\partial B}B + \bigl[(d-1)a-\chi_1-\chi_4\bigr](d-1)  \bigg\}\Punorm(B,k).
\end{align}
Normalisability of the PDF requires that
\begin{align}\label{eqn:Pnorm}
1 &= \int \od^d\eb \, \delta(|\eb|^2 -1) \int \od^d \bb{k}\, \delta(\eb\bcdot \bb{k}) \int \od B \, \Punorm(B,k)
\nonumber\\*
\mbox{} &= \frac{S_{d-1}S_{d-2}}{2}\int_0^\infty \od k \,k^{d-2} \int \od B \, \Punorm(B,k),
\end{align}
where $S_n$ is the surface area of unit $n$-sphere (e.g.,~$S_1 = 2\upi$, $S_2 = 4\upi$). Taking this normalization into consideration, we define $P(B,k) \doteq S_{d-1}S_{d-2} k^{d-2} \Punorm(B,k)/2$ and use
\begin{subequations}
\begin{align}
k^{d-2}k\frac{\partial}{\partial k} \frac{1}{k^{d-2}}P &= \left[k\frac{\partial}{\partial k} - (d-2)\right]P, 
\\*
k^{d-2}k^2 \frac{\partial^2}{\partial k^2} \frac{1}{k^{d-2}} P &= \left[k^2 \frac{\partial^2}{\partial k^2} - 2(d-2) k \frac{\partial}{\partial k} + (d-2)(d-1)\right]P 
\end{align}
\end{subequations}
to turn \eqref{eqn:dPdt_averaged} into
\begin{align}\frac{\partial P }{\partial t} &= \frac{1}{2}\kappa_2 \bigg \{(1+2a)\frac{\partial}{\partial k} k^2 \frac{\partial}{\partial k} + ( 1+2a  + \chi_1 + \chi_2 + \chi_3 + 4 \chi_4 ) \frac{\partial}{\partial B}B \frac{\partial}{\partial B}B 
\nonumber \\*
\mbox{} &\quad - 2(a+\chi_4) \frac{\partial}{\partial B}B \frac{\partial}{\partial k}k - \bigl[(d-2) + (d-3)a + \chi_1 + \chi_4\bigr] \frac{\partial}{\partial k}k 
\nonumber \\*
\mbox{} &\quad - (1+a+\chi_1 +\chi_4 )(d-1)\frac{\partial}{\partial B}B\bigg\} P +  \eta k^2 \pD{B}{} B P .\label{eq:app_almost}
\end{align}
We now enforce incompressibility. Substituting \eqref{eqn:kappa_incomp} into \eqref{eq:app_almost} leads to the final form of the evolution equation of the joint PDF:
\begin{align}\label{eq:app_there}
\frac{\partial P}{\partial t} &= \frac{\kappa_2}{2(d+1)}\bigg\{ (d-1-2\chi_4)\frac{\partial}{\partial k}k^2 \frac{\partial}{\partial k} + (d-1)(1-d\chi_4)\frac{\partial}{\partial B} B \frac{\partial}{\partial B} B
 \nonumber \\*
 \mbox{} &\quad  +2(1+d\chi_4) \frac{\partial}{\partial B} B \frac{\partial}{\partial k} k - \bigl[(d-1)^2 +4 \chi_4 + (d+1)\chi_1\bigr]\frac{\partial}{\partial k}k 
\nonumber \\*
\mbox{} &\quad -(d-1)\bigl[d(1+\chi_4) + (d+1)\chi_1\bigr] \frac{\partial}{\partial B}B   \bigg\} P +  \eta k^2 \pD{B}{} B P.
\end{align}
The magnetic-energy spectrum is $M(k) = (1/2) \int_0^\infty \od B\, B^2 P(B,k)$. Therefore, taking the $B^2$ moment of \eqref{eq:app_there} leads in three dimensions ($d=3$) to
\begin{equation}\label{eq:app_final}
\pD{t}{M} = \frac{\kappa_2}{4} \left[ (1-\chi_4) \pD{t}{} k^2 \pD{k}{} - 2(2+\chi_1 -2 \chi_4) \pD{k}{} k + 2(5+4\chi_1-3\chi_4 )\right] M - 2\eta k^2 M.
\end{equation}
Equation \eqref{eqn:mod_kazantsev} for the time evolution of the magnetic spectrum written as a function of $\gamma_\perp$, $\sigma_\perp$, and $\sigma_\parallel$ follows from \eqref{eq:app_final} after noting that
\begin{subequations}
\begin{align}
    \gamma_\perp &\doteq \int \frac{\od^3\bb{k}}{(2\upi)^3} \, k_\perp^2 \kappa_\perp = \frac{3-\chi_4}{2}\kappa_2, \\*
    \sigma_\perp &\doteq \frac{1}{\gamma_\perp}\int \frac{\od^3 \,\bb{k}}{(2\upi)^3} k_\parallel^2 \kappa_\perp = \frac{2(1+\chi_1)}{3-\chi_4},\\*
    \sigma_\parallel &\doteq \frac{1}{\gamma_\perp}\int \frac{\od^3\bb{k}}{(2\upi)^3} \,k_\parallel^2 \kappa_\parallel = \frac{1-3\chi_4}{2(3-\chi_4)},
\end{align}
\end{subequations}
where \eqref{eqn:kaz_gamma} has been used alongside \eqref{eqn:kappa_prp} and \eqref{eqn:kappa_prl}. 

\bibliographystyle{jpp}
\bibliography{refs}

\end{document}